\documentstyle[preprint,aps,rmp,epic,eepic,epsf,epsfig]{revtex}
\tightenlines
\begin{document}
\title{Real-space mesh techniques in density functional theory}
\author{Thomas L. Beck}
\address{Department of Chemistry, University of Cincinnati,
Cincinnati, OH  45221-0172}
\maketitle


\noindent
to be published in Reviews of Modern Physics. 

\begin{abstract}
This review discusses
progress in efficient solvers which
have as their foundation a representation in real space, 
either through finite-difference or finite-element 
formulations. The relationship of real-space approaches 
to linear-scaling electrostatics and
electronic structure methods is first
discussed.  Then the basic aspects of real-space representations
are presented.  Multigrid techniques for 
solving the discretized problems are covered;
these numerical schemes allow for highly efficient solution
of the grid-based equations. Applications to problems in electrostatics
are discussed, in particular numerical solutions of Poisson and 
Poisson-Boltzmann equations. Next, methods for solving
self-consistent eigenvalue problems in 
real space are presented; these techniques have been extensively
applied to solutions of the Hartree-Fock and Kohn-Sham 
equations of electronic structure, and to eigenvalue problems
arising in semiconductor and polymer physics. Finally, real-space 
methods have found recent application in computations 
of optical response and excited states in
time-dependent density functional theory, and these 
computational developments are summarized.  Multiscale
solvers are competitive with the 
most efficient available plane-wave
techniques in terms of the number of self-consistency
steps required to reach the ground state, and they
require less work in each self-consistency update on a uniform
grid.  Besides excellent efficiencies, 
the decided advantages of the real-space multiscale approach are
1) the near-locality of each function update, 2) the ability
to handle global eigenfunction constraints 
and potential updates on coarse levels, and 3)
the ability to incorporate adaptive local mesh refinements
without loss of optimal multigrid efficiencies.  
\end{abstract}

\tableofcontents

\section{INTRODUCTION}

\label{sec:intro}

The last decade has witnessed a proliferation in 
methodologies for numerically solving 
large-scale problems in 
electrostatics and electronic structure.   The 
rapid growth has been driven by several factors. First,
theoretical advances in the understanding
of localization properties of electronic systems have justified
at a fundamental level
approaches which utilize localized 
density matrices or orbitals in their
formulation (Kohn, 1996; Ismail-Beigi and Arias, 1998;
Goedecker, 1999).  
Second, a wide variety of 
computational methods have exploited that physical locality,
leading to linear scaling of the computing time 
with system size (Goedecker, 1999).  Third, general algorithms 
for solving electrostatics and eigenvalue problems
have been improved or newly developed including
particle-mesh methods (Hockney and
Eastwood, 1988; Darden {\it et al.}, 1993; Pollock and Glosli, 1996), 
fast-multipole approaches (Greengard, 1994),
multigrid techniques (Brandt, 1977, 1982, 1984; Hackbusch, 1985), and 
Krylov subspace and related 
algorithms (Booten and van der Vorst, 1996).  
Last, and perhaps not least, the ready 
availability of very fast processors for low 
cost has allowed for quantum modeling of systems
of unprecedented size. These calculations
can be performed on workstations or workstation clusters, 
thus creating opportunities for a wide range of 
researchers in fields both inside and outside of computational
physics and chemistry (Bernholc, 1999).  Several 
monographs and collections of reviews
illustrate the great variety of problems recently 
addressed with 
electrostatics and electronic structure methods 
(Gross and Dreizler, 1994; Bicout and Field, 1996;
Seminario, 1996;
Springborg, 1997; Banci and 
Comba, 1997; Von Rague Schleyer, 1998; Jensen, 1999; Hummer
and Pratt, 1999). 

This review examines one subset of these new computational
methods, namely real-space techniques.  Real-space methods
can loosely be categorized as one of three types: finite
differences (FD), finite elements (FE), or wavelets.  All three lead
to structured, very sparse matrix representations of the 
underlying differential equations on meshes in real space.  
Applications of 
wavelets in electronic structure calculations have been thoroughly
reviewed recently (Arias, 1999) and will therefore not be addressed 
here.  This article discusses the fundamentals of 
FD and FE solutions of Poisson and nonlinear Poisson-Boltzmann
equations in electrostatics and self-consistent
eigenvalue problems in electronic structure.  As 
implied in the title, the primary focus is on calculations
in density functional theory (DFT); real-space methods are
in no way limited to DFT, but since DFT calculations
comprise a dominant theme in modern electrostatics and 
electronic structure, the discussion here will mainly be restricted
to this particular theoretical approach.  

\begin{figure}
\centerline{\epsfig{file=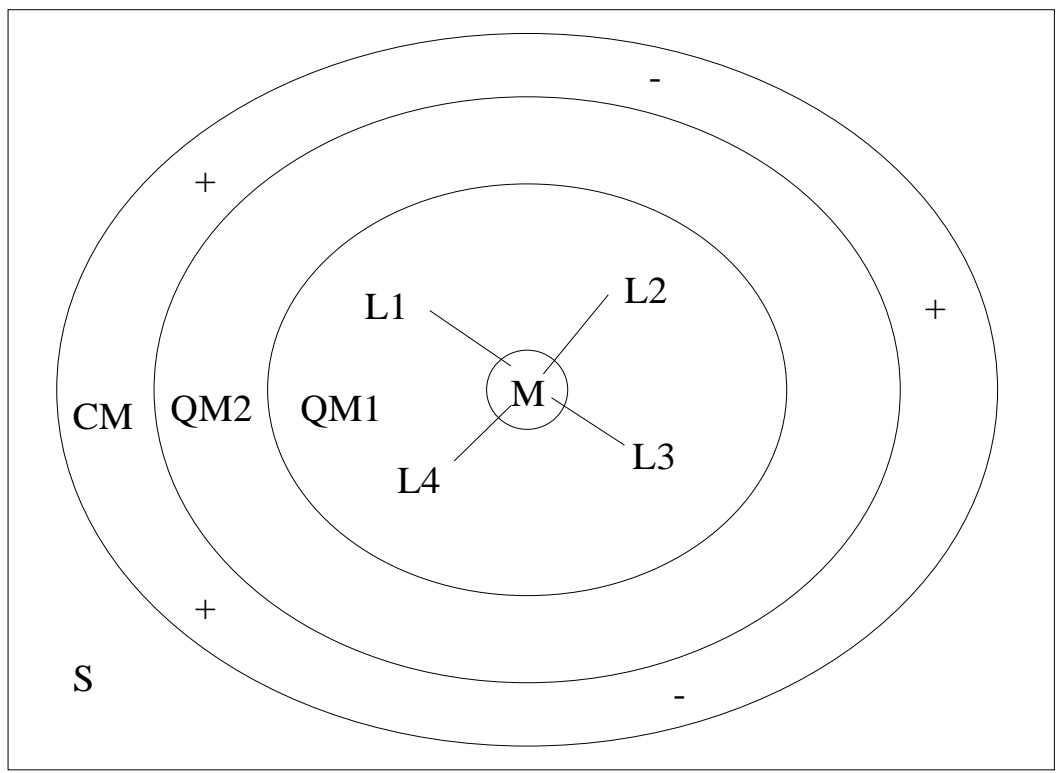,height=5.5in,width=4.in,angle=270}}
\vspace{10pt}
\caption{Schematic diagram for real-space treatment of a transition 
metal ion in a protein. The metal ion is labeled M, and the ligands
are labeled L1-4. The electronic structure is treated self-consistently
in the QM1 zone, while the orbitals are fixed in QM2. The fixed charges
on the protein are located in the CM region. The solvent (typically
water) may be
included via a continuum dielectric model in the S zone. }
\label{fig:protein}
\end{figure}

Consider a physical system for which local approaches such 
as real-space methods
are appropriate: a transition metal ion bound to several
ligands embedded in a protein.  These systems are 
of significance in a wide range of biochemical
mechanisms (Banci and Comba, 1997). Treating the entire system
with {\it ab initio} methods is presently impossible. 
However, if the primary interest is in the nature of the 
bonding structure and electronic states of the transition
metal ion, one can imagine a three-tier 
approach (Fig.\ \ref{fig:protein}). The central
region, including the metal ion and the ligands, is treated
with an accurate quantum method such as DFT.  A second neighboring
shell is represented quantum mechanically but is not allowed to
change during self-consistency iterations.  The wavefunctions
in the central zone must be orthogonalized to the fixed
orbitals in the second region. Finally, the very
distant portions of the protein are fixed 
in space and treated classically; the main
factors to include from the far locations are the electrostatic 
field from charged or partially charged groups 
on the protein and the response of  
the solvent (typically treated as a dielectric
continuum).  Real-space
methods provide a helpful language for representing such a 
problem.  The real-space grid can be refined to account for
the high resolution necessary around the metal ion and 
can be adjusted for a coarser treatment further away. There is
clearly no need to allow the metal and ligand orbitals to extend far
from the central zone, so a localized representation is
advantageous. Also, 
the electrostatic potential can be generated over the entire
domain (quantum, classical, and solvent zones)
with a single real-space solver without requiring special
techniques for matching conditions in the 
various regions. The same ideas could be applied to
defects in a covalent solid or impurity atoms in a cluster. 

In order to place the real-space methods in context,
we first briefly examine other computational approaches.
The plane-wave 
pseudopotential method has proven to be a powerful
technique for locating the electronic ground state
for many-particle systems in condensed phases
(Payne {\it et al.}, 1992). In this method the 
orbitals are expanded in the nonlocal
plane-wave basis.  The core states are removed
via pseudopotential methods which allow for
relatively smooth valence functions in the core region
even for first-row and transition elements
(Vanderbilt, 1990). Therefore, a reasonable
number of plane-waves can be used to 
accurately represent 
most elements important for materials
simulation.  Strengths of this method include the
use of efficient
fast Fourier transform (FFT) techniques
for updates of the orbitals and electrostatic
potentials, lack of dependence of the basis on atom
positions, and the rigorous control of 
numerical convergence
of the approximation with decrease in wavelength of
the highest Fourier mode.  Algorithmic advances have
led to excellent convergence characteristics of
the method in terms of the number of required 
self-consistency steps (Payne {\it et al.}, 
1992; Hutter {\it et al.}, 1994; Kresse and
Furthm\"{u}ller, 1996); only 5-10 self-consistency iterations 
are required to obtain tight convergence of
the total energy, even for metals.  
In spite of the numerous
advantages of this approach, there are 
restrictions centered around the nonlocal
basis set.  Even with the advances in 
pseudopotential methods, strong variations
in the potential occur in the core regions
(especially for first-row and transition
elements),
and local refinements would allow for 
smaller effective energy cutoffs away 
from the nuclei.  This issue has been 
addressed by the development
of adaptive-coordinate plane-wave methods
(Gygi, 1993). If any information is required
concerning the inner-shell electrons, plane-wave 
methods suffer severe difficulties. 
Of course such states can be represented with 
a sufficient number of plane waves
(Bellaiche and Kunc, 1997), but 
the short-wavelength modes required
to build in the rapidly varying local structure
extend over the entire domain to portions of
the system where such resolution is not necessary.  
Also, for localized systems like molecules, 
clusters, or surfaces, nontrivial
effort is expended to accurately 
reproduce the vacuum; the zero-density
regions must be of significant
size in order to minimize spurious effects in 
a supercell representation.  In addition, charged
systems create technical difficulties since a 
uniform neutralizing background needs to 
be properly added and subtracted in computations
of total energies.  Lastly, without 
special efforts to utilize localized-orbital
representations,
the wavefunction orthogonality step scales
as $N^3$, where $N$ is the number of electrons. 

In quantum chemistry, localized basis sets built
from either Slater-type orbitals (STOs) or
Gaussian functions have predominated in the 
description of atoms and molecules (Szabo and 
Ostlund, 1989; Jensen, 1999).  The molecular orbitals
are constructed from linear combinations
of the atomic orbitals (LCAO).  An accurate
representation can be obtained with 
less than thirty Gaussians for a first-row atom. 
In relation to plane-wave expansions, the localized nature
of these basis functions is more in line with
chemical concepts.  With
STOs or other numerical orbitals, the multicenter integrals 
in the Hamiltonian must be
evaluated numerically, while with a Gaussian
basis, the Coulomb integrals are available analytically.
The `price' for using Gaussians is that more
basis functions are required to accurately 
describe the electron states, since they 
do not exhibit the correct behavior at
either small or large distances from the 
nuclei. Techniques such as direct inversion
in the iterative subspace (DIIS) have been developed to 
significantly accelerate the convergence behavior
of basis-set self-consistent electronic structure
methods (Pulay, 1980, 1982; Hamilton and 
Pulay, 1986).\footnote{One must be cautious, however,
to properly initialize the orbitals in the 
DIIS procedure. See Kresse
and Furthm\"{u}ller (1996).}
The LCAO methods have led to a dramatic
growth in accurate calculations on molecules
with up to tens of atoms.  It is now common 
to see papers devoted to detailed comparisons
of experimental results and electronic structure
calculations on systems with more than one
hundred electrons (Rodriguez
{\it et al.}, 1998).   Often in basis-set 
calculations, care must be taken to account
for basis-set superposition errors which arise
due to overlap of
nonorthogonal atom centered functions for composite
systems. Also, linear dependence is a problem
for very large basis sets chosen to minimize
the errors.  These factors lead 
to difficulty in obtaining the basis-set limit
for a given level of theory.\footnote{Moncrieff and 
Wilson (1993) presented a comparative analysis of
FD, FE, and Gaussian basis-set computations 
for first row diatomics
to assess their relationship.}  
The scaling of
basis-set methods can be severe, but recent
developments (see Section \ref{sec:linsc}) have
brought the scaling down to linear for large
systems. 

With the successes of plane-wave and 
quantum chemical basis functions, what is the 
motivation to search for alternative algorithms?  
Ten years ago, a
review article discussing the relevance of 
Gaussian basis-set calculations for lattice
gauge theories argued for the utilization
of Gaussian basis sets in place of grids (Wilson, 1990).
The author states (concerning the growth of quantum
chemistry): ``The most important algorithmic advance was
the introduction of systematic algorithms using 
analytic basis functions in place of numerical grids,
first proposed in the early 1950s."  The point was 
illustrated by examination of core states for carbon:
only a few Gaussians are required (with variable exponential
parameters), while up to $8 \times 10^6$ grid points
are necessary for the same energy resolution
on a uniform mesh. What developments have occurred
over the last decade which could begin to overcome such 
a large disparity in computational effort?

This review seeks to answer the above question by summarizing recent
research on real-space mesh techniques. To locate them in 
relation to plane-wave expansions and LCAO methods, some
general features are introduced here and further developed
throughout the article.  The representation
of the physical problems is simple: the potential operator
is diagonal in coordinate space and the Laplacian is nearly local,
depending on the order of the approximation.
The near-locality makes real-space methods
well suited for incorporation into linear-scaling approaches.  It
also allows for relatively straightforward domain-decomposition
parallel implementation.  Finite or charged systems are
easily handled.  With higher-order FD and FE approximations,
the size of the overall domain is substantially reduced
from the estimate above.  Adaptive mesh refinements 
or coordinate transformations can be employed to gain resolution
in local regions of space, further reducing the grid overhead.
Real-space pseudopotentials result in smooth valence functions 
in the core region, again leading to smaller required grids.
As mentioned above, the grid-based matrix representation 
produces structured and highly banded matrices, in contrast to 
plane-wave and LCAO expansions (Payne {\it et al}, 1992;
Challacombe, 2000). 
These matrix equations can be
rapidly solved with efficient multiscale 
(or other preconditioning) techniques.
However, while more banded than
LCAO representations, the overall
dimension of the Hamiltonian is substantially 
higher.\footnote{To provide a crude estimate of this point,
a 4th-order FD Hamiltonian on a 65$^3$ mesh leads to
roughly 0.005\% nonzero elements or 
$\approx 3.6 \times 10^6$ total terms; with 2000
basis functions in an STO-3G/LDA water cluster calculation, about
10\% of the elements are nonzero implying $4 \times 10^5$ remaining
matrix terms. See Millam and Scuseria (1997).}
In a sense, the real-space methods are closely linked  
to plane-wave approaches: they are both `fully numerical' methods 
with one or at most a few parameters controlling the
convergence of the approximation, for example the grid spacing
$h$ or the wavevector of the highest mode $k$.\footnote{The expression 
`fully numerical' is somewhat misleading as all the
methods discussed here employ some combination
of analytical and numerical procedures. 
A more accurate statement
is that the representations are more systematic than in
the LCAO approach.} 
On the other hand, the LCAO methods 
employ a better physical representation of 
the orbitals (thus requiring fewer basis functions);
attached with this 
representation, however, are some of the 
problems discussed above related to the art of constructing
nonorthogonal, atom- or bond-centered basis sets. 
The purposes of this paper are 1) to provide a basic introduction
to real-space computational techniques, 
2) to review their recent applications to chemical
and physical problems,
and 3) to relate the methods 
to other commonly used numerical approaches 
in electrostatics and electronic 
structure. 

The numerical problems addressed in this review can be categorized
into four types in order of increasing complexity: 

\begin{eqnarray}
\nabla^2 \phi({\bf r}) & = & ?\hspace{.5cm} ;\hspace{1.cm} \mbox{Real-Space Laplacian} \\
\nabla^2 \phi({\bf r}) & = & f({\bf r})\hspace{.5cm} ;\hspace{1.cm}  \mbox{Poisson} \\
\nabla^2 \phi({\bf r}) & = & f({\bf r}, \phi)\hspace{.5cm} ;\hspace{1.cm} \mbox{Poisson-Boltzmann} \\
\nabla^2 \phi({\bf r}) + v({\bf r},\phi)\phi({\bf r}) & = & \lambda\phi({\bf r})\hspace{.5cm} ;\hspace{1.cm} \mbox{Eigenvalue}
\end{eqnarray}

\noindent
The first expression symbolizes the generation of the Laplacian 
on the real-space grid. The second is the linear, elliptic Poisson
equation. The third is the nonlinear Poisson-Boltzmann equation 
of electrostatics which describes the motion of small counterions
in the field of fixed charges.  The final equation is an 
eigenvalue equation such as the self-consistent Schr\"{o}dinger
equation occurring in electronic structure.   Note that both the
third and fourth equations are nonlinear. The Poisson-Boltzmann 
equation includes exponential driving terms on the rhs. The 
self-consistent eigenvalue problem is `doubly nonlinear': one
must solve for both the eigenvalues and eigenfunctions, and the
potential generally depends nonlinearly on the eigenfunctions. 
The multigrid method allows for solution of both linear and
nonlinear problems with similar efficiencies. 

The article is organized into several sections beginning
with background discussion and then following
the order of problems listed above.  Section \ref{sec:dft} 
introduces the central equations 
of DFT for electronic structure and charged classical  
systems.  Section \ref{sec:linsc} reviews 
developments in linear-scaling computational algorithms 
and discusses their relationship
to real-space methods. Section \ref{sec:fdfe} presents
the fundamental aspects
of representation in real space 
by examination of Poisson problems.  
Section \ref{sec:mg} discusses 
multigrid methods for efficient solution 
of the resulting matrix representations.
Section \ref{sec:es} summarizes recent advances in 
electrostatics computations
in real space including both
Poisson and nonlinear Poisson-Boltzmann solvers.  Applications
in biophysics are illustrated with
several examples.  
Section \ref{sec:ks} discusses real-space eigenvalue 
methods for self-consistent problems
in electronic structure. 
Section \ref{sec:tddft} summarizes recent
computations of optical response properties and
excitation energies with real-space methods. The 
review concludes with a short summary 
and discussion of possible future directions for research.   

\section{DENSITY FUNCTIONAL THEORY}

\label{sec:dft}

Motivated by the fundamental Hohenberg-Kohn
theorems (Hohenberg and Kohn 1964) of DFT, 
Kohn and Sham (1965) developed
a set of accessible one-electron  self-consistent 
eigenvalue equations. These equations have  
provided a practical tool for realistic 
electronic structure computations on a vast
array of atoms, molecules, and materials (Parr and Yang, 1989).  The 
Hohenberg-Kohn theorems have been extended to finite-temperature
quantum systems by Mermin (1965) and to purely classical 
fluids in subsequent work (Hansen and McDonald, 1986; Ichimaru, 1994). 
An integral formulation of electronic structure has also been discovered
in which the one-electron
density is obtained directly without the introduction
of orbitals (Harris and Pratt, 1985; Parr and Yang, 1989). This approach
is in the spirit of the original Hohenberg-Kohn theorems, but
to date this promising theory
has not been used extensively in numerical studies. 
This section reviews the basic equations of DFT 
for electronic structure 
and charged classical systems. These equations 
provide the background for discussion
of the real-space numerical methods.   

\subsection{Kohn-Sham equations}

\label{subsec:kseqns}

The Kohn-Sham self-consistent 
eigenvalue equations for electronic structure can be written
as follows (atomic units are assumed throughout):

\begin{equation}
[- \frac{1}{2} \nabla^2 + v_{eff}({\bf r}) ] \psi_i({\bf r}) = \epsilon_i \psi_i({\bf r}),
\label{eq:ks}
\end{equation}

\noindent
where the density-dependent effective potential is
 
\begin{equation}
 v_{eff}({\bf r}) = v_{es}({\bf r})+ v_{xc}([\rho({\bf r})]; {\bf r}) .
\label{eq:veff} 
\end{equation}

\noindent
The classical electrostatic potential $v_{es}({\bf r})$ is due 
to both the electrons and nuclei, and the (in principle) exact
exchange-correlation potential 
$v_{xc}([\rho({\bf r})]; {\bf r})$ 
incorporates all nonclassical effects.
The exchange-correlation potential includes a kinetic
contribution since the expectation value of the Kohn-Sham
kinetic energy is that for a set of {\it non-interacting} electrons
moving in the one-electron effective potential.   
The electron density, $\rho({\bf r})$, is obtained 
from the occupied orbitals (double occupation is assumed
here):

\begin{equation}
\rho({\bf r}) = 2 \sum_{i=1}^{{N_e}/2} |\psi_i({\bf r})|^2.
\label{eq:rho}
\end{equation}

\noindent
The electrostatic portion of the potential for a system
of electrons and nuclei (Hartree potential
plus nuclear potential) is given by

\begin{equation}
v_{es}({\bf r}) = \int \frac{\rho({\bf r'})}{| {\bf r} - 
{\bf r'} |} d {\bf r'} -
\sum_{i=1}^{N_n} \frac{Z_i}{| {\bf r} - {\bf R}_i |} .
\label{eq:ves}
\end{equation}

\noindent
This potential can be obtained by numerical
solution of the Poisson equation:

\begin{equation}
\nabla^2 v_{es}({\bf r}) = - 4 \pi \rho_{tot} ({\bf r}),
\label{eq:poisson}
\end{equation}

\noindent
where $\rho_{tot}({\bf r})$ is the total charge density  
due to the electrons and nuclei.

If the exchange-correlation potential is taken as a local
{\it function} (as opposed to functional) of the density
with the value the same as for a uniform electron gas, the 
approximation is termed the {\it local density approximation} (LDA). 
Ceperley and Alder (1980) determined the 
exchange-correlation energy for the uniform electron gas 
numerically via Monte Carlo simulation.
The data have been parametrized in various ways
for implementation in computational algorithms (see, for example,
Vosko {\it et al.}, 1980).  The LDA
theory has been extended to handle spin-polarized 
systems (Parr and Yang, 1989). 
The LDA yields results with accuracies comparable to or
often superior to Hartree-Fock, but generally leads to 
overbinding in chemical bonds among other deficiencies. 
One obtains the Hartree-Fock approximation 
if the local exchange-correlation
potential in Eq.\ (\ref{eq:veff}) is replaced by the nonlocal
exact exchange operator.  

In recent years, a great deal of effort has gone into developing
more accurate exchange-correlation
potentials (see Jensen, 1999, for a review).  
These advances involve both gradient expansions
which incorporate information from electron density
derivatives and hybrid methods which include some degree of exact 
Hartree-Fock exchange.  With the utilization of these 
modifications, results of chemical accuracy can be
obtained.  Since the main focus of this review
is on numerical approaches for solving the self-consistent
equations, we do not further examine these developments. 

Pseudopotential techniques allow for the removal 
of the core electrons. The 
valence electrons then move in a 
smoother (nonlocal) potential in the 
core region while exhibiting behavior the same as in an
all-electron calculation outside the core.  
Recently developed real-space versions of the pseudopotentials
allow for computations on meshes (Troullier and Martins, 1991a,
1991b; Goedecker {\it et al.}, 1996; 
Briggs {\it et al.}, 1996). Inclusion of 
the pseudopotentials substantially reduces the computational overhead
since fewer orbitals are treated explicitly
and the required mesh resolution can be coarser.
However, truly local mesh-refinement techniques may allow
for the efficient inclusion of core electrons when necessary
(see Sections \ref{subsubsec:mrt} and \ref{subsec:scmrt}). 

Self-consistent solution of the Kohn-Sham equations [Eq.\
(\ref{eq:ks})] for fixed nuclear locations 
is conceptually straightforward.  An initial
guess is made for the orbitals.  This yields an electron 
density from which the effective potential is 
constructed by solution of the Poisson equation and 
generation of the exchange-correlation potential. 
The eigenvalue equation is solved with the current effective
potential [Eq.\ (\ref{eq:veff})], resulting in a new
set of orbitals.  The process is repeated until the
density or total energy change only to within some
desired tolerance.  Alternatively, the total energy
can be minimized variationally using a technique
such as conjugate gradients (Payne {\it et al.}, 1992);
the orbitals at the minimum correspond to those from
the iterative process described above.  

\subsection{Classical DFT}

\label{subsec:cmdft}

The ground-state theory discussed above has been extended 
to finite-temperature quantum and classical systems and 
has found wide use in the theory of fluids 
(Rowlinson and Widom, 1982; Hansen and McDonald, 1986; Ichimaru, 1994).
Here I discuss the formulation for systems of charged point
particles (mobile ions) moving in the external potential produced by
other charged particles in the solution 
(for example, colloid spheres
or cylinders).  The solvent is assumed to be a uniform
dielectric with dielectric constant $\epsilon$
in these equations. The free energy for an ion gas 
of counterions can be
written as the sum of an ideal term,
the energy of the mobile ions in the 
external field due to the fixed colloid particles (this term
incorporates both the electrostatic potential from the fixed charges
on the colloids and an excluded-volume potential),

\begin{equation}
F_{ext} = q \int d{\bf r} \rho_m({\bf r}) V_{ext} ({\bf r},
\{ {\bf R}_j \} ) ,
\label{eq:cmdftext }
\end{equation}

\noindent
the Coulomb interaction potential energy of the mobile ions with 
each other, and a correlation free energy. 
The mobile-ion
density $\rho_m ({\bf r})$ is the number density,
not the charge density,
in the solution. The charge on the counterions is $q$, and 
the approximate correlation free energy typically 
assumes a local density approximation for a one-component plasma. 
Thus the theory includes ion correlations, but the 
approximation is not systematically refineable, just
as in the Kohn-Sham LDA equations.  L\"{o}wen (1994) utilized
this free energy functional in Car-Parrinello-type simulations
(Car and Parrinello, 1985)
of charged rods with surrounding counterions.

The equilibrium distribution is obtained by taking 
the functional derivative of the free energy with respect
to the mobile-ion density and setting it to zero.
It is clear that, if the correlation term is set to 
zero, the equilibrium density of the mobile counterions
is proportional to the Boltzmann factor of the sum
of the external and mobile ion Coulomb potentials:

\begin{equation}
\rho_m({\bf r}) \sim \exp \{ -\beta q[V_{ext}({\bf r})
+ \phi_m ({\bf r})]\}.
\label{eq:cmdftrho}
\end{equation}

\noindent
The potential $\phi_m ({\bf r})$ is that due
to the mobile ions only and $\beta = (kT)^{-1}$. 

Since the total charge (fixed charges on colloid particles
and mobile ion charges) at equilibrium must satisfy the Poisson
equation, the following nonlinear differential equation
results for the equilibrium distribution of the mobile ions
in the absence of correlations.
The treatment is generalized here to account for the possibility
of additional salt in the solution and a dielectric constant
that can vary in space (Coalson and Beck, 1998):

\begin{equation}
\nabla  \cdot
(\epsilon ({\bf r}) \nabla \phi ({\bf r})) = - 4\pi [
\rho_f ({\bf r}) + q \bar n_+ e^{-\beta q\phi({\bf r})-v({\bf r})}
- q \bar n_- e^{\beta q\phi({\bf r})-v({\bf r})} ] ,
\label{eq:pb}
\end{equation}

\noindent
where $\phi({\bf r})$ is the total potential due to the fixed 
colloid charges and mobile ions, $\rho_f({\bf r})$ is the
charge density of the fixed charges on the colloids,
$\bar n_+$ and $\bar n_-$ are
the bulk equilibrium ion densities at infinity
(determined self-consistently
so as to conserve charge in the region of interest), 
and $v({\bf r})$ is a very large positive 
excluded-volume potential which prevents penetration of the 
mobile ions into the colloids.  Fushiki (1992) performed
molecular dynamics simulations of charged colloidal
dispersions at the Poisson-Boltzmann level; the 
nonlinear Poisson-Boltzmann equation was solved
numerically at each time step with FD
techniques.  

An alternative elegant statistical mechanical theory for
the ion gas has been formulated (Coalson and Duncan, 1992).
It uses field theoretic techniques to convert the
Boltzmann factor for the ion interactions into a 
functional-integral representation of the partition
function.  The Poisson-Boltzmann-level theory results
from a saddle-point approximation to the 
functional integral.  The distinct advantage of
this theory is that correlations can be systematically
included by computing the corrections to the 
mean-field approximation via loop expansions.  
However, in practice the corrections are 
computationally costly for real-space grids 
of substantial size.  This theory was
used in simulations of colloids 
(Walsh and Coalson, 1994), and the 
deviations from mean-field theory were 
investigated.  For realistic concentrations
of monovalent background ions, the corrections are
often small in magnitude, thus justifying
the Poisson-Boltzmann-level treatment. 
Correlations must be considered, however,
for accurate computations involving
divalent ions (Guldbrand {\it et al.}, 1984;
Tomac and Gr\"{a}slund, 1998; Patra and Yethiraj, 1999).   

\section{LINEAR-SCALING CALCULATIONS}

\label{sec:linsc}

Several new methods have appeared for computations
involving systems with long-range interactions. 
In this section, developments
in linear-scaling methods for classical and quantum
systems are summarized. 
Goedecker (1999) has clearly reviewed
applications in electronic structure, so the discussion
of this topic will be limited. The purpose is to illustrate
the context in which real-space methods can be 
utilized in linear-scaling solvers for electrostatics
and electronic structure.   

\subsection{Classical electrostatics}

\label{subsec:linsccm}

Three algorithms have been most widely used in 
classical electrostatics
calculations which require consideration of long-range forces.
The first is the Ewald (1921) summation, which partitions
the Coulomb interactions
into a short-range sum handled in real space
and a long-range contribution summed in 
reciprocal space.  Both sums are 
convergent.  The partitioning is effected by
adding and subtracting localized Gaussian
functions centered about the discrete 
charges (Pollock and Glosli, 1996).  In the original
Ewald method, the Coulomb interaction of 
the Gaussians is obtained analytically:

\begin{equation}
E_{Gauss} = \frac{1}{2} \sum_{{\bf k}\neq 0}
\frac{4\pi}{\Omega} \frac{\exp (-k^2/2G^2)}
{k^2} |S({\bf k})|^2 ,
\label{eq:egauss}
\end{equation}

\noindent
once the charge structure factor,

\begin{equation}
S({\bf k}) = \sum_{i=1}^{N} Z_i \exp (i {\bf k} \cdot {\bf r}_i),
\label{eq:strfact}
\end{equation}

\noindent
is computed. In Eq.\ (\ref{eq:egauss}), $\Omega$ is the 
cell volume and $G$ is the Gaussian width.
This method scales as $N^{3/2}$ (where $N$ is the number of
particles) so long as 
an optimal exponential factor is used in the Gaussians. Discussion
of the optimization equation which yields the $N^{3/2}$ scaling
is given in Pollock (1999). 
The Ewald technique has been used extensively in simulations
of charged systems (Allen and Tildesley, 1987). An 
efficient alternative procedure for Madelung sums 
in electronic structure calculations 
on crystals was proposed 
by Harris and Monkhorst (1970). 

The scaling of the Ewald method has been reduced by an
alternative treatment for the interaction energy
of the Gaussians.   Instead of solving the problem
analytically, 1) the charge density is assigned to 
a mesh, 2) the Poisson equation is solved using
FFT methods, 3) the potential is differentiated, 
and lastly 4) the forces are interpolated to the 
particles.  These methods are termed 
particle-particle particle-mesh (Hockney and Eastwood, 1988)
or particle-mesh Ewald (Darden {\it et al.}, 1993)
[or an improved 
version called smooth particle-mesh Ewald
(Essmann {\it et al.}, 1995)].
Since the potential is generated numerically 
via FFT, the methods scale as $N \ln N$ (or 
$N\sqrt{\ln N}$ if an optimal $G$ is used;
see Pollock, 1999).  The 
above-mentioned methods differ in how the four
steps in the force generation are performed, but
all three center on the use of FFT algorithms for 
their efficiency.  Comparative studies have suggested
that the original particle-particle particle-mesh method is more
accurate than the particle-mesh Ewald versions; 
Deserno and Holm (1998) recommend its use with
modifications obtained from particle-mesh Ewald. 
See also Sagui and Darden (1999), where it 
is shown that similar accuracies can be
obtained with particle-mesh Ewald as compared to 
the particle-particle particle-mesh method.

The second algorithmic approach utilizes
the fast multipole method (FMM)
(Greengard, 1994) or related
hierarchical techniques. In these methods, 
the near-field contributions are treated
explicitly, while the far field is handled
by clustering charges into spatial cells 
and representing the field with a multipole
expansion.  The methods are claimed 
to scale linearly with system size, but 
recent work contends the scaling is 
slightly higher (P\'{e}rez-Jord\'{a} 
and Yang, 1998).  Fast multipole techniques
and the quantum chemical tree code (QCTC)
of Challacombe {\it et al.} (1996), have been
widely applied in Gaussian-based electronic
structure calculations. Since the classical 
Coulomb part of the problem is a significant
or even dominant part of the overall computational
effort, near linear scaling is required for an 
overall linear-scaling solver (Strain {\it et al.}, 1996; 
White {\it et al.}, 1996). P\'{e}rez-Jord\'{a}
and Yang (1997) have developed an alternative 
efficient recursive bisection method for obtaining the 
Coulomb energy from electron densities. 
The FMM has also been utilized extensively 
in particle simulations.  In comparative studies
of periodic systems,
Pollock and Glosli (1996) and Challacombe {\it et al.} (1997),
have shown that, for the case of discrete particles, 
the particle-mesh related techniques are more efficient than
the fast multipole method over a wide 
range of system sizes (up to 10$^5$ particles).  
However, for the case of 
continuous overlapping distributions, it is difficult
to develop systematic ways in the particle-mesh approach
to handle the charge penetration
in large-scale Gaussian calculations (Challacombe, 1999a).  
Recently, Cheng {\it et al.} (1999) have developed a more
efficient and adaptive version of the fast multipole method 
which will make the technique competitive with the 
particle-mesh method. Also, Greengard and Lee (1996) 
presented a method combining a local spectral approximation
and the fast multipole method for the Poisson equation.  

A third set of linear-scaling algorithms for classical
electrostatics employs real-space methods, which 
will be discussed in-depth in subsequent 
sections.  The problem is 
represented with FD equations, 
FE methods, or wavelets, and solved iteratively
on the mesh.  Since all operations are near-local in 
space, the application of the Laplacian to the potential
is strictly linear scaling. However, the iterative 
process on the fine mesh typically suffers from slowing
down in the solution process, so efficient preconditioning
techniques must be employed to obtain the linear 
scaling.  The multigrid method (Brandt, 1977, 1982,
1984, 1999; Hackbusch, 1985) is 
a particularly efficient method for solving the discrete
equations. Linear-scaling real-space methods 
have been developed for solution
of the Poisson problem in DFT (White {\it et al.}, 1989;
Merrick {\it et al.}, 1995, 1996;
Gygi and Galli, 1995; Briggs {\it et al.}, 1995; Modine
{\it et al.}, 1997;
Goedecker and Ivanov, 1998a). These studies have
illustrated the accuracies and efficiencies of the 
real-space approach.  One possible application of multigrid
techniques which has not received attention is in solving
for the Coulomb energy of the Gaussian charges in the 
particle-mesh algorithms.  Since the multigrid techniques
are highly efficient, scale linearly, and allow for 
variable resolution, they may provide a helpful counterpart
to the FFT-based methods currently used.  An advantageous 
feature of the multigrid solution during a charged-particle
simulation is that, once the potential is generated for a 
given configuration, it can be saved for the next 
solution process for the updated positions which have 
changed only slightly.  Thus 
the required number of iterations is likely to be
low. Tsuchida and Tsukada (1998) utilized similar
ideas in their FE method for electronic structure,
where they employed MG acceleration for rapid solution of
the Poisson equation and discussed the relation of
their method to the particle-mesh approach. 

\subsection{Electronic structure}

\label{subsec:linsces}

Electronic structure calculations involve computational
complexities which go well beyond the necessity for efficient
solution of the Poisson equation.  In order to obtain
linear scaling, physical localization properties must be
exploited either for the range of the density matrix or 
the orbitals.  Goedecker (1999) categorized the various
linear-scaling electronic structure methods as 
follows: Fermi operator expansion (FOE), 
Fermi operator projection (FOP), divide and conquer (DC),
density matrix minimization (DMM), orbital minimization (OM),
and optimal-basis density matrix minimization (OBDMM).  He 
further classified the algorithms into those 
which employ small basis sets 
(LCAO-type approaches) and ones which
utilize large basis sets (FD or 
FE).\footnote{See Section \ref{subsubsec:bfdr} 
for discussion of
the use of basis set terminology in reference to
the FD method.} Clearly the methods most 
relevant to the present discussion are those which 
can be implemented with large basis 
sets (FOP, OM, and OBDMM). The two approaches
most directly related to the FD and FE mesh techniques
considered here are the OM and OBDMM methods,
so we review their characteristics. 

The OM method obtains the localized Wannier functions by 
minimization of the functional:

\begin{equation}
\Omega = 2 \sum_n \sum_{i,j} c_i^n H'_{i,j}c_j^n -
\sum_{n,m} \sum_{i,j} c_i^n H'_{i,j}c_j^m
\sum_l c_l^n c_l^m.
\label{eq:omfunc}
\end{equation}

\noindent
The minimization is unconstrained in that no orthogonalization
is required; the orthonormality
condition is automatically satisfied at 
convergence.  In Eq.\ (\ref{eq:omfunc}), $\Omega$ 
is the `grand potential', the $c_i^n$ are the expansion 
coefficients for the Wannier function $n$ with 
basis function $i$, and the $H'_{i,j}$ are the matrix
elements of $H - \mu I$, where $\mu$ is the chemical 
potential controlling the number of electrons and $I$
is the identity matrix. The 
functional can be derived by making a Taylor expansion
of the inverse of the overlap matrix occurring in 
the total energy expression (Mauri {\it et al.},
1993). Ordej\'{o}n
{\it et al.} (1995) presented an alternative 
derivation and related the OM functionals
to the DMM approach.  Assuming no localization
restriction on the orbitals, it can be shown that 
the functional $\Omega$ gives the correct ground state
at its minimum. However, some problems arise when 
localization constraints are imposed: 1) the functional
can have multiple minima, 2) the number of required iterations 
to reach the ground state can be quite large, 3) there
may be runaway solutions depending
on the initial guess, and 4) the total charge is not 
conserved for all stages of the minimization
(although charge is accurately conserved close to the minimum).  

The original OM methods utilized 
underlying plane-wave (Mauri
{\it et al.}, 1993; Mauri and Galli, 1994) and 
tight-binding or LCAO-type bases (Kim, {\it et 
al.}, 1995; Ordej\'{o}n {\it et al.}, 1995; 
S\'{a}nchez-Portal {\it et al.}, 1997) for the 
representation of the localized orbitals.  In the
work of S\'{a}nchez-Portal {\it et al.} (1997) on 
very large systems, a fully numerical LCAO basis developed
by Sankey and Niklewski (1989) was implemented for the 
orbitals, and the Hartree problem was solved 
via FFT techniques on a real-space grid.  Lippert {\it et al.}
(1997) developed a related hybrid Gaussian and 
plane-wave algorithm which uses Gaussians in place
of the numerical atomic basis.  Also, Haynes and 
Payne (1997) formulated a new localized spherical-wave
basis which has features in common with plane waves
in that a single parameter controls the convergence.

Real-space formulations have also  
applied OM ideas; since the real-space
approach is inherently local, it 
provides a natural representation for the 
linear-scaling algorithms.  Tsuchida and
Tsukada (1998) incorporated unconstrained
minimization into their FE electronic 
structure method. Hoshi and Fujiwara (1997)
also employed unconstrained minimization
in their FD self-consistent electronic
structure solver.  Finally, Bernholc
{\it et al.} (1997) utilized the original
localized-orbital
functional of Galli and Parrinello (1992) in their
FD multigrid method to obtain linear 
scaling.  They are also investigating other
order $N$ functionals. These real-space algorithms will
be the subject of extensive discussion 
in Section \ref{sec:ks}.  

The OBDMM method is an efficient combination of density 
matrix and orbital-based methodologies.  The optimization
process to locate the ground state is divided into
two minimization steps.  In the inner loop, 
the usual DMM procedure is followed to obtain the
density matrix for a fixed contracted basis.  
The density matrix $F({\bf r},{\bf r}')$ 
is represented in terms of contracted
basis functions $\psi_i$ and a matrix $K$ which is
a purified form from the DMM method:

\begin{equation}
F({\bf r},{\bf r}') = \sum_{i,j} \psi_i^* ({\bf r}) K_{i,j}
\psi_j ({\bf r}') ,
\label{eq:dmm}
\end{equation}

\noindent
and

\begin{equation}
K = 3LOL - 2LOLOL ,
\label{eq:dmmmatrix}
\end{equation}

\noindent
where $L$ is the contracted basis density matrix
and $O$ the overlap matrix. The matrix $K$ is `purified' 
in that if the eigenvalues of $L$ are close to zero or one,
the eigenvalues of $K$ will be even closer to those values.  
The outer loop searches for the optimal basis with fixed 
$L$.  The OBDMM method was developed 
independently by Hierse and
Stechel (1994) and Hern\'{a}ndez and Gillan (1995).  
The two approaches differ in that the algorithm of
Hern\'{a}ndez and Gillan allows for a number of 
basis functions larger than the number of electrons. 
Also, Hierse and Stechel (1994) used  
tight-binding and Gaussian bases, while Hern\'{a}ndez
and Gillan (1994) employed a FD difference representation
in their original work. Later, Hern\'{a}ndez {\it et al.}
(1997) developed a blip-function basis (a local basis 
of B-splines, see Strang and Fix, 1973), very closely 
related to FE methods.  

In the quantum chemistry literature, efforts have
focused on Gaussian basis-function algorithms. 
As discussed above, the Coulomb problem is typically
solved with the FMM or other hierarchical techniques 
(White {\it et al.}, 1996; Strain {\it et al.},
1996; Challacombe {\it et al.}, 1996; Challacombe
and Schwegler, 1997). 
Additional algorithmic advances include linear 
scaling for the exchange-correlation calculation
in DFT (Stratmann {\it et al.}, 1996), for the 
exact exchange matrix in Hartree-Fock theory
(Schwegler and Challacombe, 1996; Schwegler
{\it et al.}, 1997), and for the
diagonalization operation (Millam and Scuseria, 1997;
Challacombe, 1999b). 
Alternative linear-scaling algorithms include
the early Green's function based FD method of
Baroni and Giannozzi (1992) and the finite-temperature 
real-space method of Alavi {\it et al.} (1994).   
 
It is evident from the above discussion that real-space
methods, in particular FD and FE approaches,\footnote{See
Arias (1999) and Goedecker (1999) for discussion of linear-scaling 
applications of wavelets.} are well suited 
for linear-scaling algorithms.  In classical electrostatics 
calculations, the multigrid method provides an
efficient and linear-scaling technique for solution
of Poisson problems given a charge distribution 
on a mesh (finite or periodic systems).  
In electronic structure, FD and FE representations
have been extensively employed 
in the OM and OBDMM localized-orbital 
linear-scaling contexts.  

\section{REAL-SPACE REPRESENTATIONS}

\label{sec:fdfe}

The early development of FD and FE
methods for solving partial differential equations
stemmed from engineering problems involving
complex geometries, where 
analytical approaches were not possible (Strang and Fix,
1973).  Example applications include structural mechanics
and fluid dynamics in complicated geometries.  
However, even in the early days of quantum 
mechanics, attention was paid to FD numerical solutions
of the Schr\"{o}dinger equation (Kimball and
Shortley, 1934; Pauling and Wilson, 1935).
Also, fully converged numerical solutions of 
self-consistent electronic structure calculations have
played an important role in atomic physics (see Mahan 
and Subbaswamy, 1990, for a discussion of the methodology
for spherically symmetric systems) 
and more recently in molecular
physics (Laaksonen {\it et al.}, 1985; Becke, 1989).  

Real-space calculations are performed on meshes; these
meshes can be as simple as Cartesian grids or can be
constructed to conform to the more demanding geometries 
arising in many applications. Finite-difference representations
are most commonly constructed on regular Cartesian
grids.  They result from a Taylor series expansion
of the desired function about the grid points. The 
advantages of FD methods lie in the simplicity of
the representation and resulting ease of implementation in 
efficient solvers. Disadvantages are that the
theory is not variational (in the sense of providing
and upper bound, see below), and it is 
difficult to construct meshes flexible enough to 
conform with the physical geometry of many problems. 
Finite-element methods, on the other hand, have the advantages
of significantly greater flexibility in the construction of the 
mesh and an underlying variational formulation. The 
cost of the flexibility is an increase in complexity
and more difficulty in the implementation of multiscale
or related solution methods.  
In this Section, we review the technical aspects of real-space
FD and FE representations of differential 
equations by examination of Poisson
problems.  

\subsection{Finite differences}

\label{subsec:fd}

\subsubsection{Basic finite-difference representation}

\label{subsubsec:bfdr}

The second-order FD representation of 
elliptic equations is very simple but serves to illustrate
several key features.  Consider the Poisson
equation in one dimension (the $4\pi$ is left here since
we will be considering three-dimensional problems):

\begin{equation}
\frac{d^2 \phi(x)}{dx^2} = - 4\pi\rho(x),
\label{eq:pe1d}
\end{equation}

\noindent
where $\phi(x)$ is the potential and $\rho(x)$ the charge
density. Expand the potential in the positive and negative
directions about the grid point $x_i$:

\begin{eqnarray}   
\phi(x_{i+1})&=&\phi(x_i)+\phi'(x_i)h+\frac{1}{2}\phi''(x_i)h^2
  +\frac{1}{6}\phi'''(x_i)h^3+\frac{1}{24}
   \phi^{(iv)}(x_i)h^4 \ldots \nonumber\\
\phi(x_{i-1})&=&\phi(x_i)-\phi'(x_i)h+\frac{1}{2}\phi''(x_i)h^2
  -\frac{1}{6}\phi'''(x_i)h^3+\frac{1}{24}\phi^{(iv)}(x_i)h^4 \ldots
\label{eq:taylorpot}
\end{eqnarray}

\noindent
The grid spacing is $h$, here assumed uniform. If these two equations
are added and the sum is solved for $\phi''(x_i)$, the 
following approximation results:

\begin{equation}
\frac{d^2 \phi(x_i)}{dx^2} \approx 
\frac{1}{h^2}\left[\phi(x_{i-1}) - 2\phi(x_i) + \phi(x_{i+1})\right]
 - \frac{1}{12}\phi^{(iv)}(x_i)h^2 + O(h^2)
\end{equation}

\noindent
The first contribution to the truncation
error is second order in $h$ 
with a prefactor involving the fourth 
derivative of the potential.  Depending on the nature of the 
function $\phi(x)$, the errors can be of either sign. When 
$\phi(x)$ is used to compute a physical quantity 
such as the total electrostatic energy,
the net errors in the energy can be either 
positive or negative.  In this sense, the FD approximation
is {\it not variational}.  As we will see below, the solution
can be obtained by minimizing an energy (or action) functional,
which is a variational process, but the solution does not 
necessarily satisfy the variational theorem obtained in 
a basis-set method. {\it So the FD approach is not a basis-set method.} 

In matrix form, the one-dimensional discrete Poisson equation is

\begin{equation}
\frac{1}{h^2}\left[ \begin{array}{rrrrr}
-2&1&0&0&\ldots \\
1&-2&1&0&\ldots \\
0&1&-2&1&\ldots \\
0&0&1&-2&\ldots \\
\vdots&\vdots&\vdots&\vdots&\ddots
\end{array} \right]
\left[ \begin{array}{r}
\phi(x_1) \\
\cdot \\
\cdot \\
\cdot \\
\phi(x_N)
\end{array} \right]
=
-4\pi\left[ \begin{array}{r}
\rho(x_1) \\
\cdot\\
\cdot\\
\cdot \\
\rho(x_N)
\end{array} \right] .
\label{eq:matrix}
\end{equation}

\noindent
This equation can be expressed symbolically as

\begin{equation}
L^h u_{ex}^h = f^h ,
\label{eq:luf}
\end{equation}

\noindent
where $L^h$ is the discrete Laplacian, $u_{ex}^h$ is the exact solution
on the grid, and $f^h$ is $-4\pi\rho$.  The operator $-L$ is 
positive definite.  An observation from the matrix form Eq.\
(\ref{eq:matrix}) is that the Laplacian is highly sparse and
banded in the 
FD representation; its application to the potential 
is thus a linear-scaling step. In one dimension the matrix
is tridiagonal, while in two or three dimensions it is no longer
tridiagonal but is still extremely sparse with nonzero 
values only near the diagonal. This differs from the
wavelet representation, which is sparse but includes
several bands in the matrix (Goedecker and 
Ivanov, 1998b, Fig.\ 7; Arias, 1999, Fig.\ 10).   

In addition to the truncation error,

\begin{equation}
t^h = -\frac{1}{12} \phi^{(iv)} (x_i) h^2 + O(h^4) ,
\label{eq:truncerror}
\end{equation}

\noindent
estimates can be made of the function error itself
(see Strang and Fix, 1973, p. 19):

\begin{equation}
e_a^h = u_{ex}^h - u_a = e_2 h^2 + O(h^4),
\label{eq:fncnerror}
\end{equation}

\noindent
where $u_a$ is the exact solution to the continuous differential
equation and $e_2$ is proportional to the second derivative
of the potential.   Therefore, one can test the order
of a given solver for a case with a known solution
by computing errors over the domain and taking ratios
for variable grid spacing $h$.  For example, the 
ratio of the errors on a grid with spacing $H = 2h$ 
to those on $h$ for overlapping points should be
close to 4.0 in a second-order calculation.  

The two- and three-dimensional representations are obtained
by summing the one-dimensional case along the two or
three orthogonal coordinate axes (this holds
for higher-order forms as well).  Since the Laplacian
is the dot product of two vector operators, off-diagonal
terms are not necessary.  The 
second-order two-dimensional Laplacian
consists of five terms with a weight of -4 instead of
-2 on the diagonal, and the three-dimensional case
has seven terms with a weight of -6 along the diagonal.
See Abramowitz and Stegun (1964, Sections 25.3.30 and
25.3.31), for the two-dimensional representation of
the Laplacian.  

\subsubsection{Solution by iterative techniques}

\label{subsubsec:fdsit}

Consider the action functional:

\begin{equation}
S[\phi] = \frac{1}{2}\int \left| \nabla\phi \right|^2 d^3x
- 4\pi \int \rho\phi d^3x .
\label{eq:actionibp}
\end{equation}

\noindent
If the first term on the rhs is integrated by parts (assuming 
the function and/or its derivative go to zero at infinity or
are periodic), one obtains

\begin{equation}
S[\phi({\bf r})] = -\frac{1}{2} \int \phi\nabla^2\phi d^3x 
 - 4\pi \int \rho\phi d^3x.
\label{eq:action}
\end{equation}

\noindent
Take the functional derivative of the action with respect
to variations of the potential, and a `force' term
results,

\begin{equation}
-\frac{\delta S}{\delta \phi} = \nabla^2 \phi + 4\pi \rho,
\label{eq:force}
\end{equation}

\noindent
which can be employed in a steepest-descent minimization
process:

\begin{equation}
\frac{\delta \phi}{\delta t}= -\frac{\delta S}{\delta \phi},  
\label{eq:sdeqn}
\end{equation}

\noindent
where $t$ is a fictitious time variable. 

Then discretize the problem in space and time, leading
to (for simplicity of representation a one-dimensional
form is given here)

\begin{equation}
\phi(x_i)^{t+1} = (1-\omega)\phi(x_i)^t + \frac{\omega}{2}
[\phi(x_{i-1})^t + \phi(x_{i+1})^t + 4\pi\rho(x_i)h^2],
\label{eq:update}
\end{equation} 

\noindent
The parameter $\omega$ is $2\delta t/h^2$.  The two- and 
three-dimensional expressions are easily obtained following
the same procedure.  Since the action as defined in Eq.\
(\ref{eq:action}) possesses only
a single minimum, the iterative process eventually
converges so long as a sufficiently small time step 
$\delta t$ is chosen to satisfy  the required stability
criterion (below).

Several relaxation strategies result from 
the steepest-descent scheme of Eq.\ (\ref{eq:update}).
As it is written, the method is termed 
{\it weighted-Jacobi} iteration.  If the previously updated 
value $\phi(x_{i-1})^{t+1}$ is used in place of
$\phi(x_{i-1})^{t}$ on the right hand side, the 
relaxation steps are called {\it Successive 
Over-Relaxation} or SOR. If the parameter in 
SOR is taken as $\omega = 1$, the 
result is {\it Gauss-Seidel} iteration.  Gauss-Seidel 
and SOR do not guarantee reduction in the 
action at each step since they use the previously
updated value.  Generally, 
Gauss-Seidel iteration is the best method for
the smoothing steps in 
multigrid solvers (Brandt, 1984). If one 
cycles sequentially through the lattice points, 
the ordering is termed {\it lexicographic}. 
Higher efficiencies (and vectorization) can be obtained with
{\it red-black} ordering schemes in which the grid 
points are partitioned into two interlinked sets 
and the red points are first updated, followed by
the black (Brandt, 1984; Press {\it et al.}, 1992).
Similar techniques can be used for high orders with
multicolor schemes.  Conjugate-gradient methods 
(Press {\it et al.}, 1992)
significantly outperform the above relaxation methods
when used on a single grid level. However, in a multigrid
solver the main function of relaxation is only to smooth the 
high-frequency components of the errors on each level
(see Section \ref{subsec:efmg}), and simple
relaxation procedures (especially Gauss-Seidel)
do very well for less cost.  

An important issue in iterative relaxation steps
relates to the eigenvalues of the update 
matrix defined by Eq.\ \ref{eq:update} 
(Briggs, 1987).  Solution of the Laplace
equation using weighted-Jacobi iteration 
illustrates the basic problem.  For that particular case,
the eigenvalues of the update matrix are

\begin{equation}
\lambda_k = 1 - 2\omega\sin^2 \left( \frac
{k\pi}{2N}\right) ;  1 \leq k \leq N-1 ,
\label{eq:propeval}
\end{equation}

\noindent
where $\omega$ is the relaxation parameter defined above,
$N+1$ is the number of grid points in the 
domain, and $k$ labels the mode in the Fourier
expansion of the function.  Generally, the Fourier
component of the error with wavevector $k$ is 
reduced in magnitude by a factor proportional
to $\lambda_k^t$ in $t$ iterations. 

First, it is easy to 
see that if too large an $\omega$ value (that is
`time' step for fixed $h$) is taken, the magnitude
of some modes will exceed one, leading to instability.
This shows up very quickly in a numerical solver!
Second, for the longest-wavelength modes, 
the eigenvalues are of the form:

\begin{equation}
\lambda = 1 - O(h^2).
\label{eq:longmode}
\end{equation}

\begin{figure}
\centerline{\epsfig{file=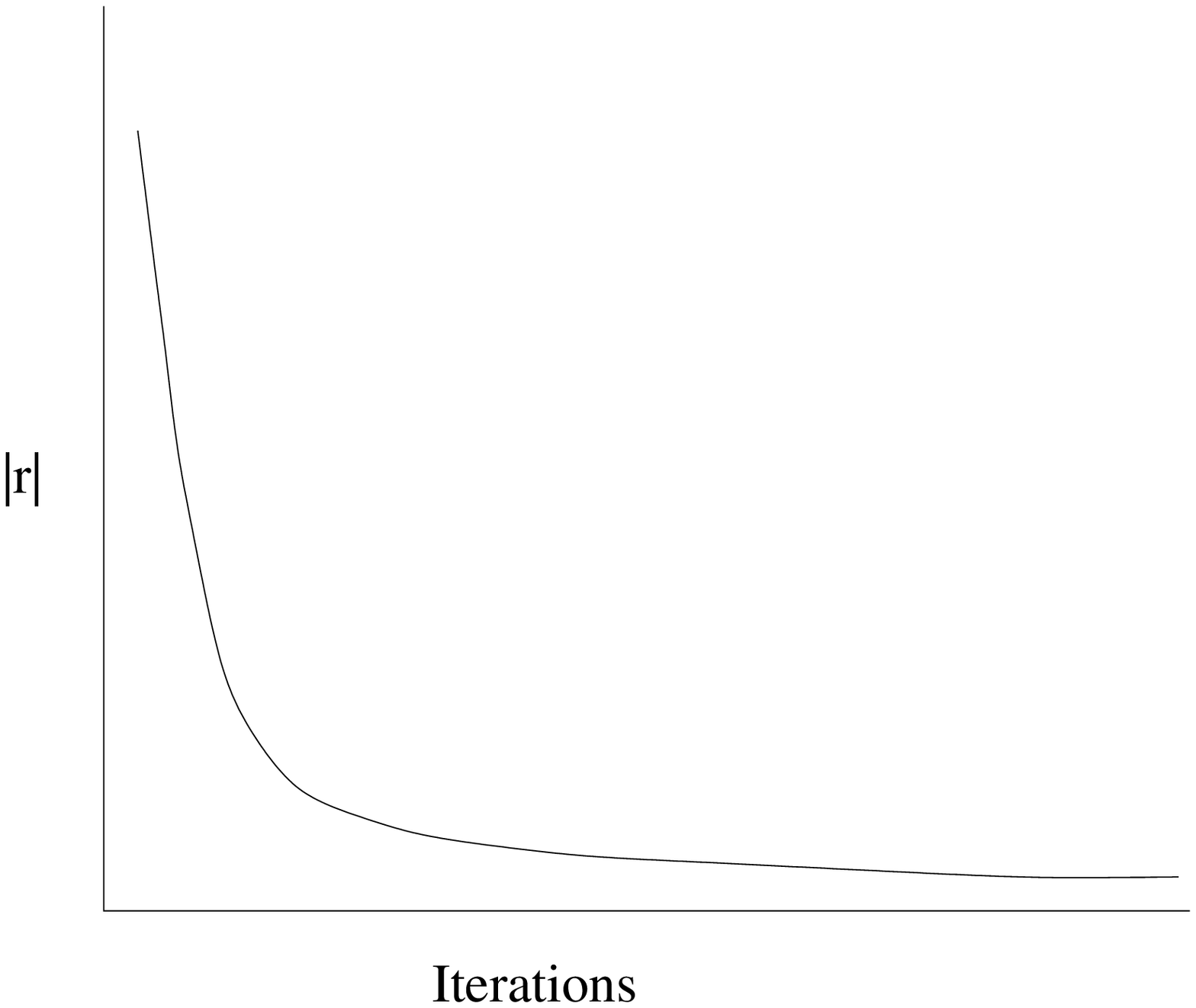,height=3.5in,width=3.in,angle=270}}
\vspace{10pt}
\caption{Typical behavior of the residual during iterations on a 
fine level only.}
\label{fig:csd}
\end{figure}

\noindent
As more grid points are used to obtain 
increased accuracy on a fixed domain, the eigenvalues of the 
longest-wavelength modes approach one. Therefore, these modes
of the error are very slowly reduced. This 
fact leads to the phenomenon of {\it critical 
slowing down} in the iterative process (Fig.\ \ref{fig:csd}), 
which motivated the development of multigrid techniques.
Multigrid methods utilize information from multiple length 
scales to overcome the critical
slowing down (Section \ref{sec:mg}).  

\subsubsection{Generation of high-order finite-difference formulas}

\label{subsubsec:hofd}

Mathematical arguments lead to the conclusion that 
the FD scheme discussed above is 
convergent in the sense that $E^h \rightarrow 0$
as $h \rightarrow 0$ (Strang and Fix, 1973; 
Vichnevetsky, 1981). 
Therefore, one only needs to 
proceed to smaller grid spacings to
obtain results with a desired accuracy.  This neglects
the practical issues of computer time and memory,
however, and it has become apparent that orders
higher than second are most often necessary to
obtain sufficient accuracy in electronic
structure calculations on reasonable-sized 
meshes (Chelikowsky, Troullier, and Saad, 1994). 

The higher-order difference formulas are
well known (Hamming, 1962; Vichnevetsky, 1981),
and can easily be generated using computer algebra
programs (see Appendix A).
Why does it pay to use high-order approximations?
Consider the three-dimensional Poisson equation
with a singular-source charge density:

\begin{equation}
\nabla^2 \phi({\bf x}) = -4\pi\delta({\bf x}) .
\label{eq:pesingular}
\end{equation}

\noindent
The Dirac delta function is approximated by a unit charge on a 
single grid point.  Let us solve the FD version of  
Eq.\ (\ref{eq:pesingular}) on a 65$^3$ domain
using 2nd- and 4th-order
Laplacians and compare the potential eight grid points
away from the origin.\footnote{The calculations in 
this section were performed with multigrid solvers discussed
in Sections \ref{sec:mg} and \ref{sec:ks}.}
In order to obtain the same numerical accuracy
with a 2nd-order Laplacian, a grid
spacing with one third that for the 4th-order case is required.
This implies a 27-fold increase in storage and roughly a 
14-fold increase in computer time, since the application 
of the Laplacian contains 7(13) terms for the 
2nd(4th)-order calculations.   

\begin{figure}
\centerline{\epsfig{file=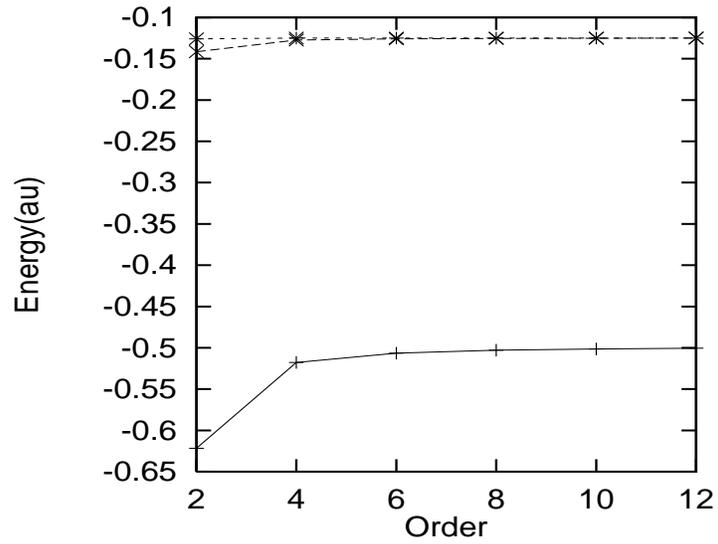,height=3.5in,width=3.in,angle=270}}
\vspace{10pt}
\caption{Effect of order on the eigenvalues for 
the H atom. The (+) symbols are for the 1s orbital,
(x) is for 2s, and the stars are for 2p. The analytical
results are -0.5, -0.125, and -0.125 respectively.}
\label{fig:evalord}
\end{figure}

\begin{figure}
\centerline{\epsfig{file=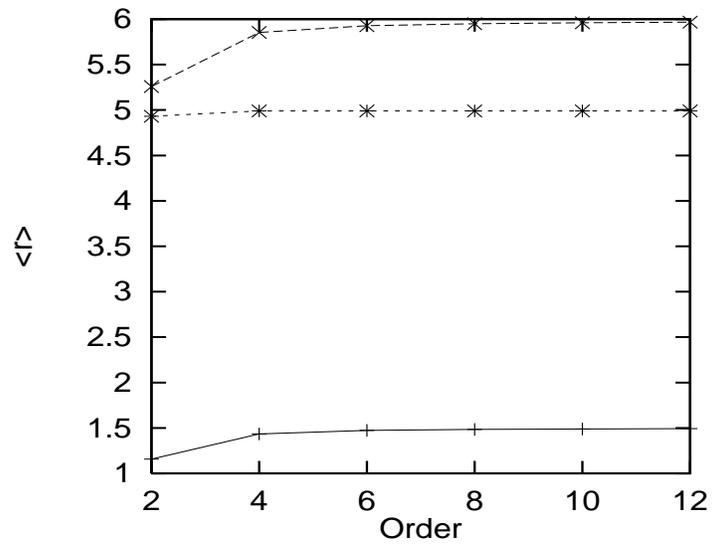,height=3.5in,width=3.in,angle=270}}
\vspace{10pt}
\caption{Effect of order on the orbital
first moments for 
the H atom. The (+) symbols are for the 1s orbital,
(x) is for 2s, and the stars are for 2p. The analytical 
results are 1.5, 6, and 5 respectively.}
\label{fig:raveord}
\end{figure}

\begin{figure}
\centerline{\epsfig{file=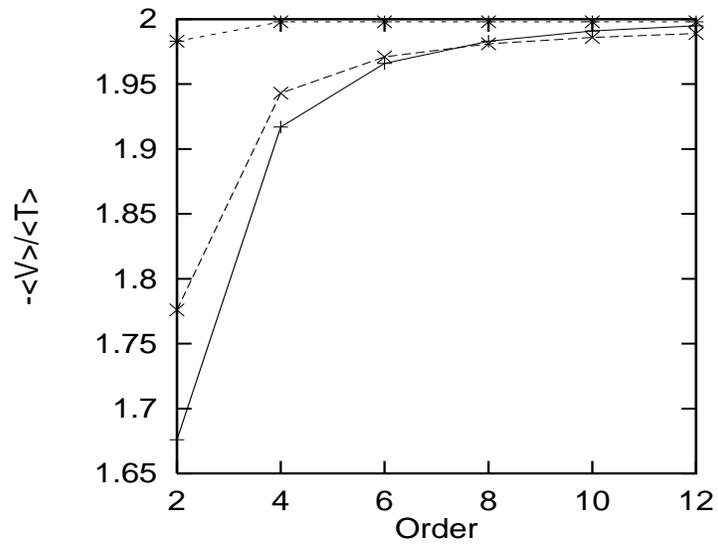,height=3.5in,width=3.in,angle=270}}
\vspace{10pt}
\caption{Effect of order on the orbital
virial ratios for 
the H atom. The (+) symbols are for the 1s orbital,
(x) is for 2s, and the stars are for 2p. The analytical 
result is 2.}
\label{fig:virord}
\end{figure}

As a second example, we solve for five states of
the hydrogen-atom eigenvalue problem 
using the fixed potential
generated in the solution of Eq.\ (\ref{eq:pesingular}).
The grid parameters are the same as those used in 
the multigrid eigenvalue computations of 
Section  \ref{subsubsec:evalapps}.
The variation of the eigenvalues, the first orbital 
moments, and the virial ratios with 
approximation order are presented in 
Figs.\ \ref{fig:evalord}, \ref{fig:raveord}, and
\ref{fig:virord}.  A possible accuracy target is the
thermal energy at room temperature ($kT \approx 0.001$ au);
this accuracy is achieved at 12th order. 
Clearly the results at 2nd order
are not physically reasonable, but
accurate results can be obtained with the higher orders. 
Merrick {\it et al.} (1995) and Chelikowsky, Trouller,
Wu, and Saad (1994) have
presented analyses of the impact of order
on accuracy in DFT electrostatics and Kohn-Sham
calculations; in the Kohn-Sham calculations, 8th
or 12th orders were required for adequate convergence.  

There exist alternative high-order discretizations such 
as the Mehrstellen form used in the work of Briggs
{\it et al.} (1996). 
This discretization is 4th order and leads to terms which 
are off-diagonal in both the kinetic and potential operators. 
The advantage of the Mehrstellen approach is that both terms only
require near-neighbor points on the lattice, while the 
high-order forms above include information from further
points (which increases the communication overhead
somewhat in parallel implementations).  However, the 
4th-order Mehrstellen operator involves 33 multiplies to 
apply the Hamiltonian to the wavefunction, while the 
standard 4th-order discretization requires only 14 (a
12th-order standard form uses 38 multiplies). Also, the 
Mehrstellen representation has only been applied to
the 4th-order case, and for some applications higher
orders may be required. The exact terms for the Mehrstellen 
representation of the real-space Hamiltonian are given 
in Briggs {\it et al.} (1996). 

\subsection{Finite elements}

\label{subsec:fe}

\subsubsection{Variational formulation}

\label{subsubsec:fevar}

Consider again the action of Eq.\
(\ref{eq:actionibp}) in one dimension:

\begin{equation}
S[\phi] = \frac{1}{2}\int \left( \frac{d\phi}{dx} \right)^2 dx
- 4\pi \int \rho\phi dx .
\label{eq:actionibp1d}
\end{equation}

\noindent
This form of the action proves useful since the appearance
of the first derivative as opposed to the second 
expands the class of functions which may be used 
to represent the potential. 
Now, expand the potential in a basis:

\begin{equation}
\phi(x) = \sum_{i=1}^{n} u_i \zeta_i (x),
\label{eq:febasisexp}
\end{equation}

\noindent 
where the $u_i$ are the expansion coefficients and  
$\zeta_i$ the basis functions. 
The action is then

\begin{equation}
S = \frac{1}{2}\int \left( \sum_{i=1}^n \frac{d\zeta_i}{dx}u_i 
\right)^2 dx
- 4\pi \int \rho\left( \sum_{i=1}^n \zeta_i u_i \right) dx .
\label{eq:actionbasis}
\end{equation}

\noindent
The variational calculation is performed by minimizing the
action with respect to variations in the expansion 
coefficients (assuming the original differential operator
is positive definite):

\begin{equation}
\frac{\partial S}{\partial u_i} = \int \left[ \frac 
{d\zeta_i}{dx} \left( \sum_{j=1}^n \frac {d\zeta_j}{dx}u_j 
\right) - 4\pi\rho\zeta_i  \right] dx = 0 .
\label{eq:varbasis}
\end{equation}

\noindent
The minimization equation leads to a matrix problem completely
analogous to Eq.\ (\ref{eq:luf}). In the present case, the 
grid index is replaced by the basis-function index.  
It is often necessary to perform the integral of the second
term (which involves the charge density) numerically.

A more general origin of the FE method is termed the 
Galerkin approach which takes as its starting point
the ``weak" formulation of the problem. This method
allows one to handle problems which cannot be cast
in the minimization format described above by requiring
only an extremum of the action functional and not a
minimum. Also, it does not require symmetric
operators. Take the action functional of 
Eq.\ (\ref{eq:actionibp1d}) and perturb it by the addition
of a small term $\epsilon v$.\footnote{The functions 
$\phi$ and $v$ exist in a 
subspace of a Hilbert space which 
becomes a finite-dimensional subspace for any 
FE basis-set numerical 
computation.} 
The action becomes

\begin{equation}
S[\phi + \epsilon v] = S[\phi] + \frac{1}{2}\epsilon^2\int 
\left( \frac{dv}{dx} \right)^2 dx + 
\epsilon\int\left(\frac{d\phi}{dx}\right)
\left(\frac{dv}{dx}\right) dx - 4\pi \epsilon\int \rho v dx.
\label{eq:actionpert}
\end{equation}

\noindent
By taking the derivative with respect to $\epsilon$,
making $\epsilon$ zero, and setting what remains
to zero, the stationary point is obtained. This 
variational form results in the following integral equation:

\begin{equation}
\int \left(\frac{d\phi}{dx}\right)
\left(\frac{dv}{dx}\right) dx = 4\pi\int\rho v dx.
\label{eq:vbvp}
\end{equation}

\noindent
This equation is valid for any 
test function $v$; solution requires
finding the function $\phi$ for which the equation holds 
for all $v$. Alternatively, Eq.\ \ref{eq:vbvp} can be derived
by simply left multiplying the differential equation by
the test function $v$ and integrating by parts.
When the functions $\phi$ and $v$ are represented
in the $\zeta_i$ basis, a matrix equation the same as
Eq.\ (\ref{eq:varbasis}) is obtained. This basis-set 
manifestation of the weak formulation is 
termed the Galerkin method.
If the test function space for $v$ is taken to include
all Dirac delta 
functions, and the problem is cast in the strong
form $<v,L\phi+4\pi\rho>=0$ (where $L$ is the differential operator,
in this case the Laplacian),
the {\it collocation} (or pseudospectral) 
approximation results when the problem
is discretized (Orszag, 1972; Vichnevetsky, 1981; 
Ringnalda {\it et al.}, 1990). 
Excellent reviews of the theory and application of 
finite elements are given in Strang and Fix (1973),
Vichnevetsky (1981), Brenner and Scott (1994), and Reddy (1998).

\subsubsection{Finite-element bases}

\label{subsubsec:febases}

Any linearly independent basis may be used to expand the potential.
One choice would be to expand in trigonometric functions which span
the whole domain. Then Fourier transform techniques could be used
to solve the equations.   
In the FE method, the basis functions are rather taken as 
piecewise polynomials which are nonzero only in a local
region of space (that is, have small support).  
The simplest possible basis consists of piecewise-linear 
functions whose values are one at the grid point
about which they are centered and zero everywhere beyond
the nearest-neighbor grid points.  Then the coefficients
$u_i$ correspond to the actual function values 
on the mesh. With this basis and
a basis-set representation of the charge density $\rho(x)$,
the resulting matrix representation of the one-dimensional 
Poisson equation is identical to 
Eq.\ (\ref{eq:matrix}), except the right hand side
is replaced by terms which are local averages
of the charge density over three points.  The
local average is idential to Simpson's rule
integration. Therefore, for uniform meshes,
there is a close correspondence between FD
and FE representations. Relaxation methods 
similar to those described 
above can be used to
solve the FE equations. 

Besides the variational foundation of the 
FE method, the key advantage over FD 
approaches is in the flexibility available
to construct the mesh to conform to the 
physical geometry.  This issue becomes
particularly relevant for two- and
three-dimensional problems.  There is 
of course an immense literature on 
development of accurate and efficient
basis sets for FE calculations in
a wide variety of engineering and 
physical applications, and that topic
cannot be covered properly here.  
Some representative bases are
mentioned from recent three-dimensional electronic
structure calculations. White {\it et al.}
(1989) employed a cubic-polynomial basis and
constructed an orthogonal basis
from the nonorthogonal set. Ackermann
{\it et al.} (1994) used a tetrahedral
discretization with orders $p = 1 - 5$.
Pask {\it et al.} (1999) utilized
piecewise cubic functions (termed
``serendipity'' elements). Yu 
{\it et al.} (1994) employed a Lobatto-Gauss
basis set with orders ranging from
five to seven. Hern\'{a}ndez {\it et al.}
(1997) developed a B-spline basis which is closely
related to traditional FE bases. 
Tsuchida and Tsukada (1998) used
piecewise third-order polynomials  
in their self-consistent electronic
structure calculations. 

In relating the FD and FE methods, two
points are worth noting. First,
the FE bases are typically nonorthogonal and this issue
must be dealt with in the formulation. Second, since the
basis is local, the representation is
banded with the width depending on
the degree of the polynomials.  For 
the FD representation, the 
high-order Laplacian includes
$3p + 1$ terms in a row of $L$ for three-dimensional
calculations. Alternatively, the FE method requires $O(p^3)$
terms along a row of $L$ in the limit of
high orders, although the
exact number of terms depends on the 
particular elements (Pask, 1999).  This
issue of scaling of the bandwidth with order
may become a significant one in development
of efficient iterative solvers of the equations. 
Due to the relative merits of the two representations,
there is no clear `answer' as to which one is preferable;
the key feature for this review is that both are 
near-local leading to structured and sparse matrix representations of
the differential equations.  The wavelet
basis method is closer in form to the FE representation but,
as mentioned above,
leads to more complicated matrix structures 
than either the FD or FE cases (Goedecker and Ivanov, 
1998b; Arias, 1999).  

\section{MULTIGRID TECHNIQUES}

\label{sec:mg}

The previous section discussed the basics of real-space
formulations.  The representations are near-local in space,
and this locality manifests itself in the stalling 
process of iterative solvers induced by Eq.\ (\ref{eq:longmode}).
The finer the resolution of the mesh, the longer it takes
to remove the long-wavelength modes of the error.
The multigrid technique was developed in order to overcome
this inherent difficulty in real-space methods. Multigrid
methods provide the optimal solvers for problems
represented in real space.

\subsection{Essential features of multigrid}

\label{subsec:efmg}

The asymptotic convergence
of an iterative solver on a given scale is controlled by
Eq.\ (\ref{eq:longmode}).  However, for shorter-wavelength
modes it is easy to show that the {\it convergence factor}

\begin{equation}
\mu = \frac{|{\tilde e}^h|}{|e^h|},
\label{eq:convfactor}
\end{equation}

\noindent
where $|e^h|$ is the norm of the difference vector between the
exact grid solution $u_{ex}^h$ and the current approximation $u^h$, and
${\tilde e}^h$ is the vector for the next step of iteration,
is of order 0.5 for Gauss-Seidel iteration on the Poisson
equation (Brandt, 1984). Those components
of the error are reduced by an order of magnitude in
only three relaxation sweeps.  Thus, relaxation steps on a
given grid level are referred to as {\it smoothing} steps;
the high-frequency components of the error are efficiently
removed while the long-wavelength modes remain.  Following
the fine-scale smoothing, the key step 
of multigrid is then to pass 
the problem to a coarser level, say $H=2h$
(with appropriate rules for the construction of the problem
on the coarse grid); smoothing steps on the coarse level
efficiently remove errors of twice the wavelength. 
Finally, the fine-grid function is corrected with the 
error interpolated from the coarse level, and further
iterations on the fine level remove 
remaining high-frequency components induced by the
coarse-grid correction.  

When this process is recursively
followed through several levels, the stalling behavior
can be completely removed and the solution is obtained in 
$O(N)$ operations, where $N$ is the number of 
unknowns. Typically, the problem can be solved to 
within the truncation errors in roughly ten total smoothing
steps on the finest level. The previous discussion rests
on a local-mode analysis of the errors (Brandt, 1977, 1984); additional
mathematical arguments confirm the excellent convergence
rates and linear scaling of multigrid solvers (Hackbusch, 1985).

\subsection{Full approximation scheme multigrid V-cycle}

\label{subsec:fas}

For linear problems, the algebraic Eq.\ (\ref{eq:luf})
can be rewritten as

\begin{equation}
L^h e^h = r^h ,
\label{eq:linear}
\end{equation}

\noindent
where $h$ is the finest grid spacing,
$e^h = u_{ex}^h - u^h$ (grid error), and 
$r^h = f^h - L^h u^h$ (residual equation). 
During the multigrid correction cycles, the 
coarse-grid iterations
only need to be performed on the error term $e^H$ which
is subsequently interpolated to the fine grid to 
provide the correction.  However,
this rearrangement is not possible for nonlinear problems.
Brandt (1977, 1984) developed the full approximation scheme (FAS)
approach for handling such problems.  Besides 
providing solutions to nonlinear
differential equations like the Poisson-Boltzmann
equation, the FAS strategy is well suited to handle
eigenvalue problems and mesh-refinement approaches. 
The FAS form of multigrid is thus
presented here due to its generality.  
In the case of linear problems, the 
FAS is equivalent to the error-iteration 
version mentioned above. 

Consider a Poisson problem discretized on
a Cartesian lattice with a
FD representation on a fine grid 
with spacing $h$ (Eq.\ \ref{eq:luf}).
Now construct a sequence of coarser grids each
with grid spacing twice the previous finer 
value. For a 4-level problem in
three dimensions, the sequence of grids will
consist of $17^3$, $9^3$, $5^3$, and $3^3$ points 
including the boundaries. If $h=1$, the coarser
grid spacings are 2, 4, and 8. The boundary values
of the potential on each level are fixed
based on the physics of the problem. For example,
if there are a set of discrete charges inside
the lattice, direct summation of the $1/r$ 
potential or a multipole expansion can be performed.
Alternatively it is easy to apply periodic boundary
conditions by wrapping the potential.
On the coarsest grid, only the one central point
is iterated during relaxation steps there.  

Assume there are $l$ levels for the general
case; each level is labeled by the index $k$
which runs from 1 (coarsest level) to $l$ (finest
level).  The operator $L^k$ is defined by the 
FD discretization on level $k$ with grid spacing
$h^k$. The goal is to obtain the solution
$u_{ex}^l$ of Eq.\ \ref{eq:luf} on the finest level.
The equations to be iterated on level $k$ take
the form:  

\begin{equation}
L^k u^{k} = f^{k} + \tau^k .
\label{eq:lufcoarse}
\end{equation}

\noindent
where one starts from a trial $u^k$ and improves it. 
The initial $u^k$ on coarse levels is obtained 
by applying the {\it full-weighting restriction}
operator $I_{k+1}^k$ to $u^{k+1}$:

\begin{equation}
u^k = I_{k+1}^k u^{k+1}.
\label{eq:restriction}
\end{equation}

\noindent
The restriction operator takes a local average of the 
finer-grid function. The average is over all 27 
fine grid points (in three dimensions)
including the central point which coincides with
the coarse grid and the 26 neighboring points. 
The weights are: 1/8 for the central point, 1/16
for the 6 faces, 1/32 for the 12 edges, and 1/64
for the 8 corners. The restriction operator 
is a rectangular matrix
of size $N_g^{k+1}$ (columns) by $N_g^k$ (rows)
where $N_g^k$ is the number of grid points on level $k$.
Of course, only the weights need be stored. 
The coarse-grid charge density
$f^k$ is obtained similary from $f^{k+1}$.  
The {\it defect correction} $\tau^k$ is defined as

\begin{equation}
\tau^k = L^k I^k_{k+1} u^{k+1} - I^k_{k+1} L^{k+1} u^{k+1} 
 + I^k_{k+1}\tau^{k+1}.
\label{eq:tau}
\end{equation}

\noindent 
The defect correction is zero on the finest level $l$. Therefore
the third term on the rhs is zero for the grid 
next-coarser to the fine scale.  It is easy to show that
if one had the exact grid solution $u_{ex}^l$ on the finest
level, the coarse-grid equations (Eq.\ \ref{eq:lufcoarse})
would also be satisfied on all levels, illustrating
zero correction at convergence.  Another point of view is
that the defect correction modifies the coarse-grid equations
to `optimally mimic' the finer scales. The defect correction
provides an approximate measure of the discretization errors 
and can be used in the construction 
of adaptive solvers (Brandt, 1984):
higher resolution is placed in regions where the defect 
correction magnitude exceeds a prescribed value. 

The solver begins with initial iterations on the finest level
(typically two or three relaxation steps are adequate on 
each level).  The problem is restricted to the next coarser level
as outlined above, and relaxation steps are performed there.
This process is repeated until
the coarsest grid is reached.  The solver then returns
to the fine level by providing corrections to each
next-finer level and applying relaxation steps there.
The correction equation for grid $k+1$ is

\begin{equation}
u^{k+1} \leftarrow u^{k+1} + I^{k+1}_k (u^k - I^k_{k+1} u^{k+1}).
\label{eq:correct}
\end{equation}

\begin{figure}
\centerline{\epsfig{file=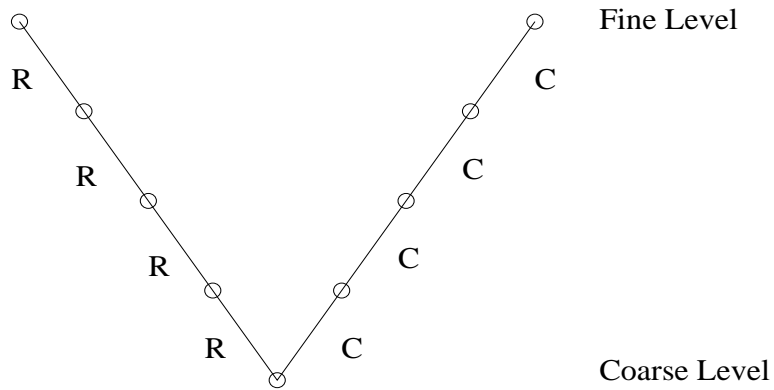,height=4.in,width=2.in,angle=270}}
\vspace{10pt}
\caption{A multigrid V-cycle. Iterations begin on the fine level
on the left side of the diagram. R indicates restriction of the 
problem to the next coarser level.  Corrections (C) begin as
the computations move from the coarsest level to the finest level.}
\label{fig:vcycle}
\end{figure}

\noindent
The additional operator $I_k^{k+1}$ is the interpolation
operator.  Most often it is acceptable to use linear interpolation,
and the easiest way to apply the operator in three dimensions is to
interpolate along the lines in each plane and finally 
to interpolate along lines
between the planes. That is, the operator can be applied by a 
sequence of one-dimensional interpolations. For linear interpolation,
the coarse-grid points which coincide with the fine grid are
placed directly into the fine-grid function, and the intermediate
points get a weight of 1/2 from each neighboring coarse-grid 
point. In the same way, high-order interpolation operators
can be applied as a sequence of one-dimensional operations;
see Beck (1999b) for a listing of the high-order interpolation
weights. The high-order weights are used in interpolating
to new fine levels in full multigrid eigenvalue solvers
(Brandt {\it et al.}, 1993) and in high-order local mesh-refinement
multigrid methods (Beck, 1999b). The interpolation operator
is a rectangular matrix of size $N_g^k$ (columns)
by $N_g^{k+1}$ rows. Only the weights need
be stored, just as for the restriction operator. 
All of the operators 
defined above can be initialized once 
and used repeatedly throughout the algorithm. The multigrid
cycle defined by the above discussion is termed a V-cycle
which is shown schematically in Fig.\ \ref{fig:vcycle}.
Alternative cycling methods have
been employed as well, such as W-cycles.
Reductions in the norm
of the residual in one V-cycle are generally an
order of magnitude. The same set of operations
is employed in a high-order solver; the second-order
Laplacian is simply replaced by the high-order
version.  The form of the multigrid solver is quite
flexible; for example, a lower-order representation could 
be used on coarse levels during the correction cycles. 
In our own work, we have observed similar optimal
convergence rates for high-order solvers as for
second-order ones, so there is no degradation in 
efficiency with order. Applications in electrostatics
and extensions for eigenvalue
problems are discussed in Sections 
\ref{sec:es} and \ref{sec:ks}.

\subsection{Full multigrid}

\label{subsec:fmg}

\begin{figure}
\centerline{\epsfig{file=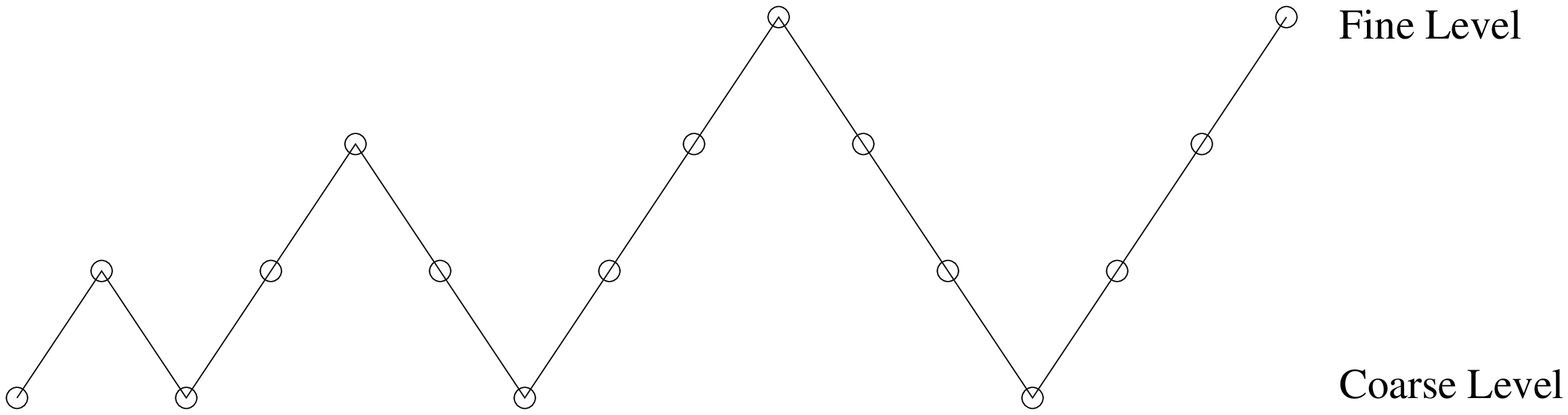,height=5in,width=1.5in,angle=270}}
\vspace{10pt}
\caption{Full multigrid cycle. Iterations begin on the left on the 
coarsest level. The solver proceeds sequentially down to the finest
level, where a good initial approximation is generated from the 
coarse-level processing.}
\label{fig:fmg}
\end{figure}

The grid solution can be efficiently obtained with
one or at most a few V-cycles described above. 
The process obeys linear scaling since the solution
is obtained with a fixed number of multigrid cycles
and each operation on the grid scales linearly 
with the number of grid points.  In three dimensions,
the total grid overhead is $N_{tot} = S_l N_{fine}$,
where

\begin{equation}
S_l = \frac{8}{7} \left( 1 - \frac{1}{8^{l}} \right),
\label{eq:gridoverhead}
\end{equation}

\noindent
and $l$ is the number of levels. In the limit of 
many levels, $N_{tot}$ thus approaches $1.143 N_{fine}$.  
Another development in the multigrid
approach, full multigrid (FMG), can even further 
accelerate the solution process beyond the 
V-cycle algorithm. The idea of FMG
is to begin iterations on the coarsest level. 
The initial approximation there is interpolated
to the next-finer level, iterated, and the new
fine-grid approximation is corrected in a 
V-cycle on that level. This process is repeated
until the finest scale is reached.  The FMG
solver for a Poisson problem is illustrated in
Fig.\ \ref{fig:fmg}. The advantage of this approach is
that a good initial (or preconditioned) 
approximation to the fine-scale function is obtained
on the left side of the final V-cycle.  With this
strategy, the solution to Poisson problems can 
be obtained with a single passage through the
FMG solver. (Self-consistent problems may 
require two or more passages through the final
V-cycle to obtain convergence.) Note that a direct passage via
iterations and interpolation from
coarse to fine scales without the correction cycles
does not guarantee multigrid 
convergence behavior since 
residual long-wavelength errors can remain from
coarser levels.  The multigrid corrections on 
each level serve to remove those errors, leading
to optimal convergence (Brandt, 1984; Hackbusch, 1985;
Briggs, 1987; Wesseling, 1991).

Multigrid solvers have been applied to many problems in
fluid dynamics, structural mechanics, 
electrostatics, eigenvalue problems, {\it etc.} 
The majority of applications have utilized FD-type 
representations, but significant effort has
gone into developing efficient solvers for FE
representations as well (Brandt, 1980; 
Deconinck and Hirsch, 1982; Hackbusch, 1985; 
Braess and Verfurth, 1990; Brenner 
and Scott, 1994).  An additional difficulty with 
FE multigrid methods is a proper representation
of the problem on coarse levels: the more regular
the fine-scale mesh, the easier is the coarsening process.

\section{ELECTROSTATICS CALCULATIONS}

\label{sec:es}

The original formulation of the multigrid method was directed
at solution of linear elliptic equations like the Poisson equation.
Subsequently, methods were developed to handle nonlinear 
problems such as the Poisson-Boltzmann equation of
ionic solution theory.  In this section, applications of
real-space methods to electrostatics problems are discussed.
First, the high efficiency of the multigrid method is demonstrated
by examination of a Poisson problem. Then, 
new mesh-refinement techniques which allow for treatment
of widely varying length scales are examined. 
Poisson-Boltzmann numerical
solvers are discussed, with presentation of some representative
applications in biophysics.   

\subsection{Poisson solvers}

\label{subsec:ps}

\subsubsection{Illustration of multigrid efficiency}

\label{subsubsec:icb}

We investigate a model atomic-like Poisson problem 
which has an analytic solution:

\begin{equation}
\nabla^2 \phi({\bf r}) = -4\pi
\left[\delta ({\bf r}) - \frac{1}{4\pi}
\frac{e^{-r}}{r} \right] .
\label{eq:arfken}
\end{equation}

\noindent
The analytic solution is $\phi(r) = e^{-r}/r$.  The source
singularity is modeled as a single discrete charge
at the origin, and the neutralizing background 
charge value at the origin
is set to give a net charge of zero summed over the 
whole domain.  Here we discretize the problem with a 
12th-order Laplacian on a $65^3$ 
lattice with fine grid spacing $h = 0.25$.  The problem
was solved with the FAS-FMG technique with a single passage
through the FMG process.  Linear interpolation 
and full-weighting restriction were employed for the grid
transfers. The potential was initially
set to zero over the whole domain. 
Three Gauss-Seidel smoothing steps were 
performed on each level.  Several additional smoothing steps
were taken for points just surrounding the singularity
to accelerate the convergence there 
(Bai and Brandt, 1987).  This requires 
virtually no additional effort since only 
few grid points are involved. 

\begin{figure}
\centerline{\epsfig{file=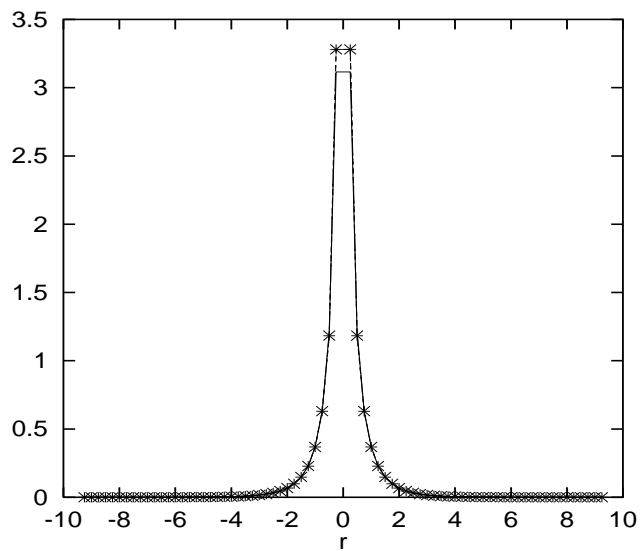,height=3.5in,width=3.in,angle=270}}
\vspace{10pt}
\caption{The electrostatic potential for the screened atomic model. The
analytic curve is the solid line, while the numerical results are the 
crosses. The numerical result deviates noticeably from the analytic 
values at points neighboring the origin due to the source singularity.
The numerical result at the origin has been omitted.}
\label{fig:arfken}
\end{figure}

The solution is obtained 
to within the truncation errors with a total of six
relaxation sweeps on the finest 
level (Fig.\ \ref{fig:arfken}).  Thus the entire
solution process only requires roughly ten times
the effort it takes to 
represent the differential equation on the grid. The total
energy of the charge distribution is $E = -S/4\pi$,
where $S$ is the action of Eq.\ (\ref{eq:action}). 
After the single FMG cycle, the energy is converged
to within 0.00029 au of the fully 
converged energy of 4.31800 au 
(obtained with repeated V-cycles on the finest level).
The final residual (using the 1-norm divided by the 
total number of points, that is the average absolute
value of the grid residuals) is $5 \times 10^{-6}$.  After
1200 Gauss-Seidel iterations on the finest level alone, the residual
is still of magnitude $9 \times 10^{-6}$.  
(With an optimal SOR parameter, the number of iterations to
obtain a residual of $5 \times 10^{-6}$ can be reduced to 
200 iterations.) 
Thus there is 
an enormous acceleration due
to the multiscale processing.  Similar efficiencies
are observed for an FAS-FMG eigenvalue solver
(Section \ref{subsec:eval}). Since the number of iterations
is independent of the number of fine grid points,
these efficiencies are quite general and can be 
routinely expected from a correctly functioning multigrid solver. 

Next we compare the operations count for generating the solution
to Eq.\ \ref{eq:arfken} from scratch using multigrid and FFT methods 
on the same $65^3$ lattice.\footnote{I thank 
Jeff Giansiracusa 
for providing the FFT results.}
The FFT solver required 
$33 \times 10^6$ floating point operations. The multigrid
solver required $27 \times 10^6$, $43 \times 10^6$, 
$75 \times 10^6$, and $106 \times 10^6$ operations for 
the 2nd, 4th, 8th, and 12th order solvers, respectively. 
Therefore, there appears to be no clear advantage in generating
the Poisson potential from scratch with multigrid as opposed 
to FFT.  However, there are some advantages to using 
the real-space multigrid approach: 1) finite and periodic
systems are handled with equal ease, 2) in a quantum simulation
where particles move only slightly from a previous configuration,
the potential can be saved from the previous configuration, 
thus reducing the number of iterations, and 3) one can 
incorporate mesh refinements to reduce the computational
overhead.  For example, if the same problem is solved
with three nested refinement patches centered on the 
singularity, the number of floating point operations is 
reduced by nearly two orders-of-magnitude while the 
accuracy is sufficient since the smooth parts of the potential
away from the singularity can be well represented 
on coarser meshes.  In addition, multigrid methods
can be used to solve nonlinear problems such as
the Poisson-Boltzmann equation with similar efficiencies. 

A situation that arises in many applied electrostatics 
computations is that of strongly varying dielectric
profiles.  Analogous problems occur in steady-state
diffusion problems with widely varying diffusion
coeffcients such as those encountered in neutron diffusion.  
If the coefficients vary by orders of magnitude, 
multigrid efficiency can be lost (Alcouffe
{\it et al.}, 1981).
The reason is that the correct continuity condition across
the boundaries is $\epsilon_1 \nabla \phi({\bf r}_1) = 
\epsilon_2 \nabla \phi({\bf r}_2)$ rather than
continuity of the gradients themselves.  Thus the gradients
vary widely across the boundaries, and the standard
smoothing steps do not properly reduce the errors in 
the function.  Alcouffe {\it et al.} (1981) developed procedures
based on the above continuity condition which
restore the standard multigrid convergence.  In biophysical
applications, the dielectric constant varies from one to 
eighty, so such modifications prove useful for that case
(Holst and Saied, 1993). 

\subsubsection{Mesh-refinement techniques}

\label{subsubsec:mrt}

Many physical problems require consideration of a wide
range of length scales.  One example given in the
Introduction is a transition
metal ion buried inside a protein.  
A protein interacting with a
charged membrane surface is another example: 
particular charged groups 
near the interaction region must be treated accurately,
but distant portions of the protein and membrane do not 
require high resolution to obtain reliable 
energetics.  In electronic structure, the electron
density is very large near the nucleus but is diffuse
further away. A significant strength of real-space methods
lies in the ability to place adaptive refinements in regions
where the desired functions vary rapidly while treating
the distant zones with a coarser description.  

Two approaches exist for such 
refinements in the FD method (FE methods allow quite
easily for grid adaptation): grid curving and local mesh
refinements.  While grid curving is an elegant procedure
for adapting higher resolution in certain regions of 
space, generally the coordinate transformations are 
global. Therefore, the higher resolution tends to spread some
distance from where the refinement is necessary 
(Modine {\it et al.}, 1997, Figs.\ 1, 4, and 5), and 
depending on the geometry of the problem it may be
difficult or impossible to design an appropriate grid
transformation.  Also, 
the transformations can be quite complex leading to 
additional difficulties in the solution process. 
Finally, the grid-curving transformations alter the
underlying spectral properties of the operators which
can in principle lead to degradation of the multigrid efficiency
in the solution process.  However, this does not appear to 
have been a problem in the methods of Gygi and Galli (1995)
and Modine {\it et al.} (1997), although the convergence
behavior of their multigrid Poisson solvers was not 
extensively discussed in those works. Mesh-curving strategies
for electronic structure calculations are 
discussed in Section \ref{subsec:scmrt}.

\begin{figure}
\centerline{\epsfig{file=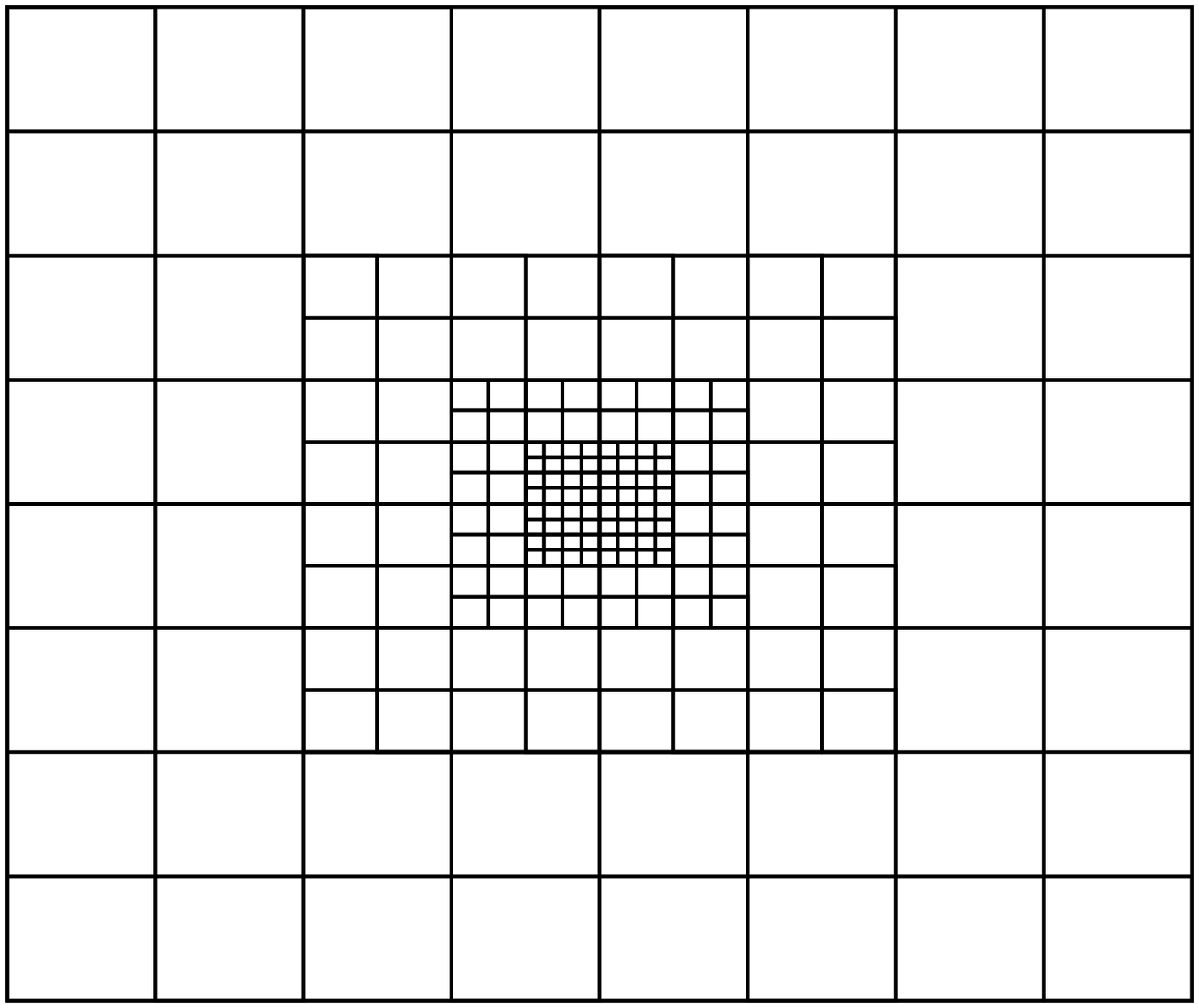,height=3.5in,width=3.5in,angle=270}}
\vspace{10pt}
\caption{Four-level local mesh-refinement grid.}
\label{fig:meshrefine}
\end{figure}

An alternative procedure is to place nested uniform
patches of refinement locally in space 
(Fig.\ \ref{fig:meshrefine}).  Then the
overall structure of the multigrid solver is the 
same,  except fine-level iterations are performed
only over the nested patches.  The same forms for 
the Laplacian, restriction, interpolation, and 
smoothing operators are maintained. This procedure
is highly flexible since the nested refinements can
be centered about any locations of space and can
move as the problem evolves.  The placement of the 
refinements can be adaptively controlled by examination of
the defect correction $\tau^H$; higher resolution should be
placed in regions where $\tau^H$ is large. If an underlying 
FD representation is employed, it is relatively easy to 
extend the method to high-order solvers since
the mesh of the refinement patch is uniform.  

Bai and Brandt (1987) developed an FAS multigrid 
mesh-refinement method for treating widely varying 
length-scale Poisson-type problems.  They first
developed a $\lambda$-FMG exchange-rate algorithm
which minimizes the error obtained for a given amount
of computational work.  Since the number of visits
to the coarse levels (which extend over the whole
domain) is proportional to the number of patches,
direct application of the multigrid
algorithm does not scale strictly linearly for
many levels; the $\lambda$-FMG process restores 
the linear scaling for a solver including the mesh refinements. 
Second, they showed that extra local relaxations around
structural singularities restore asymptotic convergence
rates which can otherwise degrade. Third, they 
developed a conservative-differencing 
technique for handling source singularities.  

To motivate the need for conservative differencing 
in the FAS-FMG mesh-refinement solver, consider Eq.\
(\ref{eq:lufcoarse}) and a two-level problem with one
nested patch.  The defect correction on the coarse
level $H$ is initially defined only over the interior
region of the patch.  However, if one examines the sum
of $\tau^H$ over the refinement, most of the terms in the
interior cancel, but nonzero values remain near the 
boundaries.  The remaining terms closely resemble
flux operators at the boundary.  The net effect 
is thus the introduction of additional sources in the Poisson
equation, which pollutes the solution severely over 
the whole domain.  By balancing the local fluxes with
additional defect correction terms on the patch boundary,
the correct source strength is restored.  Bai and 
Brandt (1987) solved this problem for second-order equations
and tested the method on a source-singularity problem 
in two dimensions. 

Recently the method has been extended to high-order 
FD approximations by Beck (1999b, 2000). The boundary 
defect correction terms were determined by examination
of the noncancelling terms for the high-order approximations. 
Without the conservative scheme, significant errors are
apparent over the whole domain.  With the inclusion of 
the boundary corrections, the sum of $\tau^H$ over the 
patch is zero to machine precision and the correct high-order
behavior is obtained over the whole domain.  The method
was tested on a source-singularity problem in three
dimensions for multiple nested patches.  Typical multigrid
efficiencies were observed. Additional
corrections will be necessary at the boundaries
for continuous charge distributions which cover the 
refinement boundaries, but these are independent of
the order of the Laplacian.  These techniques
are currently being included in high-order FD electronic
structure calculations.  They will significantly reduce
the grid overhead in comparison to uniform-grid calculations
while still maintaining the linear-scaling properties of
the multigrid method.  It is not possible to handle 
truly local refinements with the FFT approach. 

In related work, Goedecker and Ivanov (1998a) 
developed a linear-scaling multiresolution wavelet method for the 
Poisson equation which allows for treatment of widely
varying length scales.  They utilized second-generation
interpolating wavelets since the mapping from grid 
values to expansion coefficients is easy for these 
functions, and they
have a fast wavelet transform.  They solved the Poisson
equation for the challenging case of the all-electron 
uranium dimer. Their solver employed 22 hierarchical levels,
and the potential was obtained to six significant digits.   

\subsection{Poisson-Boltzmann solvers}

\label{subsec:pbs}

As discussed in Section \ref{subsec:cmdft}, the Poisson-Boltzmann
equation arises from the assumption of no ion correlations.
That is, it is a mean-field treatment.  Onsager (1933) showed that
there exists an inherent asymmetry at the Poisson-Boltzmann
level. Nevertheless, calculations performed at this level
of theory can yield accurate energetics for monovalent ions
at moderate concentrations (Honig and Nicholls, 1995;
Tomac and Gr\"{a}slund, 1998; Patra and Yethiraj, 1999).  
Linearization of the 
Poisson-Boltzmann equation restores the symmetry, but
for many cases of experimental interest the linearization
assumption is too severe.  Solution of the Poisson-Boltzmann
equation produces the electrostatic potential throughout 
space, which in turn generates the equilibrium mobile-ion charge 
densities and the total free energy of the ion gas (below). 
By computing the total energies for several macroion configurations,
the potential of mean force due to electrostatic effects 
can be approximated (Rice, 1959). 
In this section, we focus on real-space
numerical methods for solution of the nonlinear Poisson-Boltzmann
equation [Eq.\ (\ref{eq:pb})]. 

Numerical solution of nonlinear partial differential equations
is problematic.  For the Poisson-Boltzmann case, 
the nonlinearities can be severe near fixed
charges since the ratio of the potential to $kT$ can
be large.  Also, strong dielectric discontinuities at the boundary 
of a large molecular ion and the solution create technical 
difficulties.  However, it is known that there is a single stable
minimum of the action functional whose derivative yields
the Poisson-Boltzmann equation (Coalson and Duncan, 1992; Ben-Tal
and Coalson, 1994). Therefore, properly constructed
iterative processes can be expected to locate that minimum.

Early numerical work centered on FD representations.
Nicholls and Honig (1991) developed 
an efficient single-level SOR method
which included special techniques for memory allocation
and for locating the optimal relaxation parameter. For
the test cases considered, between 76 and 184 iterations
were required for convergence.  They also 
observed divergence for some
highly nonlinear cases.  Davis
and McCammon (1989) and Luty {\it et al.}
(1992) used instead a conjugate-gradients relaxation method. 
Between 90 and 118 iterations were required to 
obtain convergence. They observed a factor of at least
two improvement in efficiency in comparison with 
SOR relaxation in their test calculations.  

After the development of multigrid methods for 
solving linear Poisson-type problems, efforts focused
on nonlinear problems.  The FAS algorithm
presented above is well suited for solving nonlinear
problems (Brandt, 1984).  Two modifications are needed: the
driving term $f^h$ on a given level now includes the 
nonlinear terms, and additional terms must be included
in the defect correction to ensure zero correction at
convergence.  The defect correction 
for the Poisson-Boltzmann problem is of the form
(a single monovalent positive ion component
with uniform dielectric is considered here):

\begin{equation}
\tau^H = L^H I_h^H u^h - I_h^H L^h u^h +
\frac{4\pi}{\epsilon} \left[\bar n_+^H e^{-\beta u^H - v^H} - 
\bar n_+^h I_h^H e^{-\beta u^h - v^h} \right] ,
\label{eq:taupb}
\end{equation}

\noindent
where the additional terms reflect the differing representations
of the nonlinear terms on the two levels.   
The concentration on a given level is given by

\begin{equation}
\bar n_+^h = \frac{N_+}{h^3 \sum e^{-\beta u^h - v^h}} .
\label{eq:pbconc}
\end{equation}

\noindent
The sum is over the lattice and
$N_+$ is the number of positive ions in the computation
domain (Coalson and Duncan, 1992). This 
procedure for obtaining the bulk ion concentrations
ensures charge conservation at all steps of iteration. 
Simple smoothing steps can be taken to relax on a given level,
or Newton iterations (Press {\it et al.}, 1992) 
may also be conducted on each level.
Variable $\omega$ parameters may
be required in the relaxation steps due to the 
differing degrees of nonlinearities
on the respective levels. Related multigrid techniques for nonlinear
problems are presented in St\"{u}ben and Trottenberg (1982)
and Hackbusch (1985).  

Holst and Saied (1995) developed a highly efficient
method which combines linear multilevel techniques
with inexact-Newton iterations. 
They compared the 
convergence behavior of the inexact-Newton multigrid
method with SOR and conjugate-gradients minimization
on a single level.  Their multilevel technique converged 
robustly and
more efficiently than the relaxation 
methods on all problems
investigated including challenging source problems
with dielectric discontinuities.  Conjugate gradients 
and SOR exhibited similar convergence
rates when compared with each other. They also examined
a standard nonlinear multigrid method similar to 
that outlined above.  For some cases, the nonlinear
multigrid technique gave good convergence, but 
under certain conditions it diverged.  The authors
thus recommended caution in applying the FAS 
multigrid method directly to the Poisson-Boltzmann equation.
Coalson and Beck (1998) tested the FAS approach
on model problems in the lattice field
theory including source 
singularities and found convergence
for each case.  Oberei and Allewell (1993) have
also developed a convergent multigrid solver for the
Poisson-Boltzmann equation. It is not 
entirely clear at the 
present time whether differences in observed
convergence are due to 
the model problems investigated or to differences in
the algorithms. 

One issue that has not been
addressed to date concerns charge conservation
on the various levels.  The standard form
of the Poisson-Boltzmann equation (Honig and
Nicholls, 1995) assumes fixed
and equal concentrations of the mobile ions at
infinity where the potential is zero.  The ion
charge density can then be expressed as the 
product of a constant term involving the Debye length
of the ion gas and a sinh term involving 
the potential.  Typically the boundary
potential is fixed with the linearized Debye-H\"{u}ckel
value.  This representation conserves
charge if the system size is allowed to go to infinity
due to the infinite extent of the bath.  However, 
charge is not conserved for finite system sizes, and in a 
multilevel procedure differing charge states will
be encountered on the various levels.  
In the lattice field theory of Coalson and Duncan (1992),
on the other hand, the charge is naturally conserved (maintaining
overall charge neutrality) by 
updating the parameters $\bar n_+$ and $\bar n_-$
during each step of iteration (see Eqs.\ 
\ref{eq:pb} and \ref{eq:pbconc}).  Lack of conservation
of charge on the grid levels in the standard approach 
may impact the convergence behavior of a multilevel
solver; this issue deserves further attention.

In a recent study, Tomac and Gr\"{a}slund (1998)
extended the Poisson-Boltzmann level of theory
to include ion correlations in an approximate
way.  They solved for the Kirkwood (1934) hierarchy of
equations on a FD grid assuming  a closure
proposed by Loeb (1951).  Multigrid techniques were used to solve
the initial Poisson-Boltzmann equation and 
to implement the inclusion of ion correlations.
Coarser grids were used to estimate the 
fluctuation term, and the impact of large grid
spacing on the accuracy of the correlation term
was examined.  Excellent agreement with previous
theoretical results and Monte Carlo simulation
was obtained for divalent ion distributions 
around a central sphere.  Test calculations were
also performed on ion distributions around 
an ATP molecule. This work allows for the more accurate
treatment of systems containing multivalent ions.
The computational expense of obtaining the fluctuation
contribution is extensive, however. 
 
What is clear from the multigrid studies to date is 
that multilevel methods can yield
solutions to the Poisson-Boltzmann equation 
(and its modifications to include ion correlations) 
with efficiencies resembling those for linear 
problems and with linear-scaling
behavior. Hence, they show a great deal of promise for large-scale 
colloid and biophysical applications. Under some circumstances,
special measures may be necessary to obtain correct multigrid
convergence efficiencies. To my knowledge, 
all FD Poisson-Boltzmann calculations
so far have employed second-order Laplacians; going
to higher orders improves accuracy for little additional
cost, so higher-order solvers should be considered. 
However, high-order techniques near dielectric discontinuities 
introduce some additional complexity. 

In addition to FD-related methods for solving
the Poisson-Boltzmann equation, FE solutions
have appeared.  The FE discretization
leads to a more accurate physical representation
of complex molecular surfaces at the expense of 
additional computational overhead. You and 
Harvey (1993) developed a three-dimensional
FE method for solving the linearized Poisson-Boltzmann
equation.  More accurate results were obtained with
the FE approach compared with FD solutions in 
model problems. Potential distributions were
computed surrounding tRNA molecules and the
enzyme superoxide dismutase.  This was the first
application of the FE method to large-scale
biological macromolecular electrostatics. 
Cortis and Friesner (1997) formulated a method for
constructing tetrahedral FE meshes around
macromolecules. The authors discussed the relative
merits of FD and FE representations including
applications of multilevel methods in their solution.
They used their discretization procedure to solve
the linearized Poisson-Boltzmann equation. 
Bowen and Sharif (1997) presented a FE numerical
method for solution of the nonlinear Poisson-Boltzmann
equation in cylindrical coordinates.  Adaptive
mesh refinements were employed to gain accuracy
near curved surfaces.  They considered
applications to membrane separation processes by
examining the case of a charged spherical particle
near a cylindrical pore.   Alternative formulations
of electrostatic problems include boundary element
methods which reduce Poisson problems to calculations
involving the molecular surface (Yoon and Lenhoff, 1990, 1992;
Pratt {\it et al.}, 1997). 
They lead to dense matrix representations of the problem;
if nonlinear salt effects are to be included, volume integrals,
in addition to surface integrals, must be incorporated. 

\subsection{Computations of free energies}

\label{subsec:pbfe}

Several proposals have appeared concerning computation
of free energies of the ion gas
once the solution of the Poisson-Boltzmann
equation is obtained.  The free energies are crucial for
determining electrostatic interaction energies of 
charged macromolecules at the mean-field level. The 
energies can be obtained either by charging methods
or volume/surface integrations 
(Verwey and Overbeek, 1948; Marcus, 1955; Rice, 1959; 
Reiner and Radke, 1990).
The most commonly used volume integration
approach stems from the variational formulation of Sharp
and Honig (1990b).  They postulated a form for the free
energy which, when extremized, produces the Poisson-Boltzmann
equation.  Fogolari and Briggs (1997) critiqued this 
variational form, showing that the extremum in the 
free energy is a maximum, not a minimum, with respect
to variations of the potential. They presented another
form which is minimized. The lattice field theoretic
free energy is derived from a rigorous representation
of the grand partition function 
of the ion gas (Coalson and Duncan, 1992). 
In this section, we will derive the variational form
from the lattice field theory formulation to illustrate
differences between the two; the variational form
is obtained from the infinite system size limit of 
the lattice field theory. We assume here the case of
uniform dielectric and monovalent ions; the extensions
for variable dielectric and higher valences follow 
the same arguments.  

The mean-field lattice field theory Helmholtz free energy
is

\begin{equation}
\beta F = -S_{LFT} + N_+ \ln (\bar n_+ h^3) + N_- \ln (\bar n_- h^3) ,
\label{eq:lftfe}
\end{equation}

\noindent
where $S_{LFT}$ is an action term (defined below)
and $N_+$ and $N_-$ are the 
total numbers of positive and negative mobile ions in the
calculation domain. In order to handle periodic as well as
finite domains, we assume that the total number of mobile
and fixed charges is such that overall charge neutrality is 
maintained.  The free energy Eq.\ (\ref{eq:lftfe}) is invariant to 
a uniform shift of the potential, which is the 
correct physical result.  The concentration 
$\bar n_+$ is given by Eq.\ (\ref{eq:pbconc}), while
$\bar n_-$ is obtained by the analogous formula for
negative charges. 

Consider the action which, when minimized, results in 
the Poisson-Boltzmann equation:

\begin{equation}
S = -\frac{1}{2} \int \phi \nabla^2 \phi d^3x - \frac{4\pi}{\epsilon}
\int \left[ \rho_f \phi - \frac{\bar n_+}{\beta}
e^{-\beta\phi - v} - \frac{\bar n_-}{\beta}
e^{\beta\phi - v} \right] d^3x .
\label{eq:pbaction}
\end{equation}

\noindent 
The action of Eq.\ (\ref{eq:lftfe}) is related to $S$ by

\begin{equation}
S_{LFT} = \frac{\beta\epsilon}{4\pi} S.
\label{eq:slfts}
\end{equation}

\noindent
Then the total Helmholtz free energy on the lattice is

\begin{equation}
\beta F = \frac{\beta\epsilon h^3}{8\pi} \sum \phi^h L^h
\phi^h  + \beta h^3 \sum \rho_f^h \phi^h  
+ N_+ \ln (\bar n_+ h^3/e) + N_- \ln (\bar n_- h^3/e) ,
\label{eq:latticefe}
\end{equation}
 
\noindent
where the grid potential is $\phi^h$, and the sums are over
the lattice points. 

\begin{figure}
\centerline{\epsfig{file=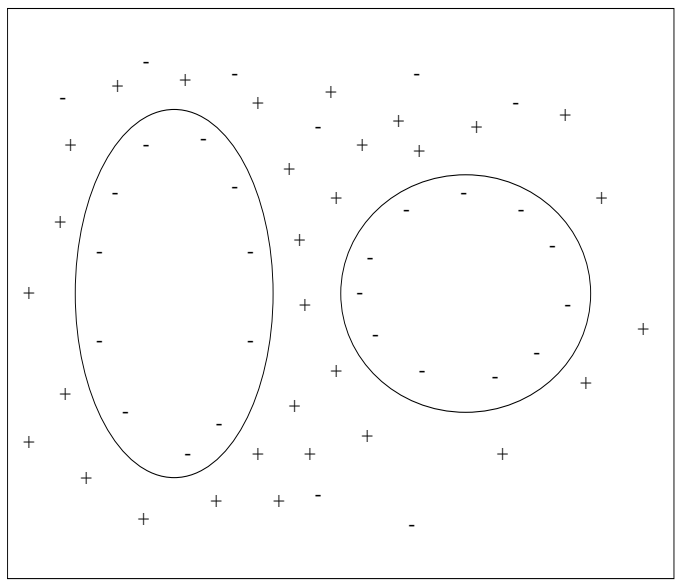,height=3.5in,width=3.in,angle=270}}
\vspace{10pt}
\caption{Schematic of two colloid particles located in a solution 
containing counterions and salt. The potential decays toward zero
at locations distant from the colloids due to exponential screening.}
\label{fig:colloid}
\end{figure}

Let us examine a process in which two macroions are moved
relative to each other (Fig.\ \ref{fig:colloid}).  The macroions
are assumed to reside in a large calculation domain which contains
counterions plus perhaps salt ions, so the potential is 
screened at large distances.  We assume that the potential
decays effectively to zero some finite distance from the ions and is
zero all the way to the boundaries. The mobile ions behave as 
an ideal gas where the potential is zero.  The numbers of 
fixed and mobile ions is maintained constant throughout 
the process.  Now, consider the free energy
change from the `activity' term for the positive mobile ions
upon moving from configuration 1 to 2:

\begin{equation}
\beta \Delta F_{a+} = N_+ \ln \frac{\sum e^{-\beta\phi^h_1 - v_1^h}}
{\sum e^{-\beta\phi^h_2 - v_2^h}}. 
\label{eq:deltafact}
\end{equation}

\noindent
Call the number of free sites in the domain where the potential
is effectively zero $N^f_1$ and $N^f_2$. The sums over
regions where the potential is nonzero are labelled
$\Sigma_1$ and $\Sigma_2$.  The free energy change
$\Delta F_{a+}$ is then

\begin{equation}
\beta \Delta F_{a+} = N_+ \ln \frac{\left[ N^f_1 + \Sigma_1 \right]}
{\left[ N^f_2 + \Sigma_2 \right]}.
\label{eq:deltafact1} 
\end{equation}

\noindent
Factor out the $N_f$ terms:

\begin{equation}
\beta \Delta F_{a+} = N_+ \ln \frac{N^f_1 \left[ 1 + \frac{\Sigma_1}{N^f_1} \right]}
{N^f_2 \left[ 1 + \frac{\Sigma_2}{N^f_2} \right]}. 
\label{eq:deltafact2}
\end{equation}

\noindent
The term involving the ratio of the free sites can
be represented as

\begin{equation}
N_+ \ln \frac{N^f_1}{N^f_2} =
 N_+ \ln \frac{N'_{tot} \left[ 1 - \frac{\Sigma_{1g}}{N'_{tot}} 
\right]}
{N'_{tot} \left[ 1 - \frac{\Sigma_{2g}}{N'_{tot}} \right]} , 
\label{eq:deltafact3}
\end{equation}

\noindent
where $N'_{tot}$ is the total number of grid points outside any 
excluded volume regions and $\Sigma_{1g}$ and $\Sigma_{2g}$ 
count the numbers of grid points outside of excluded
volume zones where the potential is nonzero. 

For very large system sizes, the above expressions can be 
approximated as

\begin{equation}
\beta \Delta F_{a+} \approx N_+ \left( \frac{\Sigma_1}{N^f_1}
 - \frac{\Sigma_2}{N^f_2} \right) 
+ \frac{N_+}{N'_{tot}} \left( -\Sigma_{1g}  
+ \Sigma_{2g} \right). 
\label{eq:deltafact4}
\end{equation}

\noindent
Analogous terms are obtained for the negative ion case. 
As the system size approaches infinity, we can make
the further approximations $N_+ \approx N_-$
and $N_1^f \approx N_2^f \approx N'_{tot}$.  

The resulting free energy change for both ionic species is then

\begin{equation}
\beta \Delta F_a = \frac{N_+}{N'_{tot}}
\left[ \sum \left( e^{-\beta\phi^h_1}
+ e^{\beta\phi^h_1} - 2 \right)e^{-v_1^h} - \sum \left(
e^{-\beta\phi^h_2} + e^{\beta\phi^h_2} -2 \right)e^{-v_2^h} \right].
\label{eq:deltafact5}
\end{equation}

If we call the grid concentration $c_g = N_+/N'_{tot} = N_-/N'_{tot}$,
then the free energy change can be written as

\begin{equation}
\beta \Delta F_a =  -2 c_g \left[
\sum \left(
\cosh (\beta\phi^h_2)  - 1 \right)e^{-v_2^h}  -
\sum \left( \cosh (\beta\phi^h_1) 
 - 1 \right)e^{-v_1^h} \right].  
\label{eq:deltafact6}
\end{equation}

The grid `activity coefficient' $\gamma$ is $c_g/h^3$ which
can be assumed to be $\gamma = \sqrt{\bar n_+ \bar n_-}$. 
In the continuum limit the free energy change due to the 
logarithmic terms in the total free energy is thus

\begin{equation}
\beta \Delta F_a =  -2 \gamma \left[
\int \left[
\cosh (\beta\phi_2)   - 1 \right]e^{-v_2} d^3x  -
\int \left[ \cosh (\beta\phi_1) 
 - 1 \right]e^{-v_1} d^3x \right].
\label{eq:deltafact7}  
\end{equation}

\noindent
The overall free energy change can then be written as the difference
of two terms, one for each configuration:

\begin{equation}
\beta F = \beta \int \rho_f \phi d^3x -2 \gamma \int
\left[ \cosh (\beta\phi) -1 \right]e^{-v} d^3x
+ \frac{\beta\epsilon}{8\pi} \int \phi \nabla^2 \phi d^3x
\label{eq:honigfe1}
\end{equation}

So long as the potential and/or its derivative go to 
zero on the boundaries, Eq.\ (\ref{eq:honigfe1}) can
be rewritten as

\begin{equation}
\beta F = \beta \int \rho_f \phi d^3x -2 \gamma \int
\left[ \cosh (\beta\phi)  -1 \right]e^{-v} d^3x
- \frac{\beta}{8\pi} \int \epsilon \left| \nabla \phi\right| ^2 d^3x
\label{eq:honigfe2}
\end{equation} 

\noindent
which is identical to Eq.\ (13) in Sharp and Honig (1990b). 
Thus the variational free energy of Sharp and Honig
is derived as the infinite system limit of the lattice
field theory expression, where the potential is assumed to 
go to zero at the distant boundaries.  Since the 
variational form is not invariant to a uniform shift 
of the potential, some arbitrariness is introduced.
In addition, charge conservation is not maintained
as discussed above. 
The issue of charge conservation is particularly 
relevant if one considers periodic boundary domains. 
Therefore, it is recommended to use the lattice field
theoretic form for computations of free energies for cases
where these considerations are deemed important.

\subsection{Biophysical applications}

\label{subsec:bioapp}

One reason for a resurgence of interest in continuum models
of solvation for large macromolecules is 
that, for many systems of interest, the
total number of particles is simply too large to accurately
model at the atomic level.  For example, consider a protein
interacting with a DNA strand: the atomistic treatment 
including solvent and salt effects would involve several tens of
thousands of atoms, and the motions occur over time scales
longer than nanoseconds.  So long as the energetics are
proven to be reasonable in testable model calculations, 
some confidence can be placed in the Poisson-Boltzmann
calculations on larger systems. The number of applications
of Poisson-Boltzmann-level theory to biological macromolecules
is now very large.  Previous reviews summarize progress in 
this area (Sharp and Honig, 1990a; Honig and Nicholls, 1995).  
A few representative studies from the main
categories of application are presented here to give a flavor
of the types of problems which are accessible. 

The first type of application concerns the computed 
average electrostatic potential and the resulting 
charge distributions. Haggerty and Lenhoff (1991) performed
FD calculations to generate the electrostatic potential
on the surfaces of proteins.  They found a clear 
correlation between retention data in ion-exchange
chromatography and the average protein surface potential.
Ion-exchange chromatography is one of the important 
techniques for separating mixtures of proteins. 
Montoro and Abascal (1998) compared Monte Carlo 
simulations and FD Poisson-Boltzmann calculations on 
distributions of monovalent ions around a model
of B-DNA.  They found good agreement between
the simulations and Poisson-Boltzmann calculations
for low to moderate ion concentrations, but 
for concentrations above 1 M, the agreement deteriorates.  
Pettit and Valdeavella (1999) compared electrostatic
potentials obtained from molecular dynamics simulations
and Poisson-Boltzmann calculations for a tetra-peptide.
They observed qualitative differences for the electrostatic
potentials around the peptide.  However, they argued that
the free energies obtained by integration over the 
entire domain include cancellation of errors and 
yield more reliable results compared with the potential
itself.  Patra and Yethiraj (1999) developed a DFT 
method for the ion atmosphere 
around charged cylinders (a model for DNA or tobacco 
mosaic virus). 
Their theory includes contributions from finite ion 
size and ion correlations beyond the mean-field level. 
Their DFT approach gave good agreement with simulations
for both monovalent and divalent ion atmospheres.
The Poisson-Boltzmann level theory does well for 
low axial charge densities on the cylinder.  
Interesting charge inversion effects were seen for 
divalent salts which are entirely absent from
the Poisson-Boltzmann calculations.  Recently
Baker {\it et al.} (1999) developed a highly adaptive
multilevel FE method for solving the Poisson-Boltzmann equation. 
By placing adaptive meshes in the regions of the dielectric
discontinuities, large reductions in overall computation
cost were observed.  Computations were performed
to obtain the electrostatic potential 
around large protein and DNA systems. 

The second utility of Poisson-Boltzmann calculations
lies in the computation of free energies and resulting
interaction energies for variable macromolecule 
conformations.  Yoon and Lenhoff (1992) 
used a boundary-elements method to compute interaction energies
for a protein and a negatively charged surface at the linearized
Poisson-Boltzmann level.  They found the most favorable
orientation with the protein active site facing 
the surface.  Zacharias {\it et al.} (1992) investigated
the interaction of a protein with DNA utilizing the 
FD Poisson-Boltzmann method. They studied the distribution
of ions in the region between the two species and
the energetics for protein binding. The interaction
energy depends strongly on the charge distributions
on the DNA and protein.  The computed number of ions
released upon complexation agreed well with experiment. 
Misra {\it et al.} (1994) performed FD Poisson-Boltzmann
calculations to study the influence of added salt on 
protein-DNA interactions.  Long-range salt effects
play a significant role in relative stabilities of
competing structures of protein-DNA complexes. 
Ben-Tal {\it et al.} (1997) examined electrostatic effects
in the binding of proteins to biological membranes. 
The binding constant for the protein-membrane complex
was successfully compared with experimental data.
Chen and Honig (1997) extended their FD Poisson-Boltzmann
method to mixed salts including both monovalent and divalent
species.  They found that, for pure salt cases, the electrostatic
contribution to binding varies linearly with the logarithm
of the ion concentration; for divalent salts, nonlinear 
effects were observed due to competitive binding of the
two ionic species.  

A third type of problem addressed with Poisson-Boltzmann 
level computations is the determination of pH-dependent 
properties of proteins.  Since the net charge of the protein
is crucial in understanding its properties, a predictive
method is desired for computing electrostatic effects
(due to other charged groups) on 
the pK$_a$'s of ionizable groups.  Antosiewicz {\it et al.}
(1994, 1996) presented extensive calculations on a large
data set for several proteins.  Somewhat surprisingly,
they obtained the best
agreement with experiment assuming an interior
dielectric constant of 20 for the protein.  Possible
explanations of this effect were discussed, including
approximate accounting for specific ion binding
and conformational relaxation of the protein. They also
found improvements if NMR structural sets were used
as opposed to single crystal X-ray structures for the proteins. 
Vila {\it et al.} (1998) recently performed 
boundary-element multigrid calculations to determine pK$_a$
shifts; they obtained excellent agreement with 
experiment for polypentapeptides.  

Fourth, Poisson-Boltzmann methods have been incorporated
into electronic structure calculations to study 
solvation effects.  As an example,
Fisher {\it et al.} (1996) performed
DFT electronic structure calculations on a 
model for the manganese 
superoxide dismutase enzyme active site. The region treated
explicitly included 37 or 38 atoms (115
valence electrons).  The surrounding
solvent was modeled as a dielectric 
continuum (water).  The
electronic structure was computed self-consistently 
by updating the reaction-field potential due to 
the solvent following calculations with fixed 
potential.  Typically, the continuum solvation
procedure converged within seven iterations. 
The authors computed redox potentials and 
pK$_a$'s for the complex.  Differences from
measured redox potentials were observed, and
the authors stressed the 
importance of explicitly including
electrostatic effects from the rest of the 
protein in the calculations. 

As a final biophysical real-space application, a  
lattice relaxation algorithm has been
developed by Kurnikova {\it et al.} (1999) to 
examine ion transport through membrane-bound 
proteins.  The coupled Poisson and steady-state
diffusion equations (Poisson-Nernst-Planck or PNP equations)
were solved self-consistently on a FD real-space grid for motion
through a membrane protein, the Gramicidin A dimer.
The charges embedded in the channel interior had a large
impact on computed diffusion rates. The computed 
current-voltage behavior agreed well with experimental
findings.  The accuracy of the continuum mean-field treatment
is encouraging for the further study of ion
transport through a wide range of membrane proteins. A
recent study (Corry {\it et al.}, 2000) has critiqued the
mean-field approach for narrow ion channels, so some 
modifications in the PNP theory may be required 
for those cases. 
 
\section{SOLUTION OF SELF-CONSISTENT EIGENVALUE PROBLEMS}

\label{sec:ks}

Eigenvalue problems arise in a wide range of applications.
Solution of the Schr\"{o}dinger equation with fixed 
or self-consistent potential is of course a dominant
one.  However, eigenvalue problems occur in several other
areas.  Included are computation of modes and 
frequencies for molecular vibrations (Jensen, 1999) 
and optical modes of  
waveguides (Coalson {\it et al.}, 1994). 
Self-consistent eigenvalue problems also
arise in polymer theory (Tsonchev {\it et al.}, 1999).  
This section reviews recent 
research on real-space methods for fixed and 
self-consistent potential eigenvalue problems. The main 
focus is on novel methods for solving the Kohn-Sham
equations in electronic structure. Additional discussion concerns 
applications in semiconductor and polymer physics.  

\subsection{Fixed-potential eigenvalue problems in real-space}	

\label{subsec:eval}

\subsubsection{Algorithms}

\label{subsubsec:algorithms}

Let us consider the problem of minimizing the total energy
for a single quantum particle subject to the constraint that the 
wavefunction must be normalized. With the inclusion of a
Lagrange multiplier term for the constraint, the energy functional reads

\begin{equation}
E[\psi({\bf r})] = -\frac{1}{2} \int \psi^*\nabla^2\psi d^3x 
 + \int \psi^* V \psi d^3x - \lambda\int \psi^*\psi d^3x ,
\label{eq:evalaction}
\end{equation}

\noindent
where $\lambda$ is the Lagrange multiplier. If multiple
states are desired, then the single Lagrange multiplier
becomes a matrix of multipliers designed to enforce
orthonormality of all the eigenfunctions. The `force'
analogous to Eq.\ (\ref{eq:force}) is then

\begin{equation}
-\frac{\delta E}{\delta \psi^*} = \frac{1}{2}\nabla^2 \psi -
V\psi + \lambda \psi .
\label{eq:evalforce}
\end{equation}

\noindent
When the force term is set to zero indicating location of
the minimum, the eigenvalue equation for the ground
state results.  Discretizing this equation on a one-dimensional 
grid leads to the second-order FD representation
of the Schr\"{o}dinger equation:

\begin{equation}
-\frac{1}{2h^2}\left[ \begin{array}{rrrr}
-2&1&0&\ldots \\
1&-2&1&\ldots \\
0&1&-2&\ldots \\
\vdots&\vdots&\vdots&\ddots
\end{array} \right]
\left[ \begin{array}{r}
\psi(x_1) \\
\cdot \\
\cdot \\
\cdot \\
\psi(x_N)
\end{array} \right]
+
\left[ \begin{array}{rrrr}
V(x_1)&0&0&\ldots \\
0&V(x_2)&0&\ldots \\
0&0&V(x_3)&\ldots \\
\vdots&\vdots&\vdots&\ddots
\end{array} \right]
\left[ \begin{array}{r}
\psi(x_1) \\
\cdot \\
\cdot \\
\cdot \\
\psi(x_N)
\end{array} \right]
=
\lambda\left[ \begin{array}{r}
\psi(x_1) \\
\cdot\\
\cdot\\
\cdot\\
\psi(x_N)
\end{array} \right] 
\label{eq:evalmatrix}
\end{equation}

\noindent
Solution of this matrix equation with standard diagonalization
routines (excluding the Lanczos and multigrid methods) results
in an $N_g^3$ scaling of the solution time, where $N_g$ is the 
number of grid points.  Since the matrix is sparse, iterative
techniques are expected to lead to increased efficiencies,
just as for Poisson problems. We can note that the solution of 
Eq.\ (\ref{eq:evalmatrix}) is a nonlinear problem since we
seek both the eigenvalues and eigenvectors.  In this 
section, we consider necessary extensions of the FAS-FMG method
for the eigenvalue problem and discuss applications 
of FD and FE real-space methods for fixed-potential cases.
Clear discussion of alternative Lanczos and related algorithms 
(such as the conjugate-gradient, GMRES, and Jacobi-Davidson algorithms)
for handling sparse matrix diagonalization is
given in Golub and van Loan (1996) and Booten and
van der Vorst (1996). 

The derivation of the FD matrix eigenvalue equation
above parallels that for the Poisson problem.
The additional complexities introduced are: 1) the
necessity of solving for multiple eigenfunctions, 2) 
computation of eigenvalues, and 3) enforcement
of orthonormality related constraints. Brandt {\it et al.}
(BMR, 1983) extended the FAS-FMG algorithm to eigenvalue
problems.  Hackbusch (1985) discussed related eigenvalue methods.
The algorithm of BMR allows for 
fully nonlinear solution of the eigenvalue problem;
due to the nonlinear treatment,
the eigenvalues and constraint equations only need to
be updated on the coarsest level where the computational
expense is small. One exception to the previous statement
in the original BMR algorithm is a
Ritz projection (below) on the finest level at the end of 
each V-cycle, preceded by a Gram-Schmidt orthogonalization.
Costiner and Ta'asan (1995a) have since extended the method
to process the Ritz projection on coarse levels as well. 

The same basic FAS-FMG procedure is followed in the
BMR eigenvalue algorithm as discussed
above for Poisson problems.  The Laplacian operator
$L^h$ in Eq.\ (\ref{eq:luf})
is replaced by the real-space Hamiltonian minus 
the eigenvalue $\lambda_i$.   There is no 
source term $f^h$, and there are $q$ equations,
where $q$ is the number of eigenfunctions.  Since the orthogonalization
constraints are global operations involving integrals
over the whole domain, these processes can be performed
on the coarse levels. The relaxation sweeps (two or three) on finer
levels smooth the high-frequency errors and do not
destroy the existing orthonormality of the functions;
of course, if many unconstrained iterations were
performed on fine levels, all wavefunctions would begin
to collapse to the ground state.  Linear interpolation
and full-weighting restriction are sufficient, but 
use of cubic interpolation results in more accurate
eigenfunctions upon entry to a new finer level. 
A direct Gram-Schmidt orthogonalization 
is not applicable on coarse levels;
if the exact grid solution is restricted to the coarse
levels, the resulting eigenfunctions are no longer 
orthonormal.  Therefore, to satisfy the zero correction
at convergence condition, a coarse grid matrix equation for the 
constraints is

\begin{equation}
\langle u_i^H, I^H_h u_j^h \rangle = \langle
I_h^H u_i^h , I_h^H u_j^h \rangle .
\label{eq:constraints}
\end{equation}

\noindent
Solution requires inversion of a $q \times q$ matrix. 
The inversion can 
be effected by direct matrix methods if $q$ is small
or iterative procedures as performed by BMR
in their solver. 
The grid overhead for 
the operation is very small since it is performed on the 
coarsest level; for example, if three levels were employed in
the eigenvalue solver, the coarse grid operations would require
1/64 the effort compared to the fine scale in three dimensions. 
An additional consideration in the eigenvalue problem is 
that the coarse grid must contain enough points to `properly
resolve' the eigenfunctions; BMR give 
a criterion of $N_{cg} = 4q$ for the required number
of points. 

The eigenvalues can also be updated on the coarse levels
by inclusion of the defect correction:

\begin{equation}
\lambda_i = \frac{<{\mathcal{H}}^Hu_i^H-
\tau_i^H,u_i^H>}{<u_i^H,u_i^H>} .
\label{eq:cgeval}
\end{equation}

\noindent
The grid Hamiltonian on the coarse level is ${\mathcal{H}}^H$. 
The same set of eigenvalues applies on all levels with this
formulation.  Relaxation steps are performed on each level
with Gauss-Seidel iterations. 

A final addition to the FAS-FMG technique in the eigenvalue
algorithm of BMR is a Ritz projection
performed at the conclusion of each V-cycle in the FMG 
solver.  The purpose of this step is to improve the occupied
subspace by making all residuals orthogonal to that
subspace.  The eigenfunctions are first orthogonalized with
a Gram-Schmidt step and the 
$q \times q$ Hamiltonian matrix in 
the space of the occupied orbitals is diagonalized.  
The orbitals are then corrected. This step improves the convergence
rate.  The Ritz projection can be written as

\begin{equation}
\omega^T {\mathcal{H}}^h \omega z_i -\lambda_i z_i = 0 ,
\label{eq:ritz}
\end{equation}

\noindent
where $\omega$ is the $q \times N_g$ ($N_g$ is the total number of
grid points) matrix of the eigenfunctions, ${\mathcal{H}}^h$ is the 
grid Hamiltonian, and the $z_i$ are the solved-for coefficients 
used to improve the occupied subspace.  We have closely followed
this algorithm in our own work with two changes: 1) we update the 
eigenfunctions simultaneously (as opposed to sequentially in 
the original algorithm) and 2) high-order approximations are
used in the FD Hamiltonian.  

In the form presented above, the algorithm exhibits 
$q^2N_g$ scaling due to the Ritz projection on 
the fine scale. The scaling of the relaxation
steps is $qN_g$ so long as the orbitals span the
entire grid.  If a localized representation of the 
orbitals is possible (Fattebert and Bernholc, 2000), 
then linear scaling of 
each step in the algorithm
results. Further discussion of the scaling
of each operation is presented in 
Wang and Beck (2000). Costiner and Ta'asan (1995a)
have generalized the BMR algorithm
in several ways.  They transferred the Ritz projection
step to coarse grids and added a backrotation to 
prevent rotations of the solutions in subspaces of
equal or close eigenvalues.  They also developed
an adaptive clustering algorithm for handling
groups of eigenfunctions with near eigenvalues. 
The scaling of their algorithm
is $qN_g$ when the eigenfunctions span the 
entire grid.   Several numerical experiments
in two and three dimensions
demonstrated the high efficiency of their method, 
and the method was extended to handle self-consistency
(Costiner and Ta'asan, 1995b).

\subsubsection{Applications}

\label{subsubsec:evalapps}

To demonstrate the efficiency of the BMR FAS-FMG eigenvalue solver,
consider the three-dimensional hydrogen atom.  While
this may seem a very simple case, it presents 
numerical difficulties for a real-space method due
to the presence of the Coulomb 
singularity in the potential.  In addition, 
the s-orbitals exhibit cusps
at the singularity and the $l > 0$ angular 
momentum states are degenerate. 
Beck (1999a) presented numerical results for 
the hydrogen atom which exhibit the excellent
convergence characteristics of the nonlinear FAS-FMG
eigensolver.  The potential was generated numerically
with a 12th-order Poisson solver as described above. 
The grid was taken as a 65$^3$ Cartesian lattice. The boundary 
potentials were set to the analytical $1/r$ values. 
The fine grid uniform spacing was $h = 0.5$ au, and a 
12th-order FD discretization was employed.  Five 
eigenfunction/eigenvalue pairs were computed. The fully 
converged eigenvalues (obtained by repeated V-cycles
on the finest scale) were -0.50050 for the 1s state,
-0.12504 for the 2s state, and -0.12496 for the three 2p
states (which are degenerate out to 10 decimal places
when fully converged), so the results are accurate to better
than $kT$.  The eigenvalues were converged to five decimal
places following one passage through the FAS-FMG 
solver with three relaxation sweeps on each level 
on each side of the V-cycles. Thus, only six fine-scale 
applications of the Hamiltonian to the wavefunctions
were required to obtain the solution.  The major 
computational cost for this system occurred during the relaxation
steps on the fine scale.  The total solution time was
roughly 90 seconds on a 350 MHz Pentium II machine. 
These results show that similar convergence behavior
can be expected for eigenvalue solvers as for Poisson
solvers so long as the nonlinear FAS-FMG methodology is followed. 
Mesh refinements will yield comparable 
accuracies with much less numerical overhead. The required
high-order methods are now in place (Beck, 1999b) and
are being incorporated into the eigenvalue solver. 

We now consider related efforts at efficient solution
of real-space fixed-potential eigenvalue problems. 
Grinstein {\it et al.} (1983) developed a 
second-order FD multigrid
method to solve for a single eigenfunction.  They
employed an FAS-FMG approach and  
used a Gauss elimination method to exactly solve
the equations on the coarsest level. Since they 
solved for single eigenfunctions, constraints 
were not necessary. The 
eigenvalue was fixed and not computed, so the 
problem was effectively linear.  Seitsonen {\it et al.}
(1995) solved fixed-potential eigenvalue problems
using a high-order FD representation and a 
conjugate-gradient method for obtaining the 
eigenfunctions and eigenvalues. They tested
their method on the P$_2$ dimer and obtained
rapid convergence of the approximation with
decreasing grid size. The representation of 
the wavefunctions was better than in corresponding
plane-wave calculations. They also computed 
eigenfunctions for positron states centered
at a Cd vacancy in CdTe.   

Extensive effort has also been applied to development
of FE methodology for fixed-potential problems. 
Hackel {\it et al.} (1993) proposed a two-dimensional
FE method in which Coulomb singularities were handled
with condensed special elements around the nuclei.  
Test calculations were
performed on the linear H$_3^{2+}$ molecule, and
highly accurate results (to 10$^{-7}$ au)
were obtained. Ackerman and Roitzsch (1993) proposed
an adaptive multilevel FE approach which utilized
high-order shape functions. Inverse iteration was used
to solve the large-dimension eigenvalue problem for
the two-dimensional harmonic oscillator and the
linear H$_3^{2+}$ molecule; accuracies comparable 
or even superior to 
the previous study were reported. Subsequently, 
they extended their method to three dimensions
(Ackerman {\it et al.}, 1994); in this work,
conjugate-gradient techniques were employed to
solve the eigenproblem. Results were presented
for the three-dimensional harmonic oscillator
and H$_3^{2+}$ in the equilateral triangle
geometry.  Sugawara (1998) presented a hierarchical
FE method in which the mesh points and polynomial
orders are generated adaptively to gain high
accuracy.  The method was tested on the one-dimensional 
harmonic oscillator.  Batcho (1998)
proposed a spectral element method which combines
ideas from FE and collocation approaches. The
Coulomb singularity was treated with a Duffy (1982) 
transformation. Pask {\it et al.} (1999)
have recently developed a FE method 
for periodic solid-state
computations. The method uses a flexible 
$C^0$ piecewise-cubic
basis and incorporates general Bloch boundary
conditions, thus allowing arbitrary sampling
of the Brillouin zone. 
Band structure results were presented which
illustrate the rapid convergence of the method
with decreasing grid size. The authors emphasized
the structured, banded, and variational properties
of the FE basis.  Sterne {\it et al.} (1999)
subsequently applied the method to large-scale
{\it ab initio} positron calculations for systems
of up to 863 atoms.

\subsection{Finite-difference methods for self-consistent problems}

\label{subsec:hofdm}

In this section, we begin our examination of real-space
methods for solving self-consistent eigenvalue problems
with a discussion of FD methods. The focus here is
mainly on the basic FD formulation and its relationship
to other numerical methods in terms of accuracy. Later
sections will discuss specialized techniques for 
solution in the real-space representation including multigrid,
mesh refinements, FE formulations, and related LCAO
methods.

One direction has been to develop atom-centered
numerical grids in order to obtain converged 
results independent of basis-set approximations. 
Becke (1989) presented a fully numerical FD
method for performing molecular orbital calculations.
In this method, the physical domain was partitioned 
into a collection of single-center components, with
radial grids centered at each nucleus. A polyatomic
numerical integration scheme was developed. This work
was the first which extended the previous 
two-dimensional methods
for diatomics (see, for example, Laaksonen 
{\it et al.}, 1985). This numerical method has allowed for accurate
computations to test various levels of DFT
approximations on small molecules without concerns
of basis-set linear-dependence effects. The 
main focus of this approach
has been on numerically converged results and not
on scaling and efficiency for large-scale problems. 

In contrast to the atom-centered grids discussed
above, recent work has focused on development of
high-order pseudopotential methods on uniform
Cartesian grids. 
Chelikowsky, Troullier, and Saad (1994) and
Chelikowsky, Troullier, Wu, and Saad (1994) 
proposed a FD pseudopotential
method in which high-order forms were utilized for the 
Laplacian (Appendix). They employed the real-space 
pseudopotentials of Troullier and Martins (1991a, 1991b).
The simplicity of the FD method in 
relation to plane-wave approaches was 
highlighted.  The Hartree potential
was obtained either by a direct summation on the grid
or by iterative subspace techniques. They also employed
iterative subspace methods for the eigenvalue problem.
A main emphasis was on the accuracy of 
the FD approximation in relation to plane-wave methods. 
A multipole expansion was used to generate the fixed
potential on the boundaries.  Three parameters 
determine the accuracy in their FD calculations:
the grid spacing, the order of the Laplacian, 
and the overall domain size. 

Results were presented concerning the convergence
of the eigenvalues with order and decreasing 
grid spacing.  The 12th-order form of the Laplacian
was found to be sufficient for well-converged
results. Accurate eigenvalues (to 0.01 au) were obtained 
for atomic states.  Extensive calculations 
on diatomic molecules were also presented. 
The high-order FD approximation gave good 
results for binding energies, bond lengths,
and vibrational frequencies. Comparisons were
made to plane-wave calculations with two 
supercell sizes, one with 12 au 
and one with 24 au on a side.  The FD calculation
box was 12 au on a side. The plane-wave
energies were not converged with the 
smaller box size, but the plane-wave
calculations approached the FD results when a supercell of 
24 au was used, suggesting that quite large
supercells must be employed (even for nonpolar
molecules) for converged orbital
energies in localized systems (see Table I). The authors
obtained a dipole value of 0.10 D for 
the CO molecule (with the C$^-$O$^+$ orientation).
The experimental value is 0.1227 D with the same
orientation, while Hartree-Fock theory yields
the wrong sign for the dipole.  However, the 
fully converged LDA dipole is 0.24 D (Laaksonen
{\it et al.}, 1985).\footnote{This paper uses
the Dirac-Slater X${\alpha}$ form for the 
exchange-correlation potential. However, the 
computed dipole is insensitive when that potential
is changed to the LDA form.  See Jensen (1999) for 
converged basis set LDA results.}
The error is most likely due to the restricted 
overall domain size in their calculation (see
Kim, St\"{a}dele, and 
Martin, 1999; Wang and Beck, 2000). 
To conclude, the authors emphasized that the FD
method is ideal for localized and charged
systems, is easy to implement, and is well suited 
for parallel computations.  Related work which 
has analyzed the impact of FD order on accuracy
for Poisson problems 
can be found in the multigrid papers of
Merrick {\it et al.} (1995),
Gupta {\it et al.} (1997), and Zhang (1998). 
Also, see Section \ref{subsubsec:hofd}.

Subsequently, Jing {\it et al.} (1994) extended the high-order
FD method to compute forces and perform molecular
dynamics simulations of Si clusters.  For most of their work,
they performed Langevin molecular dynamics simulations
with a random force component to simulate a heat 
bath. The clusters were annealed from high temperature
to room temperature and the cluster structures were
examined; the FD method gave excellent agreement with
other numerical methods. When the heat bath was turned off,
the trajectory exhibited total energy fluctuations
two orders of magnitude smaller than the potential
energy fluctuations. The fluctuations agreed 
in magnitude with those
in a plane-wave simulation to within a few percent. 
Vasiliev, {\it et al.} (1997) recently utilized the 
higher-order FD methods in computations of polarizabilities
of semiconductor clusters with finite-field methods for
the response.
The results of the high-order FD method from 
Chelikowsky's group clearly show that the FD representation
can yield results of comparable or superior accuracy
compared with plane-wave calculations on similar-sized meshes.  

In related work, 
Hoshi {\it et al.} (1995) presented a supercell FD method 
in which they used an exact form of the FD Laplacian
which spans the whole domain along each direction. 
Therefore, $3N_g^{1/3}$ points are necessary to apply the 
Laplacian to the wavefunction at each grid point; the method
is equivalent to a very high-order representation. 
Fast Fourier transform
routines were used to solve for the Hartree potential. 
A preconditioning technique similar to that of
Payne {\it et al.} (1992) was employed to improve 
convergence. Pseudopotential results were presented for the He atom
and the H$_2$ molecule. Their method required 45 
steps of iteration to converge within
10$^{-5}$ au with the preconditioning. Subsequently,
Hoshi and Fujiwara (1997) incorporated the 
unconstrained OM linear scaling scheme into their method.
Windowing functions were employed to confine the 
orbitals to localized regions of space. Test
calculations were performed on the diamond crystal
with four localized orbitals per atom. They obtained
a ground state energy of 5.602 au/atom which compared
reasonably well with their previous result of 
5.617 au/atom. 

As mentioned above, FD methods have found application
in areas outside of traditional electronic structure theory.
Abou-Elnour and Schuenemann (1993) developed 
a self-consistent FD method for computing wave functions,
carrier distributions, and sub-band energies in 
semiconductor heterostructures.  Only one-dimensional
problems were examined. They compared the FD
method to a basis set calculation and found the FD
approach to be faster. In polymer physics, self-consistent
FD methods have also appeared.  Tsonchev {\it et al.}
(1999) derived a formal field theory for the statistical mechanics
of charged polymers in electrolyte solution.  The 
theoretical development parallels the earlier work
of Coalson and Duncan (1992) for the ion gas. 
A functional-integral representation was derived for the partition 
function of the coupled polymer/ion system.  The mean-field
theory solution leads to coupled Poisson-Boltzmann (for the ion
gas moving in the field of the other ions and the polymer 
charges) and eigenvalue (for the polymer chain
distribution) equations. 
These equations were solved numerically 
with FD methods for polymers 
confined within spherical cavities. The three-dimensional
eigenvalue problem was solved with the Lanczos 
technique.  Electrostatics plays
a key role in the chain structure for high chain
charge densities and low salt concentrations in the cavities. 

\subsection{Multigrid methods}

\label{subsec:mgsc}

The finite-difference results of the previous section
show that accurate results can be obtained 
on uniform grid domains with 
high-order approximations.  Multiscale methods
allow for accelerated solution of the 
grid-based equations.  
The first application of multigrid methods to 
self-consistent eigenvalue problems in electronic
structure was by White {\it et al.} (1989). Many
of the important issues related to real-space
approaches were laid out in this early paper. 
The authors developed an orthogonal FE basis and solved
the Poisson equation numerically with multigrid. 
Due to the orthogonal basis, a standard FD solver
only required simple revisions to apply to 
the FE case.  They also presented preliminary 
results of multigrid methods applied to the 
eigenvalue portion of the problem, but only
single orbital cases were considered. They found
that the multigrid solver was faster than
a conjugate-gradient method (without 
preconditioning).  Computations
were performed on the hydgrogen atom, the 
H$_2^+$ molecular ion, the He atom, and
the H$_2$ molecule. More discussion
of their method will be given below in Section
\ref{subsec:fes} on FE methods. 
Another early method by Davstad (1992) proposed a 
two-dimensional multigrid solver for diatomic
molecules in the Hartree-Fock approximation. 
He combined multigrid and Krylov subspace methods
in the solver.
High-order FD discretization was employed. 
The Orthomin procedure (a Krylov subspace method)
was used for iterations on all coarse levels, with
Gauss-Seidel iteration as preconditioner. 
Computations were performed
on the diatomics BH, HF, CO, CuH, and the Zn atom.
Good convergence rates were observed (presented in
terms of orbital residuals), and excellent 
agreement with previous numerical work was 
obtained for total energies and orbital 
eigenvalues. 

Since this early work, several groups have
utilized multigrid solvers for 
many-orbital problems in three dimensions. 
Bernholc's group has developed a 
multigrid pseudopotential method for large systems. 
Preliminary calculations (Bernholc {\it et al.}, 1991) 
were reported
for the H atom and the H$_2$ molecule.
A grid-refinement strategy 
for adding resolution around the nuclei was
also presented. 
Subsequently, Briggs {\it et al.} (1995) included
real-space pseudopotential techniques into 
their multigrid method and presented calculations
for large condensed-phase systems
on uniform grids.  They introduced
the FD Mehrstellen discretization which leads
to a 4th-order representation. Variations of
the total energy of atoms when moved in relation
to the grid points were investigated. With 
increasing grid resolution, the errors decrease, 
so this criterion can be used to choose the 
necessary fine-grid spacing for accurate
dynamical simulations. The Hartree
potential was also generated with a multigrid
solver.  In their method, the computation time
to perform one multigrid step is comparable to 
a single propagation step in the Car-Parrinello
method. Results were presented for a 64-atom
diamond supercell, the C$_{60}$ molecule, 
and a 32-atom GaN cell.  For large systems,
the multigrid method was found to converge
to the ground state an order-of-magnitude 
faster than their Car-Parrinello code. 
For the GaN case, 240 multigrid iterations were
required to reach the ground state from 
random initial wavefunctions, while for 
an 8-atom diamond cell roughly 20 iterations
were necessary to converge the total energy
to a tolerance of 10$^{-8}$ au.  

Their multigrid algorithm was further developed in
Briggs {\it et al.} (1996), where extensive
details of the solver were presented. Calculations were performed on 
a Si supercell, bulk Al, and an AlN supercell with 
comparisons made to Car-Parrinello calculations to 
test the accuracy of the approximations. Excellent
agreement with the Car-Parrinello results was obtained. 
Their multigrid
implementation for the eigenvalue problem utilized
a double-discretization scheme; on the fine level
the Mehrstellen discretization was employed, while
on the coarse grids a seven-point central-difference
formula was used. Full-weighting restriction and
trilinear interpolation were used for the grid 
transfers, and Jacobi iterations were performed for
the smoothing steps.   The eigenvalue problem was
linearized by computing the eigenvalues only on 
the fine grid and performing coarse-grid corrections 
on each eigenvector. The constraints were imposed
on the fine scale at the end 
of the double-discretization correction cycle. Subspace
diagonalization was performed to accelerate 
convergence.   Tests of the convergence were conducted
on a 64-atom Si cell and a 64-atom diamond cell with
a substitutional N impurity. Substantial accelerations
were obtained with multigrid in comparison 
to steepest-descent iterations; 
roughly 20 self-consistency iterations were
required in the multigrid solver to obtain 10$^{-3}$ Ry
convergence in the total energy. 
While these convergence
rates are a significant improvement over steepest-descent
iterations, they are non-optimal due to the linearization 
in their method (see below). 
The overhead for implementing multigrid in addition
to steepest-descent iterations was only 10\% of the total
computing time. The authors discussed 
extensions of the multigrid method for molecular dynamics
(tested on a 64-atom Si supercell which exhibited good
energy conservation). Applications to other large-scale
systems appear in Bernholc {\it et al.} (1997). 

In an algorithm very similar to that described above,
Ancilotto {\it et al.} (1999) developed a solver
which included FMG processing to provide a good
initial guess on the finest level. The
Mehrstellen discretization was 
employed on all levels. With the 
FMG addition, the initial state of the orbitals is
irrelevant since it takes very little numerical effort
to obtain the initial fine-grid approximation during
the preliminary coarse-grid cycles.  They performed red-black
Gauss-Seidel smoothing steps on each level and used
full-weighting restriction and trilinear interpolation
for grid transfers.  Eigenvalues were computed only
on the finest level, and Ritz projections were also
performed to accelerate convergence.  They also
reported 20 self-consistency iterations 
to obtain convergence on several diatomic molecules
(C$_2$, O$_2$, CO, and Si$_2$), and
good agreement with plane-wave results was observed
for equilibrium bond lengths and vibrational
frequencies (both to within 1\%). They presented numerical results
for the C$_2$ dimer (pseudopotential calculations)
which illustrated the convergence of their algorithm 
in comparison to a Car-Parrinello 
(damped molecular dynamics) plane-wave code.  Superior
convergence was found even in relation to 
state-of-the-art Car-Parrinello algorithms (Tassone
{\it et al.}, 1994), which exhibit performance similar to 
conjugate-gradient algorithms. The method was 
tested by using simulated annealing cycles to
locate the most stable ground state
of the Al$_6$ cluster.  Then calculations were
performed to find stable minima for charged 
Li clusters with sizes $N = 9-11$. The numerical
results indicated that the fragmentation behavior
observed in experiments likely has a strong 
non-statistical component.

In addition to the two-dimensional solver of
Davstad discussed above, all-electron multigrid methods
in three dimensions have been developed.
Iyer {\it et al.} (1995) discussed a multigrid method  
for solving the Kohn-Sham equations in which the 
entire problem was discretized on a 
three-dimensional Cartesian
lattice, including all electron orbitals 
and the nuclear charge densities. 
An eighth-order FD form for the Laplacian was 
used in this work. The nuclear charge densities
were discretized as a single cube on 
the lattice, and the Poisson equation was solved
with the standard multigrid technique.  Since 
the total charge density included both the electron
and nuclear densities, all the electrostatic
interactions were handled in a single linear-scaling 
step, including the nucleus-nucleus
term.  A self energy must be subtracted from
the total energy, but this is a one-time computation
for each order of the Laplacian since the self
energy scales as $Z^2/h$.  Computations
were performed on hydrogenic atoms and the
H$_2^+$ molecule.  A simple nested procedure
was utilized for the Kohn-Sham solver in which
an initial approximation was generated on a
coarse level, smoothing steps were performed, and the
problem was interpolated to the next finer grid
followed by relaxations.  This process significantly
accelerated the convergence, but some critical
slowing down remained due to incomplete decimation
of long-wavelength modes on the coarse levels. 
These results show that a one-way multigrid 
procedure without coarse-grid corrections does not 
guarantee proper multigrid convergence. Various
relaxation procedures were compared; conjugate 
gradients gave the best convergence per step
but required more numerical effort  
than simple Gauss-Seidel iterations, so Gauss-Seidel
is equally efficient. This result
illustrates the important point that simple smoothing
iterations are enough to decimate the errors with
wavelength on the order of a few grid spacings on 
a given level, and special techniques are not necessary. 
Results were presented for the 
all-electron Ne atom which exhibited
the significant speedup due to a multiscale treatment, 
but the residual stalling on the fine levels was
used to motivate inclusion of the BMR FAS-FMG method
for the eigenvalue problem. 

Beck {\it et al.} (1997) presented the first application
of the BMR FAS-FMG algorithm 
(Section \ref{subsubsec:algorithms}) to self-consistent electronic
structure problems. In this initial effort, the BMR
algorithm was followed, except the orbitals were 
updated simultaneously during the correction cycle
as opposed to sequentially in the original method. 
Also, Gram-Schmidt orthogonalization steps were 
implemented on each level, so the constraint procedure
outlined in Section \ref{subsubsec:algorithms} was 
not followed exactly.  Convergence calculations were
performed on the Ne atom on a 33$^3$ grid; 
the FAS-FMG approach led 
to faster convergence than the one-way multigrid
calculations of Iyer {\it et al.} (1995).  
Beck (1997) extended these calculations to the
CO molecule (all electrons and three 
dimensions) and developed an FAS solver for the 
Poisson-Boltzmann equation. The convergence of
the CO molecular calculation was limited by the 
handling of the constraints discussed above.  
A relatively accurate dipole moment of 0.266 D (C$^-$O$^+$)
was obtained on a 33$^3$ mesh.  

\begin{figure}
\begin{center}
\epsfig{%
file=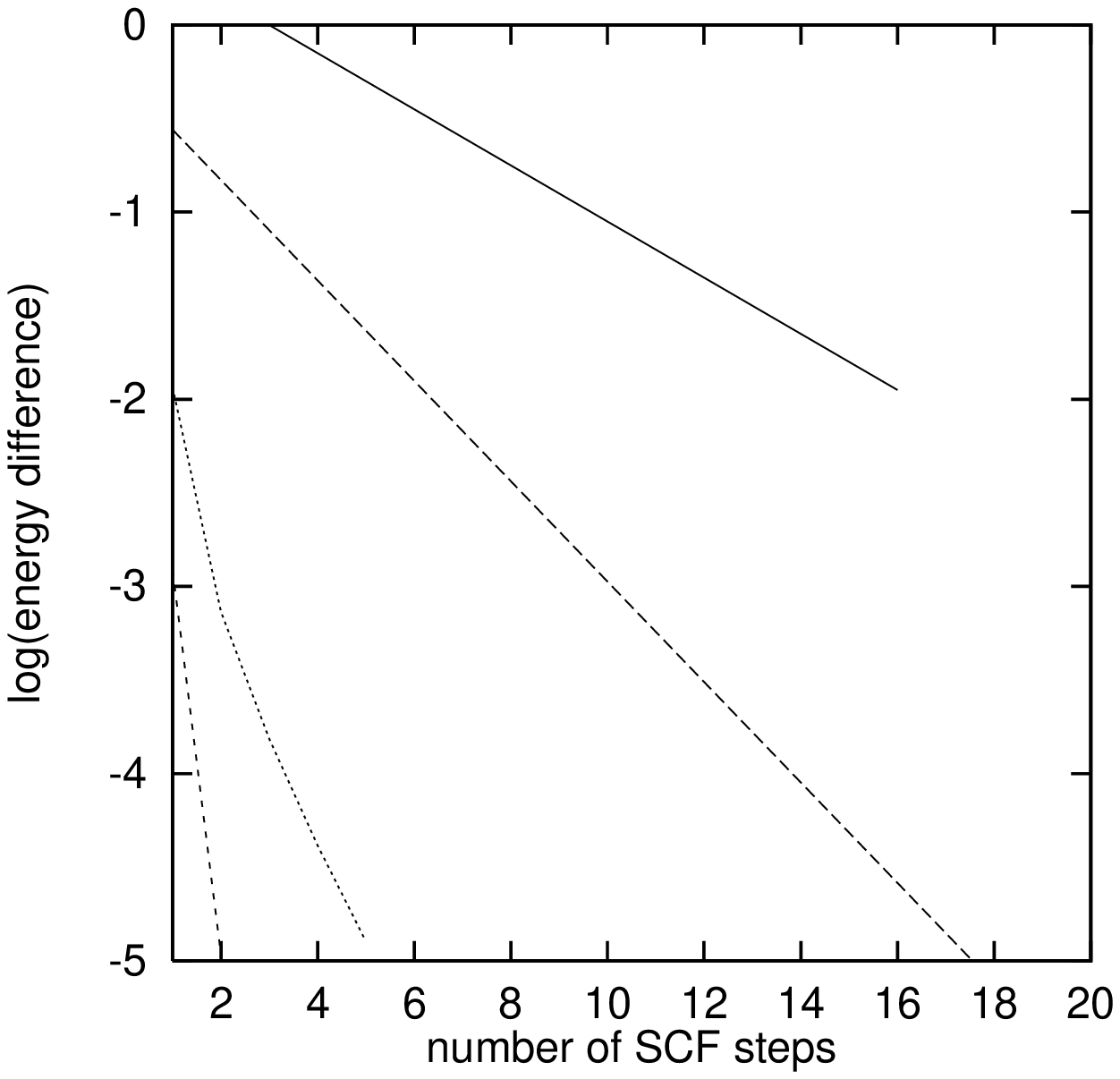,%
height=8.cm,%
width=8.cm,%
angle=270}
\end{center}
\caption{Convergence behavior.
The top curve is the Car-Parrinello 
(damped molecular dynamics) result
of Ancilotto {\it et al.} (1999). The second curve is the MG
result from that work. The next is the FAS-FMG
result of Wang and Beck (1999)
for the CO molecule with the FAS-FMG solver.
The bottom curve is the FAS-FMG result
for the Be atom.}
\label{fig:convergence}
\end{figure}

Subsequently, 
Beck (1999a) and Wang and Beck (2000) developed
a fully convergent FAS-FMG Kohn-Sham self-consistent
all-electron solver.  In this work, the eigenfunction constraint
equations [Eq. (\ref{eq:constraints})] were implemented
on the coarsest grid only, and the eigenvalues were also
updated on the coarsest level via Eq. (\ref{eq:cgeval}).
Ritz projection was performed on the finest level 
at the conclusion of each V-cycle. The 
effective potential was updated
once upon entry to the next finest level and at
the end of each V-cycle.  Both sequential and 
simultaneous updates of the orbitals were examined
to test the efficiency of each approach. The sequential
method leads to slightly more rapid convergence to 
the ground state, but it results in a $qN_g$ scaling 
in a self-consistent method since the effective 
potential is updated following coarse-grid 
corrections on each orbital.  The discretized
problem was solved on a 65$^3$ grid domain with
a 12th-order form for the Laplacian. 
Atomic ionization potential computations 
were performed to illustrate the ease of applicability
to charged, finite systems. Numerical results 
were presented for the all-electron CO 
molecule.  The CO eigenvalues were accurate
to within 0.015 au for all states above the core, and
the highest occupied $\pi$(2p) and $\sigma$(2p) states
were accurate to within 0.006 au. The computed dipole
was 0.25 D, in good agreement with previous fully
numerical results on diatomics (Laaksonen {\it et al.},
1985). Convergence data was presented for the Be
atom and the CO molecule (Fig.\ \ref{fig:convergence}).  Implementation
of the nonlinear FAS-FMG strategy leads to order-of-magnitude
efficiency improvement in relation to linearized versions of 
the multigrid algorithm (Ancilotto {\it et al.}, 1999). 
The converged ground state was obtained in only
two or three self-consistency cycles, with 
three orbital relaxation steps on each side of 
the V-cycle. Therefore, the entire self-consistent
solution process required a total of only 12-18 smoothing 
steps on the finest grid and a few updates of the 
effective potential. One self-consistency cycle 
for the 14 electron CO molecule on a 65$^3$ grid required
roughly a minute of CPU time on a 350 
MHz Pentium-II machine. 
The update of the Hartree potential involves the same effort
as the update of a single eigenfunction; it is therefore
a small contributor to the overall numerical effort. 
Due to the handling of the constraints and eigenvalues
on the coarsest level,
each self-consistency update requires less computation
than the algorithms of Briggs {\it et al.} (1996) 
and Ancilotto {\it et al.} (1999). 

Since these FAS-FMG computations included all electrons
and the nuclear singularities in three 
dimensions, the rapid convergence 
in relation to the pseudopotential computations of 
Ancilotto {\it et al.} (1999) is noteworthy (the total
energy is nearly three orders-of-magnitude larger 
than in the pseudopotential calculation). 
These results are the first
to exhibit the full power of the nonlinear BMR 
technique for solution of self-consistent 
electronic structure problems.   The slightly slower convergence 
for the CO molecule (compared with the Be atom) 
is due to the relatively poor 
treatment of the core electrons on a uniform grid; 
with a finer grid,
the convergence is even more rapid. 
Wang and Stuchebrukhov (1999) have applied
the FAS-FMG algorithm described above
to computation of tunneling 
currents in electron transfer; 
they found that real-space calculations give
a significantly more accurate representation of 
current densities than Gaussian basis-set 
calculations. 

Some simple arguments can be made concerning the 
total number of operations for the multigrid 
solution {\it vs.} the conjugate-gradient plane-wave method.
The present discussion assumes the orbitals span
the entire physical domain.
Payne {\it et al.} (1992) showed that the conjugate-gradient
method requires $6qN_{FFT} + 2q^2N_{PW}$ operations
to update all the orbitals. The second term is for
the orthogonalization constraints. 
The variable $q$ is the number of orbitals and $N_{FFT}$ 
is $16N_{PW}\ln N_{PW}$ where $N_{PW}$ is 
the number of plane waves. Thus $N_{FFT}$
is the number of 
operation counts for Fourier transformation on 
the real-space grid. The 
multigrid method requires $qN_{mgop} +
2q^2N_{g} + N_{mgop} = (q+1)N_{mgop} +
2q^2N_{g}$ operations, where $N_{mgop}$
is the number of operations to update one 
orbital with the multigrid method and $N_g$ is the number
of fine grid points. The $q^2$ dependent term 
is for the orthogonalization constraints (Gram-Schmidt
followed by Ritz projection) which are performed
once at the end of each correction cycle, and 
the second $N_{mgop}$ term is for the Poisson
solver. Since 
a multigrid update of one eigenfunction (with say
an 8th-order approximation) requires roughly four times
the number of operations count of a single
FFT (see Section \ref{subsubsec:icb}), 
the net cost for the 
multigrid update (neglecting the relative constraint costs
which are much smaller with multigrid, see below)
is slightly less than that for the conjugate-gradient
method.  Figure \ref{fig:convergence} shows that the number
of self-consistency iterations is also very low with the
multigrid solver. The study of Ancilotto {\it et al.}
(1999) compared damped molecular dynamics to their
linearized multigrid method (on diatomic molecules). 
They also compared multigrid (favorably) with
the optimized dynamics method of Tassone {\it et al.} (1994)
which in turn exhibits convergence rates 
very similar to conjugate-gradients. Since the nonlinear
FAS-FMG solver outperforms the linearized multigrid method
by an order-of-magnitude, this suggests the multigrid 
solver is more efficient than the conjugate-gradient approach.
The best available
plane-wave techniques (see, for example, Kresse
and Furthm\"{u}ller, 1996) can reduce the number 
of self-consistency iterations to 5-10, 
so the multigrid solver is at least as efficient 
as the most efficient plane-wave techniques for 
uniform-domain problems where the orbitals span 
the whole domain. 
The major benefits of the multigrid approach in 
addition to the above discussion
are: 1) all the constraint and subspace orthogonalization
operations can be removed
to coarse levels where the cost is minimal; for example
if they are performed two levels removed from the fine
level, the cost is 1/64 that on the fine level
(Costiner and Ta'asan, 1995a), 2) it is quite easy 
to impose localization constraints in the real-space
multigrid approach (Fattebert and Bernholc, 2000), and 3)
mesh refinements can be incorporated while
maintaining the same convergence rates (see Beck, 1999b,
for the Poisson version).
The mesh-refinement methods are in place and are currently being
incorporated into Kohn-Sham solvers; they should lead
to a further near order-of-magnitude reduction 
in computational cost.  Finally, Costiner and 
Ta'asan (1995b) have shown that by updating the effective
potential simultaneously with the eigenfunctions on 
coarse levels self-consistent solutions can be 
obtained in a {\it single} passage through the 
final V-cycle of the FMG process. Therefore, 
multiscale real-space approaches offer a promising
alternative to plane-wave techniques. 

Recently, Lee {\it et al.} (1999) proposed a one-way
multigrid method similar to that of Iyer {\it et al.}
(1995). Initial approximations were obtained on coarse
levels, and the solution was interpolated to the 
next finer level without multigrid correction cycles.
High-order interpolation was used
to proceed to the next finer grid. 
Conjugate-gradient techniques were employed to relax
the orbitals on each level.  The method led to a factor of five
reduction in computation time compared 
to a single-grid calculation. Computations were performed on 
a 20-electron quantum dot and charged H clusters. 
Kim, Lee, and Martin (1999) developed an object-oriented
code for implementation of the one-way multigrid
algorithm. Several other groups have utilized multigrid solvers 
as components of real-space electronic structure
algorithms; these will be discussed in the 
following sections on mesh-refinement techniques 
and FE methods. 

\subsection{Finite-difference mesh-refinement techniques}

\label{subsec:scmrt}

The previous sections have discussed FD methods
for electronic structure; the calculations
were performed primarily on uniform grids.
With the incorporation of real-space pseudopotentials,
results with accuracies comparable to plane-wave
methods (with similar grid cutoffs) can be obtained
with high-order FD techniques. 
The calculations of Beck (1999a) and Wang
and Beck (2000) are instructive in that surprisingly
accurate results are possible even in all-electron 
calculations on uniform grids; in addition, their
work shows that multigrid efficiencies are obtainable
for the challenging case of very harsh 
effective potentials which include
the nuclear singularities.  However, it is clear that 
increasing uniform grid resolution until acceptable 
accuracy is reached is a wasteful process since
small grid spacings are only required in the neighborhood
of the atomic cores. This section reviews recent work
on development of FD mesh-refinement techniques 
which address this issue for the 
eigenvalue problem.  

As discussed in Section \ref{subsubsec:mrt} which covered 
closely related methods for the Poisson equation, there are
presently two strategies for mesh refinements: grid curving
and local refinements which are included within a
coarser mesh (Fig. \ref{fig:meshrefine}). 
Gygi and Galli (1995) extended a previous plane-wave
method of Gygi (1993) to adaptive-coordinate FD calculations. 
A curvilinear coordinate system was developed which
focused resolution near the nuclei. The necessary extensions
of the standard FD method to handle the 
curvilinear Laplacian were presented.  FD 
forms of order 2 and 4 were
utilized, and norm-conserving pseudopotentials were
employed. The Poisson equation was solved with a 
multigrid method. The calculations were implemented
on a Cray-T3D massively parallel machine. Test calculations
were conducted on diatomics and the CO$_2$ molecule. 
The calculations of the total energy of the CO$_2$ 
molecule vs.\ internuclear distance exhibited a 
spurious double minimum with a uniform grid 
treatment (cutoff energy of 227 Ry).  This double
minimum is due to the numerical errors from 
a grid which is too coarse. When the 
adaptive coordinate transformation was included
(with effective cutoffs of 360 Ry for carbon
and 900 Ry for oxygen), a single minimum was
observed near the correct experimental bond length. 

Modine {\it et al.} (1997) presented another
adaptive-coordinate FD method which they
termed ACRES (adaptive-coordinate 
real-space electronic structure). 
They first discussed the goals of their real-space 
method: 1) sparsity, 2) parallelizability,
and 3) adaptability. The real-space approach
satisfies these criteria, while the plane-wave method
does not.  Extensive details were given concerning 
the construction of their grid-curved meshes 
and the resulting Laplacian. One issue to note is that the
FD Laplacian in curvilinear coordinates contains 
off-diagonal terms and the number of terms 
scales as $3((2n)^2+4n+1)$, where $n$ is the order.
Therefore, high-order derivative forms add
complexity to the 
adaptive-coordinate approach.  Computations were performed
on atoms and molecules at both the all-electron
and pseudopotential levels. 
The authors discussed the limitations of the 
Lanczos method for the eigenvalue problem; the
width of the real-space spectrum is dominated
by the largest eigenvalue which in turn is 
determined by the minimum grid spacing, so
the method slows with increasing resolution. 
Instead, they used a modified inverse iteration
eigensolver. The equations were solved with a
conjugate-gradient algorithm.  Conjugate-gradient 
techniques were also employed for
the Poisson equation, with multigrid used
for preconditioning.  Highly accurate
all-electron results were obtained for
the O atom and the H$_2$ and O$_2$
molecules; computed bond lengths for O$_2$
agreed with both the previous
calculations of Chelikowsky, Trouller, Wu,
and Saad (1994) and experiment to within 0.02$\AA$.
To conclude, they discussed
the high efficiency of ACRES in relation to uniform
grid computations. 

Two works have appeared which utilize nested mesh
refinements as opposed to grid-curving techniques
for increased resolution. Fattebert (1999) developed
an algorithm to treat a single grid refinement 
placed inside a coarser-level grid domain. A FD
Mehrstellen discretization was employed over the 
whole domain, with nonuniform difference stencils
at the boundaries between the fine
and coarse levels.  The discretization is 
4th order over the uniform regions and 2nd order
at the boundaries.  The impact of this
nonuniformity of the representation order 
on the solution order was not examined.  
The eigenvalue problem
was solved with a block Galerkin inverse
iteration in which multigrid methods were used 
to solve the linear systems. Smoothing iterations
were enacted with the GMRES algorithm (Golub
and van Loan, 1996). Pseudopotential 
calculations were performed
on the furan molecule which requires treatment
of 13 eigenfunctions. Excellent convergence rates
were observed, especially on the finer composite
meshes; the coarse grid convergence was not 
as rapid. The author also presented results for
the total energy of the CO molecule which are
similar to those of Gygi and Galli (1995) described 
above. Incorporation of the grid refinements led
to smooth variations of the energy, while the 
coarser-grid computation resulted in irregular
variations.  Ono and Hirose (1999) 
proposed another double-grid method in which
the inner products of the wavefunctions and
pseudopotentials are treated on a fine grid. 
The double-grid treatment leads to smooth
forces without the necessity of Pulay (1969) 
corrections (which are required in the 
adaptive-coordinate method).  

\subsection{Finite-element solutions}

\label{subsec:fes}

Just as for the FD formulation, the application 
of FE methods to self-consistent
eigenvalue problems has followed two different
tracks. In the first, the FE basis has been utilized
to obtain highly accurate results for atoms and
small molecules.  The FE method can achieve very
high accuracies since it does not suffer from
the linear-dependence problems of LCAO approximations,
and the mesh can be arbitrarily refined. The second
type of application concerns development of 
efficient methods for large-scale electronic 
structure problems. We begin with methods designed
to obtain high accuracies. 

Levin and Shertzer (1985) performed FE calculations
on the He atom ground state. The problem 
reduces to  
three-dimensional for the $s$ state.  
A basis of cubic Hermite polynomials
was employed. They computed both the ground-state
energy and the moments $\langle r^n \rangle$ of
the wavefunction. An energy within
0.0005 au of the numerically exact result was
obtained. Also, the orbital moments were substantially
more accurate than those computed in basis-set 
calculations.  This occurs since the LCAO basis functions
are global; if the functions are
optimized to give a good wavefunction near the 
nucleus (where the largest contribution to 
the total energy occurs), they cannot be adjusted 
simultaneously to give a good representation far
from the origin. The FE basis overcomes this
difficulty.  Heinemann {\it et al.} (1987)
and Heinemann {\it et al.} (1988) developed
a two-dimensional FE method and applied it to 
computations on the H$_2$, N$_2$, BH, and CO
molecules. Using a 5th-order basis, accuracies
to better than $10^{-8}$ au for the total 
energies were observed, which exceeds by 
two orders the 
accuracy of the FD calculations by Laaksonen
{\it et al.} (1985).  Yu {\it et al.}
(1994) implemented an order 5 or 6 
Lobatto-Gauss FE basis
and employed a block Lanczos algorithm
to solve the eigenvalue problem. A Duffy (1982)
transformation allowed for handling of the Coulomb
singularity. Calculations
were performed on diatomic and triatomic
hydrogen molecules and ions; these
three-dimensional results were not as accurate
as in the two-dimensional study of Heinemann {\it et al.}
(1987), differing by .00051 au in the total 
energy of H$_2$.  More recently, Kopylow {\it et al.} (1998)
incorporated an FMG solver into 
their two-dimensional method for diatomics. Conjugate-gradient
smoothing steps were employed on each level. 
Excellent convergence rates were obtained for
the solver which was tested on the Be$_2$ 
molecule; only 5 self-consistency iterations
were required to obtain 10$^{-6}$ au convergence
in the energy.  D\"{u}sterh\"{o}ft {\it et al.} (1998)
combined the LCAO and FE methods in a defect
correction approach which allowed for a more
rapid attainment of the ground state due to 
a better representation around the nuclei. 

Next, we consider methods directed toward larger systems.
The FE method of White {\it et al.} (1989) was 
discussed above related to multigrid methods for
self-consistent problems.  They utilized a high-order
FE basis and constructed orthogonal functions from
the nonorthogonal basis. The cost of this construction
is the requirement of more functions to obtain the same
level of completeness.  The three-dimensional basis
functions were products of the one-dimensional 
functions on a Cartesian grid. 
The Coulomb singularity was handled with an integral
transform representation of $1/r$. The Hamiltonian is
sparse in their basis since only near-neighbor overlaps
need to be considered. To solve the Poisson equation, multigrid
techniques were employed with a double-discretization procedure
similar to that of Briggs {\it et al.} (1996); on coarser
levels, the problem was represented with a FD form rather than
with a FE basis. As discussed above, multigrid solution of
the eigenvalue problem was faster than conjugate gradients. 
To conclude, they emphasized the importance of developing 
new grid methods for refinements around the nuclei, where
the largest errors occur. 

Gillan and coworkers have developed a general method
for linear-scaling electronic structure (of the OBDMM
form discussed in Section \ref{subsec:linsces}). 
Closely related is the work of Hierse and 
Stechel (1994), which differs in the choice of basis
and the number of basis functions. In
their initial work (Hern\'{a}ndez and Gillan, 1995),
the OBDMM strategy was developed, and the calculations
were performed directly on a real space grid with 
second-order FD techniques. The total energy was minimized
with conjugate-gradients iterations. Typically, 50
iterations were required to obtain energy convergence
to within 10$^{-4}$ eV/atom.  

Hern\'{a}ndez {\it et al.} (1997) 
developed a blip-function basis instead of
the previous FD representation.  This method is general
in the sense that any local function (that is, completely restricted
to a finite volume) can be used for the basis; however,
we examine this method in relation to FE bases since
it is so closely related.  The actual basis employed
in their work is a set of B-splines (see, for example,
Strang and Fix, 1973, p.\ 60).  The basis was
implemented on a Cartesian mesh as products of
three one-dimensional functions.  The kinetic and
overlap terms were treated analytically, but 
the matrix elements of the potential were evaluated numerically
on a grid different from the blip grid. The blip-function
basis agreed very well with plane-wave results 
in calculations on Si solids; a discrepancy of only
0.1 eV/atom was observed between the two different
approaches. Goringe {\it et al.} (1997) discussed 
implementation of the algorithm on very large systems
(up to 6000 atoms) on parallel machines. The essential
features of the OBDMM method were reviewed.  Fast
Fourier transform
methods were used to solve for the electrostatic
potential on a grid.  Complete discussion was
given of the steps in parallelizing 
every portion of the code
using real-space domain decomposition. The 
numerical results on a Cray-T3D parallel machine
exhibited linear scaling of CPU time with the
number of atoms using between 32 and 512 processors.  

As discussed in Bowler {\it et al.} (1999) and
reviewed in Goedecker (1999), three forms of 
ill-conditioning can lead to degradation of convergence
to the ground state in the OBDMM method: length-scale,
superposition, and redundancy ill-conditioning. 
The first is an inherent feature of any real-space
solver (Section \ref{subsubsec:fdsit}).
The second form results from the localization constraints
imposed in the method, and is similar to problems
in OM methods. The third is related to the fact that
their method includes more basis functions than 
occupied orbitals; the localization constraints
lead to small but nonzero occupation numbers of
the higher-lying states, and they have little 
influence on the total energy. Bowler and Gillan
(1998) addressed the length-scale ill-conditioning
problem. They developed a preconditioning technique
related to the plane-wave method of Payne 
{\it et al.} (1992). The blip-function preconditioning matrix
only needs to be calculated once.  Test calculations
were performed for a Si crystal with significant
accelerations of the convergence due to
the preconditioning. However, the convergence 
efficiency of their method
decreased both with decreasing grid
spacing and with increasing localization radius. 
It is an interesting question whether multigrid
methods might lead to higher efficiencies
in the context of the OBDMM method. 

An alternative FE method for large-scale electronic
structure has been developed by Tsuchida and
Tsukada (1995, 1998).  In their original method,
the authors utilized first- and second-order
shape functions and derived the appropriate 
variational expression for the total energy
at the LDA level. The Hartree potential was
generated by conjugate-gradient iteration. 
They also implemented the OM  
linear-scaling method. Nonuniform meshes
were employed to focus resolution around
the nuclei in the H$_2$ molecule. Test
calculations were also performed on an
8-atom Si solid with 16$^3$ uniform elements
of the second order.  Good agreement with plane-wave
calculations and experiment were obtained for the
lattice constant, cohesive energy, and bulk 
modulus.  Due to the integral formulation of the 
total energy, the Coulomb singularity in the potential
becomes finite. In Tsuchida and Tsukada (1998), the 
method was substantially extended for large-scale
condensed-phase systems.  Third-degree polynomials
were used as FE basis functions.  They implemented
the grid-curving method of Gygi and Galli (1995)
to adapt for higher resolution near the nuclei. 
Pulay (1969) corrections were computed to obtain accurate
forces on the ions. A multigrid procedure was 
followed to solve the FE Poisson equation.  The
multigrid aspect was used as preconditioner to 
final conjugate-gradient iterations on the finest
scale. Again, OM techniques were used to obtain
linear scaling. They also utilized a one-way
multigrid-type approach for the eigenvalue problem, 
where a good initial approximation was obtained
on the finest level from previous iterations on
a coarser level.  With this approach, 20 to 30
self-consistency iterations were required for convergence on the 
fine level. Calculations were limited to the
$\Gamma$ point; for the treatment of general Bloch
boundary conditions, see Pask {\it et al.} (1999).
A parallel code was written using
real-space domain decomposition. Many applications
were considered in this work, including computations
on diamond lattices, cubic BN, the C$_{60}$ molecule,
molecular dynamics simulations, and parallel 
implementations. Pseudopotential 
calculations on systems with
up to 512 carbon atoms were presented. The final statement
from this paper captures well the rapid development of 
real-space methods in the last decade: ``About ten
years ago, the FE method was described to be in its
infancy for electronic structure calculations
(White {\it et al.}, 1989). We have shown in this paper
that it can be routinely used for large systems today."

As a final application of FE methods to self-consistent
eigenvalue problems, Lepaul {\it et al.} (1996) 
considered semiconductor quantum nanostructures.
They solved the two-dimensional Schr\"{o}dinger 
equation self-consistently with updates of the 
Poisson equation (variable dielectric case) to 
obtain carrier densities, conduction bands,
and the potential distribution at finite 
temperatures. The image potential and 
exchange-correlation energies were neglected. 
The carrier confinement was due to heterojunction
discontinuities and the electrostatic potential.
By varying the potential bias, a bidimensional
quantum gas was observed. The real-space approach
allowed for treatment of realistic device
geometries.  

\subsection{Orbital-minimization methods}

\label{subsec:rsdmomm}

To conclude our review of real-space self-consistent
eigenvalue problems, we consider a related OM 
linear-scaling algorithm which uses LCAO bases
(S\'{a}nchez-Portal {\it et al.}, 1997).  
The reason for its inclusion here is that the bases
(Sankey and Niklewski, 1989) are 1) numerical and 2)
confined to a local region of space.  Therefore, the method 
shares features in common with FD and FE approaches. 
The authors discussed construction of the Hamiltonian
matrix elements and the total energy in the
numerical bases. The total energy
was re-expressed in a form which has terms involving
only two-centered integrals which are interpolated
from calculated tables (one-time calculation)
and other terms computed 
entirely on a real-space grid (involving screened
neutral-atom potentials
and the Hartree and exchange-correlation potentials). 
The Hartree potential was computed via FFT methods. 
Rapid convergence of the approximations with decreasing
grid spacing was observed.  The OM functional of 
Kim {\it et al.} (1995) was employed to obtain linear
scaling. 

The method was applied in calculations on
several diatomics and triatomics where various-quality
basis sets were tested in computations of bond lengths,
bond angles, and binding energies.  Gradient corrections
to the LDA approximation were also considered. Finally, 
large-scale computations were performed on a turn of
the DNA double helix consisting of ten guanine-cytosine
base pairs in periodic boundaries (650 atoms). 
The equilibrium geometry was obtained in 200 minimization
steps, requiring 5 days of computation on an HP C110
workstation.   The number of 
self-consistency iterations required for 
each minimization step was not given.  Also, it is not
entirely clear what is the sparsity of the Hamiltonian
in the numerical localized LCAO basis in relation to FD and
FE methods. 
The spherical-wave basis set of Haynes and
Payne (1997) should also prove useful since it is localized
in space and its truncation is controlled by a single parameter, the 
kinetic energy cutoff (similar to plane-wave methods).  Hybrid
basis-set/grid type methods such as discrete variable representations
(DVR) and distributed approximating 
functionals (DAFs) also exist which yield accurate local
representations (Light {\it et al.}, 1985; Marchioro
{\it et al.}, 1994; 
Schneider and Feder, 1999).

\section{TIME-DEPENDENT DFT CALCULATIONS IN REAL SPACE}	

\label{sec:tddft}

The Kohn-Sham method for electronic structure
lies on solid theoretical ground
due to the Hohenberg-Kohn theorems.  Extensions of 
DFT to excited states and/or frequency-dependent 
polarizabilities present
more difficult challenges, but significant progress
has been made in this area in the last few years. The developments
include real-space computations of excitation energies and 
response properties (Yabana and Bertsch, 1996;
Vasiliev {\it et al.}, 1999; Kim, St\"{a}dele, and Martin, 
1999).   Thorough reviews
of the foundations of time-dependent 
DFT (TDDFT) methods are 
available (see, for example, Gross and Kohn, 1990;
Casida, 1996).  The starting point for practical computations
is typically the solution of
the time-dependent LDA (TDLDA) equations:

\begin{equation}
[- \frac{1}{2} \nabla^2 + v_{eff}({\bf r}, t) ] \psi_i({\bf r}, t)
= i\frac{\partial\psi_i({\bf r}, t)}{\partial t} , 
\label{eq:tdks}
\end{equation}

\noindent
where the density-dependent effective potential 
is just the Kohn-Sham LDA
potential (Eqs.\ \ref{eq:veff} and \ref{eq:ves})
for the set of orbitals at time $t$. The TDLDA
method includes dynamic screening effects which
modify the excitation frequencies away from the Kohn-Sham 
LDA eigenvalue differences toward the physical ones. 
Inclusion of gradient corrections does not significantly
improve the results (Bauernschmitt and Ahlrichs, 1996; 
Casida {\it et al.}, 1998).
There are two important approximations 
involved in Eq.\ (\ref{eq:tdks}):
1) the static LDA potential exhibits the incorrect
asymptotic behavior at long range (exponential rather
than $-1/r$), and 2)
no time dependence is incorporated in the exchange-correlation
potential (adiabatic approximation). 
It is generally recognized that the first 
approximation is the most severe (Van Gisbergen
{\it et al.}, 1998); with improvements to the LDA which
yield the correct asymptotic large-$r$ behavior, quite
accurate results are obtainable even for Rydberg
states (Casida, 1996; Jamorski
{\it et al.}, 1996; Casida {\it et al.}, 1998; Van Gisbergen 
{\it et al.}, 1998; Tozer and Handy, 1998). 
The adiabatic approximation makes physical sense for slow
processes, and integrals (over frequency) of the response
in the mean-field theory 
obey rigorous sum rules which 
are satisfied for the small-amplitude 
TDLDA (Yabana and Bertsch, 1999). 
Proper modeling of the long-range behavior of the effective potential
is important for the higher-lying Kohn-Sham states,
which in turn are crucial for obtaining accurate excitation
energies above the highest occupied Kohn-Sham LDA
eigenvalue.  These states are also important for computing
accurate polarizabilities.  For low-lying excitations, 
the TDLDA-level of theory is remarkably accurate
(Casida {\it et al.}, 1998; Yabana and Bertsch, 1999). 
Observed errors in excitation energies 
are on the order of one or a few tenths
of an eV for small molecules in comparison with experiments.  

Two main approaches have been followed in development of
the TDDFT method.  In the first (Yabana and Bertsch, 1996),
Eq.\ (\ref{eq:tdks}) is solved directly in real time by 
propagating the orbitals on a real-space grid. The 
frequency-dependent polarizability and the strength
function are obtained by Fourier transformation of
the time-dependent dipole moment computed on the grid. In the 
second approach (Petersilka {\it et al.}, 1996; Casida, 1996),
the problem is recast in the energy representation
by calculating the response at the linear-response 
level. Solution of an eigenvalue problem 
involving the Kohn-Sham energy differences and
a coupling matrix yields the 
excitation energies and oscillator strengths and from
them the frequency-dependent polarizabilities.
Applications of the second theoretical approach
have employed both basis-set (Casida, 1996; Van Gisbergen 
{\it et al.}, 1998; Tozer and Handy, 1998)
and real-space (Vasiliev {\it et al.}, 1999) formulations.
The real-time and energy representations should give equivalent
results for physical situations which allow a linear-response
treatment.  
In this section, we review recent real-space computations
in TDLDA theory. 

\subsection{TDDFT in real time and optical response}

\label{subsec:tddftrt}

The real-time approach directly integrates Eq.\ (\ref{eq:tdks})
once an initial impulse has been given to the one-electron
orbitals (obtained from a previous ground-state calculation). 
Yabana and Bertsch (1996) propagated the wavefunctions in time with
a 4th-order Taylor expansion of the TDLDA equation. The 
procedure followed the previous time-dependent Hartree-Fock
method of Flocard {\it et al.} (1978) in nuclear physics. That
method was shown to conserve the energy and 
wavefunction norms to high accuracy.\footnote{Alternative 
accurate methods for propagating wavefunctions
developed in the chemical physics community are discussed
in Leforestier {\it et al.} (1991). See also Yu and 
Bandrauk (1995) which discusses a FE method for propagating
wavefunctions in real time.  The method was used to examine
molecules in intense laser fields.}
A predictor-corrector method was implemented to fix the density
at times between successive wavefunction evaluations. 
The Hamiltonian was represented with a FD form on a 
uniform Cartesian mesh. 
An 8th-order expression was employed for the Laplacian operator,
and the real-space pseudopotentials of Troullier and
Martins (1991a, 1991b) were utilized to remove the 
core electrons. The method scales as $qN_g$ since it only requires
repeated applications of the 
Kohn-Sham Hamiltonian to the wavefunctions,
here assumed to cover the whole domain. If the orbitals
could be confined to local regions of space, the 
method would scale linearly. 
Roughly $10^4$ time-propagation steps are required
to obtain the frequency-dependent response. 
As mentioned above, the physical quantities
generated are the frequency-dependent polarizability and
the closely related dipole strength function.  The entire
spectrum is produced in a single calculation without 
computations of excited-state Kohn-Sham orbitals,
and the method is not restricted to the linear-response
level of theory.  In addition, the method only requires
storage of the occupied states. 

Computations were first performed on the jellium
model for Li$_{138}$ to compare with previous numerical results; the 
dipole strength function agreed well with that
computed using a Green's function technique. 
Then calculations were performed on more physically realistic
models of large charged Na clusters and C$_{60}$.  The strength
function yields the polarizability; for the C$_{60}$ case,
a value of $\alpha = 80 \AA^3$
was computed compared with the experimental value of $85 \AA^3$ obtained
from the dielectric constant. A tight-binding model predicted 
a much lower polarizability of $45 \AA^3$. In a second study, 
Yabana and Bertsch (1997) applied the method to carbon chains and
rings which are found in interstellar matter.  For the C$_7$ chain,
the lowest TDLDA mode occurs at roughly twice the frequency 
of the HOMO-LUMO gap in the Kohn-Sham LDA states. 
The size dependence of the transitions was modeled as the
classical resonance of electrons in a conducting needle. 
The ring and chain geometries led to widely different
frequencies for the lowest collective mode. Yabana and 
Bertsch (1999) presented further computations on conjugated
hydrocarbons including polyenes, retinal 
(C$_{20}$H$_{28}$O), benzene,
and C$_{60}$. In this work, the scaling of the method was
displayed vs.\ system size and was found to be even
below $N^2$. The TDLDA dipole strength was compared to
precise experiments for the benzene 
molecule, and excellent agreement for the dipole strength
was obtained.  The computed lowest 
$\pi\rightarrow\pi^*$ sharp transition
was at 6.9 eV, the same as the experimental value, and a 
broad feature above 9 eV due to $\sigma\rightarrow\sigma^*$
transitions was also relatively accurately reproduced. 
Discussion was
given of the applicability of the H\"{u}ckel Hamiltonian; 
the H\"uckel treatment performed well for the 
$\pi\rightarrow\pi^*$ manifold.  In light of the 
large systems already addressed with the real-time
TDLDA method, it holds significant promise for examining
such problems as solvation effects on electronic excitations
in condensed phases. 

\subsection{TDDFT calculation of excited states}

\label{subsec:tddftesrs}

In the spin-unrestricted
linear-response energy representation (Casida, 1996), the excitation energies
are obtained from an eigenvalue equation:

\begin{equation}
\Omega {\bf F}_I = \omega_I^2 {\bf F}_I ,
\label{eq:erepeval}
\end{equation}

\noindent
where the excitation energy differences are $\omega_I$ and the matrix
$\Omega$ is 

\begin{equation}
\Omega_{ij\sigma,kl\tau} = \delta_{\sigma,\tau}\delta_{i,k}\delta_{j,l}
(\epsilon_{l\tau} - \epsilon_{k\tau})^2 + 2
\sqrt{(f_{i\sigma}-f_{j\sigma})(\epsilon_{j\sigma}-\epsilon_{i\sigma})}
K_{ij\sigma,kl\tau}
\sqrt{(f_{k\tau}-f_{l\tau})(\epsilon_{l\tau}-\epsilon_{k\tau})} .
\label{eq:omega}
\end{equation}

\noindent
The $(f_{i\sigma}-f_{j\sigma})$ terms are the occupation differences 
between the $i$ and $j$ $\sigma$-spin states, $(\epsilon_{j\sigma}-
\epsilon_{i\sigma})$ are the corresponding Kohn-Sham energy differences,
and the response matrix is

\begin{equation}
K_{ij\sigma,kl\tau} = \frac{\partial v_{ij\sigma}^{SCF}}
{\partial P_{kl\tau}} ,
\label{eq:response}
\end{equation}

\noindent
where $P_{kl\tau}$ is the linear response of the Kohn-Sham density
in the basis of the unperturbed orbitals. The resulting full expression
for $K_{ij\sigma,kl\tau}$ in terms of the Kohn-Sham orbitals
can be found in Casida (1996) and 
Vasiliev {\it et al.} (1999); it involves the unperturbed orbitals
only and (in the adiabatic approximation) the second derivatives
of the static exchange-correlation
functional $E_{xc}$ with respect to the spin densities. 
Therefore, at this level of theory 
$K_{ij\sigma,kl\tau}$ is time- and frequency-independent.  However,
it includes screening effects which alter the spectrum 
toward the correct physical result. The eigenvalues of 
Eq.\ (\ref{eq:erepeval}) give the transition energies, and the 
eigenvectors yield the oscillator strengths from which the
dynamic polarizability can be computed. The oscillator strengths
in this formulation satisfy the same sum rule as for the
real-time version presented above.  The method scales as
$N^3$ [where $N$ is the number of electrons, see
Casida (1996)]; however, linear-scaling methods can be applied
just as for the ground-state Kohn-Sham theory. 

Vasiliev {\it et al.} (1999) utilized the high-order FD
pseudopotential method of Chelikowsky, Troullier, and
Saad (1994) in solving Eq.\ (\ref{eq:erepeval}) for 
excitation energies. They considered the exact form
for $\Omega$ [Eq.\ (\ref{eq:omega})] and two approximate
forms, one of which was employed by Petersilka {\it et al.}
(1996) in their work.  They first examined excitations 
in closed-shell atoms and found that the exact expression
resulted in the highest accuracies.  Errors for 
low-lying excitations attributed
to the LDA exchange-correlation potential in Petersilka
{\it et al.} (1996) were corrected by using the exact
expression.  Computed energies were in error by only
a few tenths of an eV in comparison with experiment
for singlet excitations.
They also found that transition energies 
for singlet and triplet excitations computed with
TDLDA theory are in better agreement with experiment
than optimized effective potential (OEP, see 
Talman and Shadwick, 1976) or ordinary
self-consistent field methods due to the approximate inclusion
of correlation effects. The authors proceeded to apply
the TDLDA method to computations of absorption spectra 
for Na clusters. Only computations using the 
exact formulation resulted in spectra that agreed with
experiment (to within 0.2 eV). This indicates the importance
of collective excitations since the 
approximate forms neglect these contributions. 
Finally, they computed the static polarizabilities
of Na and Si clusters with the exact and approximate
formulations and found that only the exact representation yielded 
good agreement with finite-field calculations. In related work,
\"{O}\u{g}\"{u}t {\it et al.} (1997) 
computed {\it ab initio} optical gaps for very 
large Si nanocrystals (up to Si$_{525}$H$_{276}$)
with high-order FD methods.
Kim, St\"{a}dele, and Martin (1999) have 
recently utilized the high-order
FD pseudopotential method in calculations on small molecules
at the Krieger-Li-Iagrate (KLI, 1992) level for the effective
potential.  This potential is an approximation to the OEP
theory which is computationally tractable and has the 
correct $-1/r$ tail in the effective potential. The calculations
yielded better approximations to excited-state energies in relation
to the Kohn-Sham LDA values,
but the full TDDFT energy-representation method was not
employed for corrections to the Kohn-Sham KLI levels. 

\section{SUMMARY}

\label{sec:future}

Real-space methods for solving electrostatics and eigenvalue
problems involve either local Taylor expansions of the desired 
functions about a point or localized basis-set representations. 
Higher accuracy is obtained by increasing the order of 
the approximation and/or the resolution of 
the mesh. However, standard iterative processes become less
efficient on finer meshes  
due to the difficulty of reducing the long-wavelength
modes of the errors. Multigrid methods provide a remedy
for this slowing-down phenomenon inherent 
in real-space numerical methods.  Many of the early
limitations of real-space methods (such as very large
required meshes) have been overcome in recent years 
with the development of efficient high-order finite-difference
and finite-element methods.  This review has surveyed a wide
range of physical applications of real-space numerical techniques
including biophysical electrostatics, ground-state 
electronic structure, and computations of electronic 
response and excitation energies. Recent real-space computations
have tackled problems with hundreds to thousands of atoms
at a realistic level of representation.  The discussion 
presented in this review leads to several conclusions:

\begin{itemize}
\item The underlying representation is relatively 
simple in real space.
The finite-difference method is particulary straightforward,
while the finite-element and wavelet methods 
involve some increased
complexity. As an example, a self-consistent 
Kohn-Sham LDA multigrid program using the high-order
finite-difference method requires less than 
5000 lines of computer code.   
\item With the incorporation of high-order methods, accuracies
comparable to plane-wave calculations are obtained on 
similar-sized meshes. 
\item The Laplacian and Hamiltonian operators require information
only from close lattice points; that is, the operators are near-local
in space. Therefore, the matrices are sparse, highly
banded, and very structured. Each application of the operators scales
linearly with system size, and the method is readily implemented 
on parallel computers by partitioning the problem in space. 
The locality also allows for incorporation
into linear-scaling electronic structure methods. 
\item Multigrid methods provide the optimal solvers for problems
represented in real space.  For Poisson problems, the multigrid
method scales
linearly with system size and requires only about 10 iterations
on the finest level to obtain convergence. Eigenvalue solvers 
scale as $q^2N_g$ (where $q$ is the number of eigenfunctions and
$N_g$ the number of fine grid points) if the eigenfunctions span
the whole space.\footnote{With algorithmic improvements, this
scaling can be reduced to $qN_g$. See Costiner 
and Ta'asan (1995).} If a localized orbital 
representation is possible,
the multigrid eigenvalue methods scale linearly with size
due to the locality of each operation in the algorithm. 
\item Nonlinear multigrid methods 
require fewer operations per self-consistency update than
plane-wave methods on uniform grids with orbitals that span
the physical domain. In addition, the multigrid method is 
at least as efficient as the best plane-wave methods in terms of the
number of self-consistency steps to reach the ground state. 
The multigrid solution requires at most a few self-consistency 
iterations. The solution involves 10-20 total applications of the 
Hamiltonian to the wavefunctions on the finest level
and a few updates of the 
effective potential (one for each
self-consistency cycle); each update of the 
Hartree potential requires the same effort as the update
of one orbital.  
\item Real-space methods allow for higher resolution in space without 
loss of efficiency. That is, they are readily adaptable and thus
can handle problems with a wide range of length scales. 
\item The eigenfunction constraint and subspace orthogonalization
operations can be performed on coarse levels where the cost 
is very low. Also, the effective potential can be updated on 
coarse levels leading to the possibility of complete solution in a 
single self-consistency cycle. These developments, along with 
the mesh-refinement techniques, will lead to reductions in 
computational cost of an order-of-magnitude compared with 
existing algorithms. 
\item The flexibility of the representation has been utilized 
both in very high accuracy computations and in applications 
to large systems. The real-space methods do not suffer from
linear dependence problems which occur in LCAO methods. Typically,
the numerical convergence is controlled by a few
parameters such as grid spacing, domain size, and order
of the representation.  
\item Real-space algorithms very similar to those for electrostatics and 
ground-state electronic structure can be employed to solve
time-dependent problems. 
\end{itemize}

In the view of the author, the most promising areas for
future work on real-space methods concern the 
development of highly
{\it adaptive} and {\it efficient} numerical techniques which focus
resolution in key regions of space 
as the iterative process moves towards the ground-state
solution or evolves in real time.  
There will always
exist a tradeoff between the simplicity of the representation (where
finite differences are best) and the flexibility and accuracy of
local basis functions (where finite element methods are
superior).  The related local
LCAO methods allow for significantly smaller
overall basis-set size in relation to real-space formulations, 
but the Laplacian and Hamiltonian operators
are not as well structured and banded. 
The intersection between the simple structured approaches
on the one hand and the more physical local bases on the other 
should provide for a fruitful growth of new ideas in computational
materials science. Multiscale methods for solving the problems
will figure prominently since they allow for flexibility in
the representation while maintaining high efficiency.
A brief survey of physical and chemical problems which have 
already been addressed serves to illustrate the wide range 
of length scales accessible with real-space techniques: 
electrostatics of proteins interacting with nucleic acids,
charged polymers in confined geometries,
large-scale electronic structure of materials, and computation
of spectroscopic quantities for large molecules 
in the gas phase.  One 
can imagine a time in the not-too-distant future when
it is possible to simulate the motion of a solute
molecule in a liquid with the inclusion of all the 
electrons and model the solvent influence on the 
electronic absorption spectra.  Real-space methods 
possess many of the features that would be required to 
address such a challenging problem.  

\section*{ACKNOWLEDGMENTS}

I would like to thank Matt Challacombe, 
Rob Coalson, and John Pask for helpful discussions, and
Victoria Manea for a critical reading of the manuscript. I 
gratefully acknowledge the significant contributions from the 
members of my research group: Karthik Iyer, Michael Merrick,
and Jian Wang. I especially thank Achi Brandt for his
advice on multigrid methods.  The research  
was partially supported by the National Science Foundation. 

\section*{Appendix A}
\label{app:hofd}

As an example of the ease of generating a high-order form for
the Laplacian operator, the following Mathematica script 
for the 10th-order case is included:

\begin{verbatim}
g[x_]:=

Evaluate[InterpolatingPolynomial[{{x0-5,ym5},{x0-4,ym4},{x0-3,ym3},
  {x0-2,ym2},{x0-1,ym1},{x0,y0},{x0+1,yp1},{x0+2,yp2},{x0+3,yp3},
  {x0+4,yp4},{x0+5,yp5}},x]]

gp[x_]:=Evaluate[D[g[x],{x,2}]]

r=Simplify[Expand[Collect[gp[x0+0],{ym5,ym4,ym3,ym2,ym1,y0,yp1,yp2,
  yp3,yp4,yp5}]]]

OUTPUT:
Out[1]=(-73766 y0 + 42000ym1 - 6000ym2 + 1000ym3 - 125ym4 + 8ym5 +
   42000yp1 - 6000yp2 + 1000yp3 - 125yp4 + 8yp5)/25200

\end{verbatim}

\noindent
The weights obtained for the FD Laplacian 
up through 12th order are
presented in Table II. For
the three-dimensional case, the $p$th order approximation
requires $3p+1$ terms. Hamming (1962) also discusses procedures
for generating other high-order formulas such as interpolation
and integration.



\begin{table}[]
\begin{center}
\caption{Orbital energies for the oxygen
dimer, from Chelikowsky, Troullier, Wu, and
Saad (1994). FD-12 refers to high-order FD 
calculations in a 12 au box. PW-12 and PW-24 
refer to plane-wave calculations with supercells
of 12 and 24 au on a side. Energies are in eV.}
{\begin{tabular}{lrrr}
Orbital & FD-12 & PW-12 & PW-24 \\ \tableline
$\sigma_s$ & -32.56 & -32.09 & -32.60 \\
$\sigma_s^*$ & -19.62 & -19.11 & -19.57 \\
$\sigma_p$ & -13.63 & -12.93 & -13.37 \\
$\pi_p$ & -13.24 & -12.54 & -12.98 \\
$\pi_p^*$ & -6.35 & -5.53 & -5.98 \\
\end{tabular} }
\end{center}
\label{tab:cheldata}
\end{table}      

\begin{table}[]
\begin{center}
\caption{Coefficients for the Laplacian.  One side plus
the central point are shown.
Each coefficient term should be divided by the prefactor.
The Laplacian is symmetric about the central point.}
{\begin{tabular}{llrrrrrrrr}
Points & Order & Prefactor &
\multicolumn{7}{c}{Coefficients} \\ \tableline
N=3 & 2nd & 1 & & & && & 1 & -2  \\ 
N=5 & 4th & 12 & & & & & -1 & 16 & -30  \\ 
N=7 & 6th & 180 & & && 2 & -27 & 270 & -490  \\ 
N=9 & 8th & 5040 && &-9 &128 & -1008 & 8064 & -14350  \\ 
N=11&10th & 25200& &8 &-125 &1000 &-6000 &42000 &-73766 \\
N=13&12th&831600& -50 &864 &-7425 &44000& -222750 &1425600 &-2480478\\ 
\end{tabular} }
\end{center}
\label{tab:laplacian}
\end{table}


\begin{references}



\bibitem{}
Abramowitz, M., and I. A. Stegun, 1964, {\it Handbook of 
Mathematical Functions} (National Bureau of Standards,
Washington, DC).

\bibitem{}
Abou-Elnour, A., and K. Schuenemann, 1993, J. Appl. Phys.
{\bf 74}, 3273.

\bibitem{}
Ackermann, J., B. Erdmann, and R. Roitzsch, 1994,
J. Chem. Phys. {\bf 101}, 7643. 

\bibitem{}
Ackerman, J., and R. Roitzsch, 1993, Chem. Phys. Lett.
{\bf 214}, 109.

\bibitem{}
Alavi, A., J. Kohanoff, M. Parrinello, and D. Frenkel, 1994,
Phys. Rev. Lett. {\bf 73}, 2599.

\bibitem{}
Alcouffe, R. E., A. Brandt, J. E. Dendy, and J. W. Painter,
1981, SIAM J. Sci. Stat. Comput. {\bf 2}, 430.

\bibitem{}
Allen, M. P., and D. J. Tildesley, 1987, {\it Computer Simulation
of Liquids} (Oxford, New York). 


\bibitem{}
Ancilotto, F., P. Blandin, and F. Toigo, 1999,
Phys. Rev. B {\bf 59}, 7868.

\bibitem{}
Antosiewicz, J., J. A. McCammon, and M. K. Gilson, 1994,
J. Mol. Biol. {\bf 238}, 415.

\bibitem{}
Antosiewicz, J., J. A. McCammon, and M. K. Gilson, 1996,
Biochemistry {\bf 35}, 7819.

\bibitem{}
Arias, T. A., 1999, Rev. Mod. Phys. {\bf 71}, 267.


\bibitem{}
Bai, D., and A. Brandt, 1987, SIAM J. Sci. Stat. Comput.
{\bf 8}, 109.

\bibitem{}
Baker, N., M. Holst, and F. Wang, 1999, J. Comput. Chem.
(submitted).

\bibitem{}
Banci, L., and P. Comba, 1997, Editors, {\it Molecular Modeling and
Dynamics of Bioinorganic Systems} (Kluwer, Boston). 

\bibitem{}
Baroni, S., and P. Giannozzi, 1992, Europhys. Lett.
{\bf 17}, 547.

\bibitem{}
Batcho, P. F., 1998, Phys. Rev. A {\bf 57}, 4246.

\bibitem{}
Bauernschmitt, R., and R. Ahlrichs, 1996, Chem. Phys.
Lett. {\bf 256}, 454.

\bibitem{}
Beck, T. L., 1997, Intl. J. Quantum Chem. {\bf 65}, 
477.

\bibitem{}
Beck, T. L., 1999a, in 
{\it Simulation and Theory of Electrostatic Interactions
in Solution}, AIP Conference Proceedings No. 492,
edited by G. Hummer and
L. R. Pratt (AIP, New York). 

\bibitem{}
Beck, T. L., 1999b, J. Comput. Chem. {\bf 20}, 1731.

\bibitem{}
Beck, T. L., 2000, in {\it Multiscale Computational 
Methods in Chemistry and Biology}, edited by J. Bernholc,
K. Binder, and A. Brandt, NATO Proceedings of Eilat
Workshop, in press. 

\bibitem{}
Beck, T. L., K. A. Iyer, and M. P. Merrick, 1997,
Intl. J. Quantum Chem. {\bf 61}, 341.

\bibitem{}
Becke, A. D., 1989, Intl. J. Quantum Chem.: Quantum Chem.
Symp. {\bf 23}, 599.

\bibitem{}
Bellaiche, L., and K. Kunc, 1997, Intl. J. Quantum Chem. 
{\bf 61}, 647.

\bibitem{}
Ben-Tal, N., and R. D. Coalson, 1994, J. Chem. Phys.
{\bf 101}, 5148.

\bibitem{}
Ben-Tal, N., B. Honig, C. Miller, and S. McLaughlin, 1997,
Biophys. J. {\bf 73}, 1717.

\bibitem{}
Bernholc, J., 1999, Physics Today, September issue, page 30.

\bibitem{}
Bernholc, J., E. L. Briggs, D. J. Sullivan, C. J. Brabec,
M. Buongiorno Nardelli, K. Rapcewicz, C. Roland, and M. Wensell,
1997, Intl. J. Quantum Chem. {\bf 65}, 531.

\bibitem{}
Bernholc, J., J.-Y. Yi, and D. J. Sullivan, 1991,
Faraday Disc. Chem. Soc. {\bf 92}, 217.

\bibitem{}
Bicout, D., and M. Field, 1996, Editors, {\it Quantum
Mechanical Simulation Methods for Studying Biological
Systems} (Springer, New York).

\bibitem{}
Booten, A., and H. van der Vorst, 1996, Computers in 
Physics {\bf 10}, 239.

\bibitem{}
Bowen, W. R., and A. O. Sharif, 1997, J. Coll. Int. Sci.
{\bf 187}, 363.

\bibitem{}
Bowler, D. R., I. J. Bush, and M. J. Gillan, 1999, preprint
xxx.lanl.gov/abs/cond-mat/9902343. Submitted to Intl. J.
Quantum Chem.

\bibitem{}
Bowler, D. R., and M. J. Gillan, 1998, Comput. Phys. Commun.
{\bf 112}, 103.

\bibitem{}
Braess, D., and R. Verfurth, 1990, SIAM J. Numer. Anal.
{\bf 27}, 979.

\bibitem{}
Brandt, A., 1977, Math. Comput. {\bf 31}, 333.

\bibitem{}
Brandt, A., 1980, in {\it Special Topics of Applied Mathematics},
ed. J. Frehse, D. Pallaschke, and U. Trottenberg (North-Holland,
New York).

\bibitem{}
Brandt, A., 1982, in {\it Multigrid Methods}, edited by W. Hackbusch
and U. Trottenberg (Springer-Verlag, New York).

\bibitem{}
Brandt, A., 1984, {\it Multigrid Techniques: 1984 Guide, with
Applications to Fluid Dynamics}; GMD-Studien Nr. 85; Available
from GMD-AIW, Postfach 1316. D-53731, St. Augustin 1, Germany.

\bibitem{}
Brandt, A., 1999, {\it Multiscale Scientific Computation:
Six Year Research Summary}. Available at
www.wisdom.weizmann.ac.il/$\sim$achi.

\bibitem{}
Brandt, A., S. F. McCormick, and J. Ruge, 1983, SIAM
J. Sci. Stat. Comput. {\bf 4}, 244. 

\bibitem{}
Brenner, S. C., and L. R. Scott, 1994, {\it The Mathematical
Theory of Finite Element Methods} (Springer, New York).

\bibitem{}
Briggs, E. L., D. J. Sullivan, and J. Bernholc, 1995,
Phys. Rev. B {\bf 52}, R5471.

\bibitem{}
Briggs, E. L., D. J. Sullivan, and J. Bernholc, 1996,
Phys. Rev. B {\bf 54}, 14362.

\bibitem{}
Briggs, W. L., 1987, {\it A Multigrid Tutorial} (SIAM
Books, Philadelphia).


\bibitem{}
Car, R., and M. Parrinello, 1985, Phys. Rev. Lett.
{\bf 55}, 2471.

\bibitem{}
Casida, M. E., 1996, in {\it Recent Developments and Applications of
Modern Density Functional Theory}, edited by J. M. Seminario
(Elsevier, New York).

\bibitem{}
Casida, M. E., C. Jamorski, K. C. Casida, and D. R. Salahub,
1998, J. Chem. Phys. {\bf 108}, 4439.

\bibitem{}
Ceperley, D. M., and B. J. Alder, 1980, Phys. Rev. Lett.
{\bf 45}, 566.

\bibitem{}
Challacombe, M., 1999a, personal communication.

\bibitem{}
Challacombe, M., 1999b, J. Chem. Phys. {\bf 110},
2332.

\bibitem{}
Challacombe, M., 2000, Comput. Phys. Commun.,
in press.

\bibitem{}
Challacombe, M., and E. Schwegler, 1997, J. Chem. Phys.
{\bf 106}, 5526.

\bibitem{}
Challacombe, M., E. Schwegler, and J. Alml\"{o}f,
1996, J. Chem. Phys. {\bf 104}, 4685.

\bibitem{}
Challacombe, M., C. White, and M. Head-Gordon, 1997,
J. Chem. Phys. {\bf 107}, 10131.

\bibitem{}
Chelikowsky, J. R., N. Troullier, and Y. Saad, 1994,
Phys. Rev. Lett. {\bf 72}, 1240.

\bibitem{}
Chelikowsky, J. R., N. Troullier, K. Wu, and Y. Saad, 1994,
Phys. Rev. B {\bf 50}, 11355.

\bibitem{}
Chen, S. W., and B. Honig, 1997, J. Phys. Chem. B {\bf 101},
9113.


Cheng, H., L. Greengard, and V. Rokhlin, 1999, J. Comput. Phys.
{\bf 155}, 468.


\bibitem{}
Coalson, R. D., and T. L. Beck, 1998, in {\it Encyclopedia
of Computational Chemistry}, edited by P. von Rague Schleyer, Vol. 3,
p. 2086 (John-Wiley, New York).

\bibitem{}
Coalson, R. D., and A. Duncan, 1992, J. Chem. Phys. 
{\bf 97}, 5653.

\bibitem{}
Coalson, R. D., D. K. Pant, and D. W. Langer, 1994,
J. Lightwave Tech. {\bf 12}, 1015.

\bibitem{}
Corry, B., S. Kuyucak, and S.-H. Chung, 2000, Biophys. J.
{\bf 78}, 2364.

\bibitem{}
Cortis, C. M., and R. A. Friesner, 1997, J. Comput. Chem.
{\bf 18}, 1570.

\bibitem{}
Costiner, S., and S. Ta'asan, 1995a, Phys. Rev. E
{\bf 51}, 3704.

\bibitem{}
Costiner, S., and S. Ta'asan, 1995b, Phys. Rev. E
{\bf 52}, 1181.


\bibitem{}
Darden, T., D. York, and L. Pedersen, 1993, J. Chem. Phys.
{\bf 98}, 10089.

\bibitem{}
Davis, M. E., and J. A. McCammon, 1989, J. Comput. Chem.
{\bf 10}, 386.



\bibitem{}
Davstad, K., 1992, J. Comput. Phys. {\bf 99}, 33.


\bibitem{}
Deconinck, H., and C. Hirsch, 1982, in {\it Multigrid
Methods}, edited by W. Hackbusch and U. Trottenberg
(Springer-Verlag, New York).

\bibitem{}
Deserno, J., and C. Holm, 1998, J. Chem. Phys. {\bf 109}, 7678.

\bibitem{}
Duffy, M. G., 1982, SIAM J. Numer. Anal. {\bf 19}, 6.

\bibitem{}
D\"{u}sterh\"{o}ft, C., D. Heinemann, and D. Kolb,
1998, Chem. Phys. Lett. {\bf 296}, 77.


\bibitem{}
Essmann, U., L. Perera, M. L. Berkowitz, T. Darden,
H. Lee, and L. Pedersen, 1995, J. Chem. Phys.
{\bf 103}, 8577.

\bibitem{}
Ewald, P., 1921, Ann. Phys. {\bf 64}, 253

\bibitem{}
Fattebert, J.-L., 1999, J. Comput. Phys. {\bf 149}, 75. 

\bibitem{}
Fattebert, J.-L., and J. Bernholc, 2000, preprint. 

\bibitem{}
Fisher, C. L., J.-L. Chen, J. Li, D. Bashford, and L. Noodleman,
1996, J. Phys. Chem. {\bf 100}, 13498.

\bibitem{}
Flocard, H., S. E. Koonin, and M. S. Weiss, 1978, Phys. Rev. C 
{\bf 17}, 1682.

\bibitem{}
Fogolari, F., and J. M. Briggs, 1997, Chem. Phys. Letts.
{\bf 281}, 135.

\bibitem{}
Fushiki, M., 1992, J. Chem. Phys. {\bf 97}, 6700.

\bibitem{}
Galli, G., and M. Parrinello, 1992, Phys. Rev. Lett.
{\bf 69}, 3547.

\bibitem{}
Goedecker, S., 1999, Rev. Mod. Phys. {\bf 71}, 1085 .

\bibitem{}
Goedecker, S., and O. V. Ivanov, 1998a, Solid State Commun.
{\bf 105}, 665.

\bibitem{}
Goedecker, S., and O. V. Ivanov, 1998b, Computers in Phys.
{\bf 12}, 548.

\bibitem{}
Goedecker, S., M. Teter, and J. Hutter, 1996, Phys. Rev. B
{\bf 54}, 1703.

\bibitem{}
Golub, G. H., and C. F. van Loan, 1996, {\it Matrix
Computations} (Johns Hopkins University, Baltimore).

\bibitem{}
Goringe, C. M., E. Hern\'{a}ndez, M. J. Gillan, and
I. J. Bush, 1997, Comput. Phys. Commun. {\bf 102}, 1.

\bibitem{}
Greengard, L., 1994, Science {\bf 265}, 909.

\bibitem{}
Greengard, L., and J.-Y. Lee, 1996, J. Comput. Phys.
{\bf 125}, 415.

\bibitem{}
Grinstein, F. F., H. Rabitz, and A. Askar, 1983, J. Comput. Phys.
{\bf 51}, 423.

\bibitem{}
Gross, E. K. U., and R. M. Dreizler, 1994, Editors, {\it Density 
Functional Theory} (Plenum, New York). 

\bibitem{}
Gross, E. K. U., and W. Kohn, 1990, in {\it Advances in Quantum 
Chemistry: Vol. 21}, edited by S. B. Trickey
(Academic Press, New York).

\bibitem{}
Guldbrand, L., B. J\"{o}nsson, H. Wennerstr\"{o}m,
and P. Linse, 1984, J. Phys. Chem. {\bf 80}, 2221.

\bibitem{}
Gupta, M. M., J. Kouatchou, and J. Zhang, 1997, J. Comput. Phys.
{\bf 132}, 226.

\bibitem{}
Gygi, F., 1993, Phys. Rev. B {\bf 48}, 11692.

\bibitem{}
Gygi, F., and G. Galli, 1995, Phys. Rev. B {\bf 52}, R2229.

\bibitem{}
Hackbusch, W., 1985, {\it Multigrid Methods and Applications}
(Springer-Verlag, New York).

\bibitem{}
Hackel, S., D. Heinemann, D. Kolb, and B. Fricke, 1993,
Chem. Phys. Lett. {\bf 206}, 91.

\bibitem{}
Haggerty, L., and A. M. Lenhoff, 1991, J. Phys. Chem.
{\bf 95}, 1472.

\bibitem{}
Hamilton, T. P., and P. Pulay, 1986, J. Chem. Phys.
{\bf 84}, 5728.

\bibitem{}
Hamming, R. W., 1962, {\it Numerical Methods for Scientists and
Engineers} (Dover, New York).

\bibitem{}
Hansen, J.-P., and I. R. McDonald, 1986, {\it Theory
of Simple Liquids} (Academic Press, New York).

\bibitem{}
Harris, F. E., and H. J. Monkhorst, 1970, Phys.
Rev. B {\bf 2}, 4400; (errata) 1974, {\bf 9}, 3946.

\bibitem{}
Harris, R. A., and L. R. Pratt, 1985, J. Chem. Phys.
{\bf 82}, 856.

\bibitem{}
Haynes, P. D., and M. C. Payne, 1997, Comput. Phys. Commun.
{\bf 102}, 17.

\bibitem{}
Heinemann, D., B. Fricke, and D. Kolb, 1988, Chem. Phys. Lett.
{\bf 145}, 125.

\bibitem{}
Heinemann, D., D. Kolb, and B. Fricke, 1987, Chem. Phys. Lett.
{\bf 137}, 180.

\bibitem{}
Hern\'{a}ndez, E., and M. Gillan, 1995, Phys. Rev. B {\bf 51}, 10157.

\bibitem{}
Hern\'{a}ndez, E., M. J. Gillan, and C. M. Goringe, 1997, Phys. Rev. B
{\bf 55}, 13485.

\bibitem{}
Hierse, W., and E. B. Stechel, 1994, Phys. Rev. B {\bf 50},
17811.


\bibitem{}
Hockney, R. W., and J. W. Eastwood, 1988, {\it Computer
Simulation Using Particles} (Adam Hilger, New York). 

\bibitem{}
Hohenberg, P., and W. Kohn, 1964, Phys. Rev.
{\bf 136}, B864.

\bibitem{}
Holst, M. J., and F. Saied, 1993, J. Comput. Chem. {\bf 14}, 105.

\bibitem{}
Holst, M. J., and F. Saied, 1995, J. Comput. Chem.
{\bf 16}, 337.

\bibitem{}
Honig, B., and A. Nicholls, 1995, Science {\bf 268}, 1144.

\bibitem{}
Hoshi, T., M. Arai, and T. Fujiwara, 1995, Phys. Rev. B
{\bf 52}, R5459.

\bibitem{}
Hoshi, T., and T. Fujiwara, 1997, J. Phys. Soc. Jpn.
{\bf 66}, 3710.

\bibitem{}
Hummer, G., and L. R. Pratt, 1999, Editors, 
{\it Simulation and Theory of Electrostatic Interactions
in Solution}, AIP Conference Proceedings No. 492 (AIP, New York). 

\bibitem{}
Hutter, J., H. P. L\"{u}thi, and M. Parrinello, 
1994, Comput. 
Mat. Sci. {\bf 2}, 244.

\bibitem{}
Ichimaru, S., 1994, {\it Statistical Plasma Physics Volume II: Condensed
Plasmas} (Addison-Wesley, New York).


\bibitem{}
Ismail-Beigi, S., and T. Arias, 1998, Phys. Rev. B 
{\bf 57}, 11923.

\bibitem{}
Iyer, K., M. P. Merrick, and T. L. Beck, 1995, J. Chem. Phys.
{\bf 103}, 227.

\bibitem{}
Jamorski, C., M. E. Casida, and D. R. Salahub, J. Chem. Phys.
{\bf 104}, 5134.

\bibitem{}
Jensen, F., 1999, {\it Introduction to Computational Chemistry}
(Wiley, New York).

\bibitem{}
Jing, X., N. Troullier, D. Dean, N. Binggeli, J. R. Chelikowsky,
K. Wu, and Y. Saad, 1994, Phys. Rev. B {\bf 50}, 12234.

\bibitem{}
Kim, J., F. Mauri, and G. Galli, 1995, Phys. Rev. B
{\bf 52}, 1640.

\bibitem{}
Kim, Y.-H., I.-H. Lee, and R. M. Martin, 1999, preprint
http://xxx.lanl.gov/abs/physics/9911031.

\bibitem{}
Kim, Y.-H., M. St\"{a}dele, and R. M. Martin, 1999,
preprint xxx.lanl.gov/abs/physics/9909006.

\bibitem{}
Kimball, G. E., and G. H. Shortley, 1934, Phys. Rev.
{\bf 45}, 815.

\bibitem{}
Kirkwood, J. G., 1934, J. Chem. Phys. {\bf 2}, 767.


\bibitem{}
Kohn, W., 1996, Phys. Rev. Lett. {\bf 76}, 3168.

\bibitem{}
Kohn, W., and L. J. Sham, 1965, Phys. Rev.
{\bf 140}, A1133.

\bibitem{}
Kopylow, A. v., D. Heinemann, and D. Kolb, 1998, J. Phys. B
{\bf 31}, 4743. 



\bibitem{}
Kresse, G., and J. Furthm\"{u}ller, 1996, Phys. Rev. B {\bf 54}, 
11169.

\bibitem{}
Krieger, J. B., Y. Li, and G. J. Iafrate, 1992, Phys.
Rev. A {\bf 45}, 101.

\bibitem{}
Kurnikova, M. G., R. D. Coalson, and A. Nitzan, 1999, 
Biophys. J. {\bf 76}, 642.

\bibitem{}
Laaksonen, L., D. Sundholm, and P. Pyykko, 1985, Intl. J.
Quantum Chem. {\bf 27}, 601.

\bibitem{}
Lee, I.-H., Y.-H. Kim, and R. M. Martin, 1999, preprint
http://xxx.lanl.gov/abs/physics/9911030.

\bibitem{}
Leforestier, C., R. H. Bisseling, C. Cerjan, M. D. Feit, R. Friesner,
A. Guldberg, A. Hammerich, G. Jolicard, W. Karrlein, H.-D. Meyer,
N. Lipkin, O. Roncero, and R. Kosloff, 1991, J. Comput. Phys. {\bf 94},
59.

\bibitem{}
Lepaul, S., A. de Lustrac, and R. Bouillault, 1996, IEEE
Trans. Magn. {\bf 32}, 1018.

\bibitem{}
Levin, F. S., and J. Shertzer, 1985, Phys. Rev. A 
{\bf 32}, 3285.


\bibitem{}
Light, J. C., I. P. Hamilton, and J. V. Lill, 1985,
J. Chem. Phys. {\bf 82}, 1400.

\bibitem{}
Lippert, G., J. Hutter, and M. Parrinello, 1997,
Mol. Phys. {\bf 92}, 477.

\bibitem{}
Loeb, A. L., 1951, J. Coll. Sci. {\bf 6}, 75.

\bibitem{}
L\"{o}wen, H., 1994, J. Chem. Phys. {\bf 100}, 6738.


\bibitem{}
Luty, B. A., M. E. Davis, and J. A. McCammon, 1992,
J. Comput. Chem. {\bf 13}, 1114.

\bibitem{}
Mahan, G. D., and K. R. Subbaswamy, {\it Local Density
Theory of Polarizability} (Plenum, New York, 1990).

\bibitem{}
Marchioro, T. L., M. Arnold, D. K. Hoffman, W. Zhu, Y. Huang, 
and D. J. Kouri, 1994, Phys. Rev. E {\bf 50}, 2320.

\bibitem{}
Marcus, R. A., 1955, J. Chem. Phys. {\bf 23}, 1057.

\bibitem{}
Mauri, F., and G. Galli, 1994, Phys. Rev. B {\bf 50}, 4316.

\bibitem{}
Mauri, F., G. Galli, and R. Car, 1993, Phys. Rev. B
{\bf 47}, 9973.

\bibitem{}
Merrick, M. P., K. A. Iyer, and T. L. Beck, 1995,
J. Phys. Chem. {\bf 99}, 12478.

\bibitem{}
Merrick, M. P., K. A. Iyer, and T. L. Beck, 1996, in {\it Quantum
Mechanical Simulation Methods for Studying Biological
Systems}, edited by D. Bicout and M. Field (Springer, New York).

\bibitem{}
Mermin, N. D., 1965, Phys. Rev. {\bf 137}, A1441.

\bibitem{}
Millam, J. M., and G. E. Scuseria, 1997, J. Chem. Phys.
{\bf 106}, 5569.

\bibitem{}
Misra, V. K., J. L. Hecht, K. A. Sharp, R. A. Friedman, 
and B. Honig, 1994, J. Mol. Biol. {\bf 238}, 264.

\bibitem{}
Modine, N. A., G. Zumbach, and E. Kaxiras, 1997, Phys. Rev.
B {\bf 55}, 10289.


\bibitem{}
Moncrieff, D., and S. Wilson, 1993, Chem. Phys. Lett.
{\bf 209}, 423.

\bibitem{}
Montoro, J. C. G., and J. L. F. Abascal, 1998, J. Chem. Phys.
{\bf 109}, 6200.


\bibitem{}
Nicholls, A., and B. Honig, 1991, J. Comput. Chem. 
{\bf 12}, 435.


\bibitem{}
Oberoi, H., and N. M. Allewell, 1993, Biophys. J.
{\bf 65}, 48.

\bibitem{}
\"{O}\u{g}\"{u}t, S., J. R. Chelikowsky, and S. G. Louie,
1997, Phys. Rev. Lett. {\bf 79}, 1770.

\bibitem{}
Ono, T., and K. Hirose, 1999, Phys. Rev. Lett. 
{\bf 82}, 5016.

\bibitem{}
Onsager, L., 1933, Chem. Rev. {\bf 13}, 73.


\bibitem{}
Ordej\'{o}n, P., D. A. Drabold, R. M. Martin, and M. P. Grumbach,
1995, Phys. Rev. B {\bf 51}, 1456.

\bibitem{}
Orszag, S. A., 1972, Stud. Appl. Math. {\bf 51}, 253.

\bibitem{}
Parr, R. G., and W. Yang, 1989, {\it Density Functional Theory 
of Atoms and Molecules} (Oxford, Oxford). 

\bibitem{}
Pask, J. E., 1999, personal communication.

\bibitem{}
Pask, J. E., B. M. Klein, C. Y. Fong, and P. A. Sterne, 1999,
Phys. Rev. B {\bf 59}, 12352.

\bibitem{}
Patra, C. N., and A. Yethiraj, 1999, J. Phys. Chem. B
{\bf 103}, 6080.

\bibitem{}
Pauling, L., and E. B. Wilson, 1935, {\it Introduction to
Quantum Mechanics} (Dover, New York), p. 202.

\bibitem{}
Payne, M., M. Teter, D. Allan, T. Arias, and J. Joannopoulos, 1992,
Rev. Mod. Phys. {\bf 64}, 1045.

\bibitem{}
P\'{e}rez-Jord\'{a}, J. M., and W. Yang, 1997, J. Chem. Phys. 
{\bf 107}, 1218.

\bibitem{}
P\'{e}rez-Jord\'{a}, J. M., and W. Yang, 1998, Chem. Phys. Lett. 
{\bf 282}, 71.

\bibitem{}
Petersilka, M., U. J. Gossmann, and E. K. U. Gross, 
1996, Phys. Rev. Lett.
{\bf 76}, 1212. 

\bibitem{}
Pettit, B. M., and C. V. Valdeavella, 1999, in 
{\it Simulation and Theory of Electrostatic Interactions
in Solution}, AIP Conference Proceedings No. 492,
edited by G. Hummer and
L. R. Pratt (AIP, New York). 

\bibitem{}
Pollock, E. L., 1999, in {\it Simulation and
Theory of Electrostatic Interactions in Solution}, AIP
Conference Proceedings No. 492, edited
by L. R. Pratt and G. Hummer (AIP, New York).

\bibitem{}
Pollock, E. L., and J. Glosli, 1996, Comput. Phys. Commun.
{\bf 95}, 93.

\bibitem{}
Pratt, L. R., G. J. Tawa, G. Hummer, A. E. Garcia, and
S. A. Corcelli, 1997, Intl. J. Quantum Chem. {\bf 64}, 121.


\bibitem{}
Press, W. H., S. A. Teukolsky, W. T. Vetterling, and B. P. Flannery,
1992, {\it Numerical Recipes in C: The Art of Scientific Computing}
(Cambridge, New York).

\bibitem{}
Pulay, P., 1969, Mol. Phys. {\bf 17}, 197.

\bibitem{}
Pulay, P., 1980, Chem. Phys. Lett. {\bf 73}, 393.

\bibitem{}
Pulay, P., 1982, J. Comput. Chem. {\bf 3}, 556.



\bibitem{}
Reddy, B. D., 1998, {\it Introductory Functional Analysis with
Applications to Boundary Value Problems and Finite Elements}
(Springer, New York).

\bibitem{}
Reiner, E. S., and C. J. Radke, 1990, J. Chem. Soc. 
Faraday Trans. {\bf 86}, 3901.

\bibitem{}
Rice, S. A., 1959, Rev. Mod. Phys. {\bf 31}, 69.

\bibitem{}
Ringnalda, M. N., M. Belhadj, and R. A. Friesner, 1990, J. Chem. Phys.
{\bf 93}, 3397.

\bibitem{}
Rodriguez, J. H., D. E. Wheeler, and J. K. McCusker, 1998,
J. Am. Chem. Soc. {\bf 120}, 12051.

\bibitem{}
Rowlinson, J. S., and B. Widom, 1982, {\it Molecular Theory
of Capillarity} (Oxford, New York).

\bibitem{}
Sagui, C., and T. A. Darden, 1999, in {\it Simulation and
Theory of Electrostatic Interactions in Solution}, AIP
Conference Proceedings No. 492, edited
by L. R. Pratt and G. Hummer (AIP, New York).

\bibitem{}
S\'{a}nchez-Portal, D., P. Ordej\'{o}n, E. Artacho, and J. M. Soler, 1997,
Intl. J. Quantum Chem. {\bf 65}, 453.

\bibitem{}
Sankey, O. F., and D. J. Niklewski, 1989, Phys. Rev. B
{\bf 40}, 3979.

\bibitem{}
Schneider, B. I., and D. L. Feder, 1999, Phys. Rev. A
{\bf 59}, 2232.

\bibitem{}
Schwegler, E., and M. Challacombe, 1996, J. Chem. Phys.
{\bf 105}, 2726.

\bibitem{}
Schwegler, E., M. Challacombe, and M. Head-Gordon, 1997, 
J. Chem. Phys. {\bf 106}, 9708.

\bibitem{}
Seminario, J. M., 1996, Editor, {\it Recent Developments
and Applications of Modern Density Functional Theory}
(Elsevier, New York).

\bibitem{}
Seitsonen, A. P., M. J. Puska, and R. M. Nieminen, 1995,
Phys. Rev. B {\bf 51}, 14057.

\bibitem{}
Sharp, K. A., and B. Honig, 1990a, Annual Rev. Biophys.
Biophys. Chem. {\bf 19}, 301.

\bibitem{}
Sharp, K. A., and B. Honig, 1990b, J. Phys. Chem.
{\bf 94}, 7684.


\bibitem{}
Springborg, M., 1997, Editor, {\it Density-Functional Methods
in Chemistry and Materials Science} (Wiley, New York). 


\bibitem{}
Sterne, P. A., J. E. Pask, and B. M. Klein, 1999, 
Appl. Surf. Sci. {\bf 149}, 238.

\bibitem{}
Strain, M. C., G. E. Scuseria, and M. J. Frisch, 1996,
Science {\bf 271}, 51.

\bibitem{}
Strang, G., and G. J. Fix, 1973, {\it An Analysis of the 
Finite Element Method} (Prentice-Hall, Englewood Cliffs).

\bibitem{}
Stratmann, R. E., G. E. Scuseria, and M. J. Frisch,
1996, Chem. Phys. Lett. {\bf 257}, 213.

\bibitem{}
St\"{u}ben, K., and U. Trottenberg, 1982, 
in {\it Multigrid Methods}, ed. W. Hackbusch
and U. Trottenberg (Springer-Verlag, New York).

\bibitem{}
Sugawara, M., 1998, Chem. Phys. Lett. {\bf 295}, 423.

\bibitem{}
Szabo, A., and N. S. Ostlund, 1989, {\it Modern
Quantum Chemistry} (McGraw-Hill, New York).

\bibitem{}
Talman, J. D., and W. F. Shadwich, 1976, Phys. Rev. A
{\bf 14}, 36.


\bibitem{}
Tassone, F., F. Mauri, and R. Car, 1994, Phys. Rev. B
{\bf 50}, 10561.

\bibitem{}
Tomac, S., and A. Gr\"{a}slund, 1998, J. Comput. Chem. 
{\bf 19}, 893.

\bibitem{}
Tozer, D. J., and N. C. Handy, J. Chem. Phys. {\bf 109},
10180 (1998).

\bibitem{}
Troullier, N., and J. L. Martins, 1991a, Phys. Rev. B
{\bf 43}, 1993.

\bibitem{}
Troullier, N., and J. L. Martins, 1991b, Phys. Rev. B
{\bf 43}, 8861.

\bibitem{}
Tsonchev, S., R. D. Coalson, and A. Duncan, 1999,
Phys. Rev. E {\bf 60}, 4257.


\bibitem{}
Tsuchida, E., and M. Tsukada, 1995, Phys. Rev. B
{\bf 52}, 5573.

\bibitem{}
Tsuchida, E. and M. Tsukada, 1998, J. Phys. Soc. Jpn.
{\bf 67}, 3844.


\bibitem{}
Van Gisbergen, S. J. A., F. Kootstra, P. R. T. Schipper, 
O. V. Gritsenko, J. G. Shijders, and E. J. Baerends,
1998, Phys. Rev. A {\bf 57}, 2556.



\bibitem{}
Vanderbilt, D., 1990, Phys. Rev. B {\bf 41}, 7892.

\bibitem{}
Vasiliev, I., S. \"{O}\u{g}\"{u}t, 
and J. R. Chelikowsky, 
1997, Phys. Rev. Lett.
{\bf 78}, 4805.

\bibitem{}
Vasiliev, I., S. \"O\u{g}\"{u}t, 
and J. R. Chelikowsky, 
1999, Phys. Rev. Lett.
{\bf 82}, 1919.

\bibitem{}
Verwey, E. J., and J. Th. G. Overbeek, 1948, {\it Theory of 
the Stability of Lyophobic Colloids} (Elsevier, New York).

\bibitem{}
Vichnevetsky, R., 1981, {\it Computer Methods for Partial 
Differential Equations: Volume I} (Prentice-Hall, Englewood
Cliffs).

\bibitem{}
Vila, J. A., D. R. Ripoll, Y. N. Vorobjev, and H. A. Scheraga,
1998, J. Phys. Chem. B {\bf 102}, 3065.


\bibitem{}
Von Rague Schleyer, P., 1998, Editor, {\it Encyclopedia
of Computational Chemistry} (John-Wiley, New York).

\bibitem{}
Vosko, S. H., L. Wilk, and M. Nussair, 1980, Can. J. Phys.
{\bf 58}, 1200.

\bibitem{}
Walsh, A. M., and R. D. Coalson, 1994, J. Chem. Phys. 
{\bf 100}, 1559.

\bibitem{}
Wang, J., and T. L. Beck, 2000, J. Chem. Phys.
{\bf 112}, 9223 (2000).

\bibitem{}
Wang, J., and A. A. Stuchebrukhov, 1999, preprint.


\bibitem{}
Wesseling, P., 1991, {\it An Introduction to Multigrid
Methods} (John Wiley, New York).

\bibitem{}
White, C. A., B. G. Johnson, P. M. W. Gill, and M. Head-Gordon,
1996, Chem. Phys. Lett. {\bf 253}, 268. 

\bibitem{}
White, S. R., J.W. Wilkins, and M. P. Teter, 1989, Phys. Rev. B
{\bf 39}, 5819.

\bibitem{}
Wilson, K. G., 1990, Nucl. Phys. B (Proc. Suppl.) {\bf 17}, 82.


\bibitem{}
Yabana, K., and G. F. Bertsch, 1996, Phys. Rev. B {\bf 54}, 4484.

\bibitem{}Yabana, K., and G. F. Bertsch, 1997, Z. Phys. D {\bf 42},
219.


\bibitem{}Yabana, K., and G. F. Bertsch, 1999, Intl. J. Quantum Chem.
{\bf 75}, 55.


\bibitem{}
Yoon, B. J., and A. M. Lenhoff, 1990, J. Comput. Chem. 
{\bf 11}, 1080.

\bibitem{}
Yoon, B. J., and A. M. Lenhoff, 1992, J. Phys. Chem.
{\bf 96}, 3130.

\bibitem{}
You, T. J., and S. C. Harvey, 1993, J. Comp. Chem.
{\bf 14}, 484.

\bibitem{}
Yu, H., and A. D. Bandrauk, 1995, J. Chem. Phys.
{\bf 102}, 1257.

\bibitem{}
Yu, H., A. D. Bandrauk, and V. Sonnad, 1994, Chem. Phys. Lett.
{\bf 222}, 387.


\bibitem{}
Zacharias, M., B. A. Luty, M. E. Davis, and J. A. McCammon, 1992,
Biophys. J. {\bf 63}, 1280.

\bibitem{}
Zhang, J., 1998, J. Comput. Phys. {\bf 143}, 449.



\normalsize

\end{references}
\end{document}